\documentclass[a4paper,10pt]{fullarticle}
\usepackage[british]{babel}
\usepackage{csquotes}
\usepackage[algoruled,algosection,vlined,shortend,linesnumbered]{algorithm2e}
\usepackage{sciencestuff}

\input{arxiv.def}
\hypersetup{%
    pdftitle={Functional Renormalization for Signal Detection},
    pdfkeywords={%
        renormalization group, data analysis, signal detection, random matrix theory, information theory
        },
    pdfauthor={Riccardo Finotello and Vincent Lahoche and Dine Ousmane Samary},
    pdfsubject={signal detection}
}

\newtheorem{remark}{Remark}
\newtheorem{definition}{Definition}

\DeclareRobustCommand{\ipd}[1]{\partial_{#1}}  

\DeclareRobustCommand{\eaa}{EAA\xspace}        
\DeclareRobustCommand{\eft}{EFT\xspace}        
\DeclareRobustCommand{\iid}{i.i.d.\xspace}     
\DeclareRobustCommand{\ir}{IR\xspace}          
\DeclareRobustCommand{\kl}{KL\xspace}          
\DeclareRobustCommand{\lod}{LOD\xspace}        
\DeclareRobustCommand{\lpa}{LPA\xspace}        
\DeclareRobustCommand{\mpdistr}{MP\xspace}     
\DeclareRobustCommand{\rg}{RG\xspace}          
\DeclareRobustCommand{\uv}{UV\xspace}          
\DeclareRobustCommand{\wf}{WF\xspace}          

\DeclareRobustCommand{\frg}{FRG\xspace}        
\DeclareRobustCommand{\bbp}{BBP\xspace}        
\DeclareRobustCommand{\ai}{AI\xspace}          
\DeclareRobustCommand{\dof}{DOF\xspace}        
\DeclareRobustCommand{\pca}{PCA\xspace}        
\DeclareRobustCommand{\qft}{QFT\xspace}        
\DeclareRobustCommand{\snr}{SNR\xspace}        
\DeclareRobustCommand{\sota}{SOTA\xspace}      

\title{%
    Functional Renormalization for Signal Detection \\ \relax
    {\Large Dimensional analysis and dimensional phase transition for nearly continuous spectra effective field theory}
}

\author[1]{Riccardo Finotello\emailfoot{riccardo.finotello@cea.fr}}
\author[2]{Vincent Lahoche\emailfoot{vincent.lahoche@cea.fr}}
\author[3]{Dine Ousmane Samary\emailfoot{dine.ousmanesamary@cipma.uac.bj}}

\affil[1]{%
  Universit\'{e} Paris Saclay, CEA,
  \protect \\
  Service de G\'{e}nie Logiciel et de Simulation (SGLS),
  \protect \\
  Gif-sur-Yvette, F-91191, France
}
\affil[2]{%
  Universit\'{e} Paris Saclay, CEA,
  \protect \\
  Gif-sur-Yvette, F-91191, France
}
\affil[3]{%
  Facult\'{e} des Sciences et Techniques (ICMPA-UNESCO Chair),
  \protect \\
  Universit\'{e} d'Abomey-Calavi,
  \protect \\
  072 BP 50, Benin
}

\date{}

\begin{document}

\newgeometry{top=2cm,bottom=3cm}

\maketitle

\begin{abstract}
    Signal detection in high-dimensional data is a critical challenge in data science.
    While standard methods based on random matrix theory provide sharp detection thresholds for finite-rank perturbations, such as the known Baik-Ben Arous-P\'ech\'e (\bbp) transition, they are often insufficient for realistic data exhibiting nearly continuous (extensive-rank) signal distributions that merge with the noise bulk.
    In this regime, typically associated with real-world scenarios such as images for computer vision tasks, the signal does not manifest as a clear outlier but as a deformation of the spectral density's geometry.
    We use the functional renormalisation group (\frg) framework to probe these subtle spectral deformations.
    Treating the empirical spectrum as an effective field theory, we define a scale-dependent ``canonical dimension'' that acts as a sensitive order parameter for the spectral geometry.
    We show that this dimension undergoes a sharp crossover, interpreted as a ``dimensional phase transition'', at signal-to-noise ratios significantly lower than the standard \bbp threshold.
    This dimensional instability is shown to correlate with a spontaneous symmetry breaking in the effective potential and a deviation of eigenvector statistics from the universal Porter-Thomas distribution, confirming the consistency of the method.
    Such behaviour aligns with recent theoretical results on the ``extensive spike model'', where signal information persists inside the noise bulk before any spectral gap opens.
    We validate our approach on realistic image datasets, demonstrating that the \frg flow consistently detects the onset of this bulk deformation.
    Finally, we explore a formalisation of this methodology for analysing nearly continuous spectra, proposing a heuristic criterion for signal detection and a method to estimate the number of independent noise components based on the cyclic stability of these canonical dimensions.
\end{abstract}


\highlights{%
  We use the renormalization group to detect weak signals in nearly continuous spectra and give an agnostic definition of limit of detection.
}

\clearpage

\restoregeometry

\tableofcontents

\clearpage

\section{Introduction}

Complex system physics aims to extract the relevant features emerging from systems containing a very large number of interacting degrees of freedom (\dof)~\cite{Complex1}.
From this point of view, we can easily establish a direct connection with modern \emph{big data}-oriented data science, which tries to extract large-scale regularities.
It should not come as a surprise that techniques borrowed from complex systems and statistical physics have successfully been used in data science~\cite{DataPhy1}.
Moreover, the recent surge of interest in artificial intelligence (\ai) techniques has further strengthened the connections with statistical physics.
In the toolbox of physicists, the renormalisation group (\rg), introduced in the second part of the 1920s, is one of the powerful utilities used for discussing emergent phenomena and universality in the presence of a very large number of \dof.
The main feature of \rg is that microscopic (i.e.\ large momenta) \dof become irrelevant for the macroscopic (i.e.\ low momenta) scale as we coarse-grain the microscopic theory a large enough number of times.
In other words, only a small number of effective interactions survive macroscopically.
They correspond to the so-called \emph{relevant and marginal operators}, which are enough to describe long-range physics.
This is nowadays the simplest mechanism to explain the apparent simplicity of the effective macroscopic laws and their mysterious insensitivity to the microscopic details~\cite{Zinn1}.
A famous historical example is provided by the $\phi^4_d$ scalar field theory.
It well describes the Ising phase transition, without direct connection with the discrete nature of the fundamental degrees of freedom or direct symmetries of the lattice~\cite{Wilson1,Zinn1,Zinn2}.
It could even be that all the fundamental laws of physics are only effective theories, masking a microscopic reality not only unknown but largely irrelevant at the energy scale of current experiments.

The objective of this paper is to harness the \rg as a robust tool for data analysis in complex scenarios characterised by ``extensive-rank'' signals, that is nearly continuous spectra, where the macroscopic \dof are not isolated (i.e.\ low-rank) spikes.
Standard techniques such as Principal Component Analysis (\pca) or outlier detection based on the Baik-Ben Arous-P\'{e}ch\'{e} (\bbp) transition~\cite{math3} excel in the ``spiked covariance'' regime, where the signal consists of a few isolated eigenvalues separated from the noise bulk.
However, in many real-world applications, such as hyperspectral imaging, biological networks, or financial correlations, the signal is not sparse but extensive (proportional to the system size) and merges with the noise bulk, rendering standard outlier detection ineffective.
Our goal is to show that the \rg flow, specifically the running of canonical dimensions, can detect these subtle deformations of the spectral geometry even when no spectral gap opens.

In statistical physics, the \rg coarse-graining procedure reveals that most microscopic details are irrelevant, and the macroscopic behaviour is governed by a few relevant operators.
We apply this principle to data analysis by treating the empirical spectrum in the framework of an effective field theory (\eft).
Just as the $\phi^4$ theory describes the Ising phase transition regardless of the lattice details, the effective potential of our data-driven field theory captures the universal properties of the noise, typically in the Marchenko-Pastur (\mpdistr) universality class.
A signal, in this framework, acts as a perturbation that drives the system away from this Gaussian fixed point.
Crucially, we identify signal detection not with an outlier eigenvalue, but with a ``dimensional phase transition'': a scale where the relevant operators (specifically the scaling dimension of the coupling) deviate from their noise-dominated fixed-point values.

This work continues the programme initiated in~\cite{RG1,RG2,RG3,RG4,RG5,RG6} to establish a field-theoretic framework for signal detection.
In~\cite{RG5}, the foundations of this approach were laid by deriving the \frg flow equations for a general empirical spectrum and showing that the presence of information modifies the relevance of interactions.
Here, we significantly refine this analysis by focusing on the quantitative behaviour of the ``canonical dimension'' as a function of the signal-to-noise ratio (\snr).
We show that this dimension acts as an order parameter for the spectral geometry.
For pure noise, the dimension remains stable (rigid) and close to the Gaussian value.
In the presence of an extensive signal, this rigidity breaks down at a critical \snr, leading to a sharp crossover.
This allows us to define a detection threshold based on the stability of the \rg flow rather than the position of the largest eigenvalue, typical of \bbp detection.
The numerical evidence presented in this work serves to validate this theoretical framework, specifically confirming the presence of a dimensional phase transition correlated with the signal strength.
We also present extensive proof of consistency by looking at the evolution of the Wilson-Fisher (\wf) fixed point and the statistics of eigenvectors, which deviate from the universal Porter-Thomas distribution in the presence of a signal.

This dimensional perspective offers a dual description to the recent random matrix theory results on the ``extensive spike model''~\cite{Landau2023}, where the authors demonstrated that for extensive-rank signals, the spectrum undergoes a transition from a ``unimodal'' noise bulk to a ``bimodal connected'' phase before eventually separating into disconnected bulks.
Most importantly, they showed that signal information is statistically recoverable even in the unimodal phase, where standard \bbp detection fails due to the absence of outliers.
Our \frg analysis detects exactly this regime: the dimensional crossover corresponds to the onset of the bulk deformation (the bimodal connected phase), enabling detection at \snr levels significantly lower than the standard \bbp threshold.
In this sense, the \rg provides a natural probe for the internal geometry of the bulk, effectively sensing the information that remains hidden to outlier-based methods.

The article is organised as follows:

\begin{itemize}
    \item Section~\ref{sec:related-work} contextualises our contribution within the state-of-the-art (\sota), explicitly contrasting the finite-rank (\bbp) and extensive-rank (Landau) regimes, and reviewing the connection between \rg and data analysis.
    \item Section~\ref{sec:renorm_group} introduces the \frg formalism, defining the effective average action, the flow equations, and the operational definition of the canonical dimension as our primary observable.
    \item Section~\ref{sec:data_analysis} details the numerical methodology, including the treatment of the empirical spectrum and the robust integration of the flow equations.
    \item Section~\ref{sec:numeric} presents the core results: the dimensional phase transition, the symmetry breaking in the effective potential, and the deviation of eigenvector statistics from the Porter-Thomas distribution, all validated on realistic image datasets.
    \item Section~\ref{sec:conclusions} summarises the findings and proposes a heuristic criterion for estimating the number of independent noise components based on the cyclic stability of the flow.
\end{itemize}

\section{Related Work and Novel Contribution}\label{sec:related-work}

Modern machine learning methods have to deal with increasingly complex data structures, characterised by non-trivial interactions between a very large number of \dof.
One research area strongly impacted by these developments is computer vision, driven by the ubiquity of high-definition and multi-camera systems (large number of pixels), as well as spectral imaging for engineering and scientific applications (more information per pixel).
These data types represent a significant challenge for signal detection in analytical science, as the very notion of ``signal'' can become ill-defined in the presence of complex noise structures.
While researchers often employ methods originally developed for ``spiked covariance models'', where the signal consists of (finite) low-rank perturbation of the noise, most realistic applications do not fall into this simple regime.
Instead, they exhibit what we may call \emph{nearly continuous spectra} (or extensive-rank signals), where the information is distributed across a macroscopic fraction of the eigenvalues and merges with the noise bulk.
This mismatch is hinted at by the frequent failure of standard \pca to provide a clean separation between signal and noise~\cite{PCA1}, particularly for data exhibiting such quasi-continuous spectra.
For instance, \pca is widely adopted in hyperspectral imaging and spectroscopy to reduce dimensionality and extract relevant features~\cite{Moncayo:ExplorationMegapixelHyperspectral:2018, Finotello:HyperPCAPowerfulTool:2021}, though its limitations in lower \snr regimes are well-documented.
Similar challenges arise in finance, where the exploitation of large datasets is critical and the reliability of analysis methods remains a constant concern.
In their seminal article~\cite{Bouchaud1,Bouchaud2}, the authors developed a theory of ``dressed noise'', seeking to reconstruct the noise using random matrix theory predictions, and concluded with pessimistic remarks on the efficiency of historical covariance spectra and Markowitz's mean-field theory in that context~\cite{Marko}.
Our approach aims to improve upon these conclusions using \rg arguments, specifically targeting the regime where the dimension of the data is large and the complexity of the signal makes traditional techniques fail.

Note that, on the mathematical side, finding a rigorous signal detection threshold for spiked models remains an open research question~\cite{math1,math2,math4,math5}.
Additionally, datasets formed by tensors (e.g.\ hyperspectral images or multivariate time series), rather than matrices, pose many mathematical challenges related to spin-glass physics~\cite{seddik2024random}.
From a technical perspective, the \sota approach for high-dimensional and complex data thus remains \pca~\cite{PCA1}.
The algorithm relies on the spectral decomposition of the empirical correlation matrix to identify relevant \dof.
The matrix is constructed from the covariance $C$, in turn built using the mean-shifted data matrix $X \in \mathbb{R}^{N \times P}$, where $N$ is the sample size and $P$ the number of independent variables (often called \emph{features} in data science):
\begin{equation}
    C \coloneqq \frac{1}{N-1} X^T X \in \mathbb{R}^{P \times P}.
    \label{eq:C0def}
\end{equation}
The entries $C_{ij}$ are defined as:
\begin{equation}
    {C}^{(\text{corr})}_{ij} = \frac{C_{ij}}{\sqrt{C_{ii} C_{jj}}}.
\end{equation}
Provided that the signal is sufficiently strong, the eigenvectors of $C^{(\text{corr})}$ (or $C$) associated with large eigenvalues are expected to align with the principal directions in the data, which are then used to project the data onto a lower-dimensional space spanned by these eigenvectors.
However, the performance of \pca critically depends on the presence of well-separated eigenvalues (spikes) corresponding to the signal, which needs to be low-rank compared to the noise bulk.
This scenario is typically modelled by the so-called \emph{spiked covariance model}:
\begin{equation}
    X = \sum\limits_{k = 1}^{r} \beta_k u_k v_k^T + Z,
    \label{eq:spiked_model}
\end{equation}
where $r \ll \min(N, P)$ is the rank of the signal, $\beta_k$ represents the signal strength, $u_k \in \mathbb{R}^N$ is a uniformly distributed unit vector, $v_k \in \mathbb{R}^P$ is the vector representing the signal direction (i.e.\ the spike), and $Z \sim \mathcal{N}(0, \mathbb{I})$ is a random matrix with \iid entries following a standard normal distribution.
The \bbp transition~\cite{math3} then provides a theoretical benchmark threshold for signal detection, stating that an outlier eigenvalue emerges from the noise bulk only when the signal strength exceeds a critical value $\beta_c^2 = \sqrt{q}$, where $q = P / N$.
Notice that discreteness of the sum over the low-rank components is the key feature allowing for outlier detection via the \bbp transition.

\begin{figure}[t]
    \centering
    \includegraphics[width=0.45\textwidth]{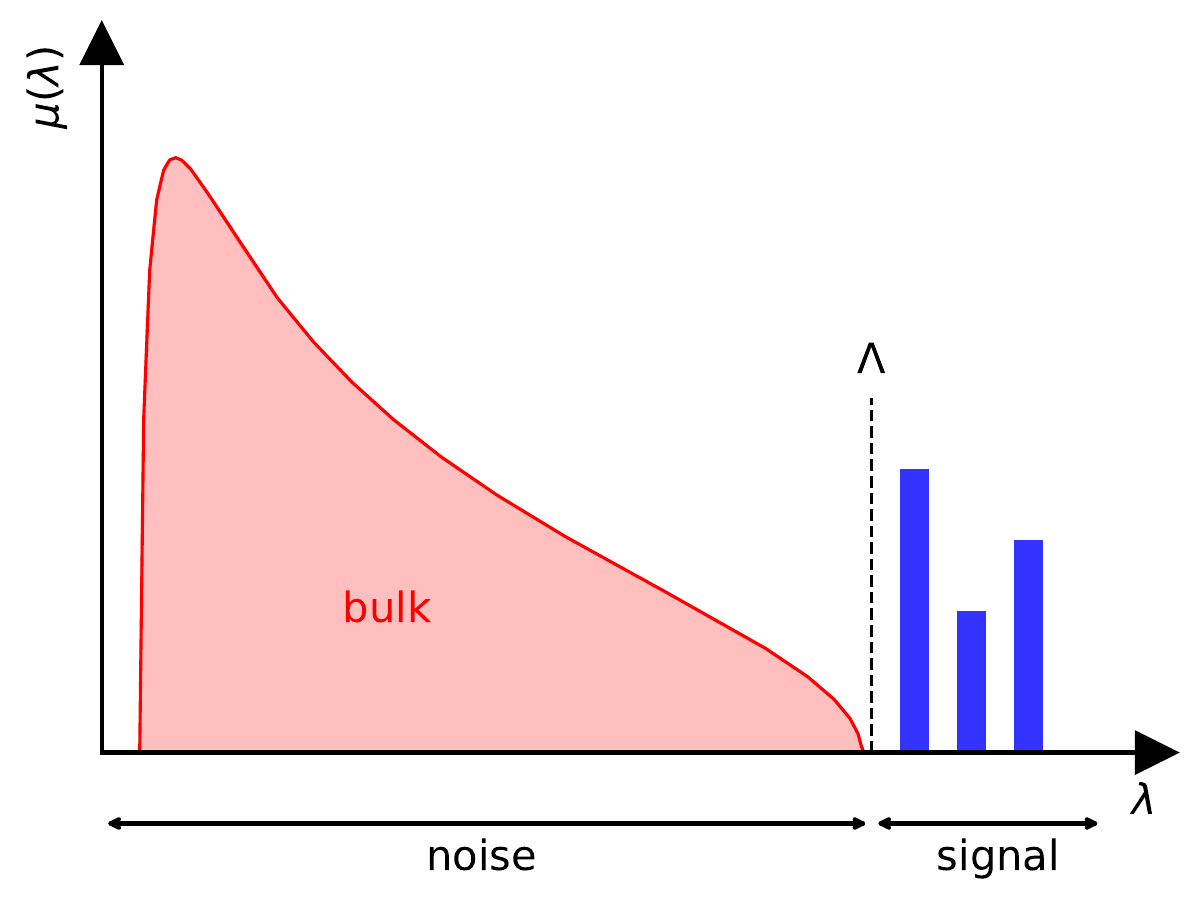}
    \hfill
    \includegraphics[width=0.45\textwidth]{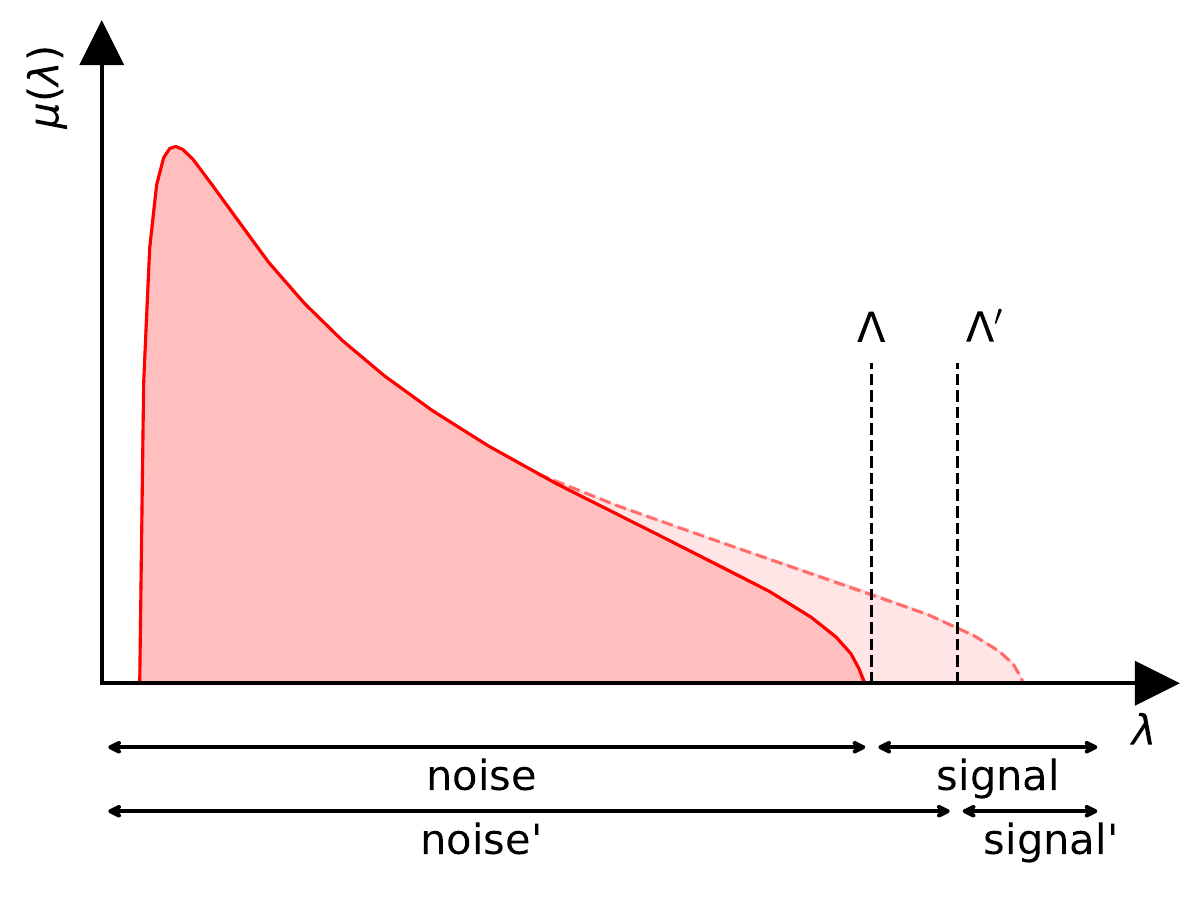}
    \caption{%
        (left) Empirical spectra can exhibit some localised spikes out of a bulk (i.e.\ noise, in red) made of non-localised eigenvectors (i.e.\ \emph{low-rank} relevant information, in blue), in which case the cut-off $\Lambda$ provides a clean separation.
        (right) For nearly continuous spectra (more realistic data), the position of the cut-off $\Lambda$ is difficult to define.
    }\label{fig:nearly_continuous_spectra}
\end{figure}

For the extensive-rank regime typical of the complex data described above, the situation is both quantitatively and qualitatively different.
Recent work on the ``extensive spike model''~\cite{Landau2023} has shown that as the signal strength increases, the spectrum does not simply eject an outlier.
Instead, the noise bulk itself undergoes a topological transition from a unimodal to a ``bimodal connected phase'', where a (nearly) continuous distribution of signal eigenvalues emerges from the bulk, as shown in Figure~\ref{fig:nearly_continuous_spectra}.
The authors demonstrated that signal recovery is possible even in the unimodal phase, implying that the true limit of detection lies below the spectral gap opening predicted by standard \bbp theory.
This provides a theoretical foundation for our \frg approach: by probing the geometry of the bulk rather than searching for outliers, we aim to detect the onset of this bulk deformation, which serves as the field-theoretic dual to the bimodal phase transition.

The failure of standard \pca to provide a clean separation between high- and low-variance components in continuous distributions suggests a connection with other approaches, such as the application of the \rg in information theory~\cite{Ben1,Ben2}.
The \rg is, in fact, able to distinguish the nature of the spectral modes depending on whether they are likely to be classified as noise or information, given the universal properties of the former.
Moreover, in ordinary quantum field theory (\qft), it is well-known that, given the form of the interactions, the distribution of momenta $\rho\qty(p^2)$ determines the relevance or irrelevance of the couplings.
For instancem and in a standard setting of a field theory defined on a Euclidean background of dimension $D$, the momentum distribution scales as $\rho\qty(p^2) \propto {\qty(p^2)}^{\frac{D-2}{2}}$.
However, we do not strictly need a background space-time to define a non-trivial notion of scale and, ultimately, of ``relevance''.
In our context, we can instead investigate whether a given distribution supports relevant interactions by relating it to the matrix universality classes that govern the large-scale behaviour of an arbitrary field whose correlations are fixed by the empirical correlation matrix.
Although this approach may seem surprising, similar ideas appear in field theories for background-independent quantum gravity~\cite{LahocheBeyond}, where the spectra dimension effectively replaces the geometric one by providing the necessary scaling laws for the renormalisation flow.
However, the definition of the \dof and the field used to construct the \rg remain to be addressed.
In a recent series of articles~\cite{RG0,RG1,RG2,RG3,RG4,RG5,RG6}, we proposed to consider an estimate of the maximum entropy distribution, constrained by the empirical distribution of correlations, at least in the tail of the spectrum.
By focusing mainly on spectra close to the \mpdistr distribution, we observed two key properties:
\begin{itemize}
    \item the presence of a signal in the spectrum modifies the relevance of the interactions (interactions become less relevant as the signal strength increases), thus significantly altering the flow behaviour in the vicinity of the Gaussian fixed point;
    \item this effect is accompanied by a spontaneous breaking of the $\mathbb{Z}_2$ reflection symmetry of the effective theory.
\end{itemize}
In turn, this provides an interesting physical justification for the effectiveness of Gaussian process in data modelling: as the signal strength increases, the effective field theory naturally flows towards a non-interacting limit, validating the Gaussian assumption at the basis of these methods.
It also suggests a physical interpretation of the failure of Gaussian processes in low \snr regimes, where interactions become relevant again.

Building on the results of the programme initiated in~\cite{RG1,RG2,RG3,RG4,RG5,RG6}, this article refines the detection mechanism by quantifying the stability of the \rg flow and establishing precise thresholds for signals merging with the noise bulk.
Our specific contributions are the following:
\begin{enumerate}
    \item we define precise detection thresholds based on the stability of the canonical dimensions: the limit of detection (\lod) $\beta_t$ (where the spectral rigidity breaks down), the critical threshold $\beta_c$ (where the canonical dimension of $u_4$ vanishes), and the optimal threshold $\beta_O$ (the first dimensional minimum). We show that these thresholds detect signals at signal-to-noise ratios significantly lower than the standard \bbp limit. This sensitivity confirms the bulk deformation predicted by the extensive spike model, effectively providing a field-theoretic dual description of the topological phase transition highlighted in~\cite{Landau2023};
    \item we characterise the specific nature of this transition using established theoretical arguments, such as the stability of the Wilson-Fisher fixed point, the contraction of the symmetric phase, and the variance of the eigenvector distribution. We complement these arguments with extensive numerical simulations, confirming the spontaneous symmetry breaking of the effective potential and the deviation from the universal Porter-Thomas distribution;
    \item we propose a model-agnostic definition of the \lod based on the \rg flow. While \lod is well-defined for univariate calibration models~\cite{Zorn1999, Thomsen2003, Armbruster2008}, its definition for multivariate analysis usually requires complex heuristics~\cite{Bauer1991, SINGH1993205, BOQUE1999397, BOQUE200041, ostra2008detection}; we show that the \frg provides a physically grounded multivariate \lod that remains valid even when the signal is inextricably mixed with the noise bulk;
    \item we introduce a novel notion of distance between spectral distributions, based on the canonical dimensions of the effective field theory. This metric offers a quantitative measure of the deviation from the random matrix universality class, complementing standard information-theoretic measures like the Kullback-Leibler divergence;
    \item we propose a heuristic criterion for estimating the number of independent noise components in complex datasets, exploiting the cyclic stability of the \rg flow that emerges in the presence of multiple confounding sources.
\end{enumerate}

\section{Renormalisation Group for Data Analysis}\label{sec:renorm_group}

The theoretical framework of our analysis is the \frg, applied to an \eft constructed directly from the data.
To address the challenge of signal detection in nearly continuous spectra, we construct this \eft by leveraging the maximum entropy principle~\cite{Jaynes1}.
Specifically, we seek the most general probability distribution for an auxiliary field $\varphi$ that reproduces the observed empirical spectrum $\rho(\lambda)$ as its 2-point correlation function, while remaining unbiased regarding higher-order correlations~\cite{RG5}.
This approach ensures that the auxiliary field possesses the same spectral properties as the data, grounding the theory in the actual correlations of the system.

This construction is fully consistent with the modern perspective on field theory established since the introduction of the \rg~\cite{Zinn2,Wilson1}: a field theory is effectively a tool for reproducing relevant long-range correlations between macroscopic degrees of freedom, where the coupling constants quantify the intensity of these correlations.
From this perspective, our approach is a direct application of the \eft formalism as a statistical inference framework, designed to reconstruct the probability distribution governing the data from empirical observations.
We note that this conceptual duality between field theory and statistical learning is not unprecedented, as similar ideas have notably appeared in the correspondence between neural networks and quantum field theory (\textsc{nn-qft}) proposed in~\cite{NNQFT1,NNQFT2,NNQFT3,NNQFT4}.

This work continues the programme initiated in the series of articles~\cite{RG1,RG2,RG3,RG4,RG5,RG6}, which introduce and explain the details of this procedure at length.
We refer the reader to those works, and particularly to the review~\cite{RG5}, for the full mathematical proofs and rigorous justification of the formalism.
In this section, we provide a sketch of the derivation for the sake of self-consistency, establishing the operational definitions required for our analysis.

A central component of this construction is the identification of the appropriate Gaussian fixed point, which serves as the reference ``null hypothesis'' for our signal detection framework: the universality class of the noise determines the reference scaling against which signal-induced deviations are measured.
For the extensive-rank covariance matrices considered here, this universal background is described by the \mpdistr distribution~\cite{Bouchaud3}:
\begin{equation}
    \mu_{\sigma^2,q}\qty(\lambda)
    \coloneqq
    \frac{1}{2 \pi \sigma^2 q}
    \frac{\sqrt{\qty(\lambda_+-\lambda)\qty(\lambda-\lambda_-)}}{\lambda},\label{MPdistribution}
\end{equation}
where $\sigma^2$ is the variance of the \iid entries of the matrix $X \in \mathbb{R}^{N \times P}$, $q = P/N$, and $\lambda_\pm = \sigma^2 {(1 \pm \sqrt{q})}^2$.
The \mpdistr distribution serves as the Gaussian fixed point of our theory.
Our argument relies on the concept of universality: just as the \eft for the critical Ising model captures the universal physics of phase transitions regardless of the microscopic lattice structure, the definition of this field theory is agnostic to the true nature of the data, provided the noisy degrees of freedom remain in the basin of attraction of a universal random matrix law.
The same field theory will thus be able to detect the presence of a signal across all nearly continuous spectra in the vicinity of the same universality class, identified through the scale-dependent running of the canonical dimensions.

\subsection{Field theory framework}\label{fieldtheory}

\begin{figure}[t]
    \centering
    \includegraphics[width=0.9\textwidth]{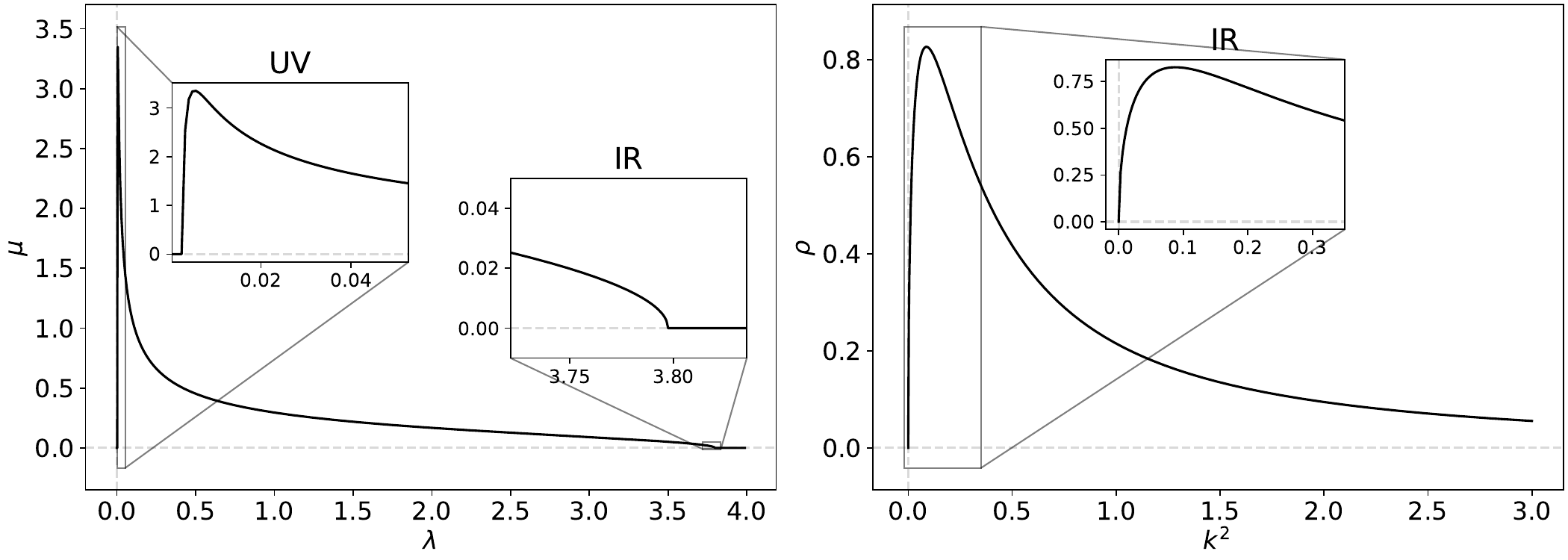} \\
    \includegraphics[width=0.9\textwidth]{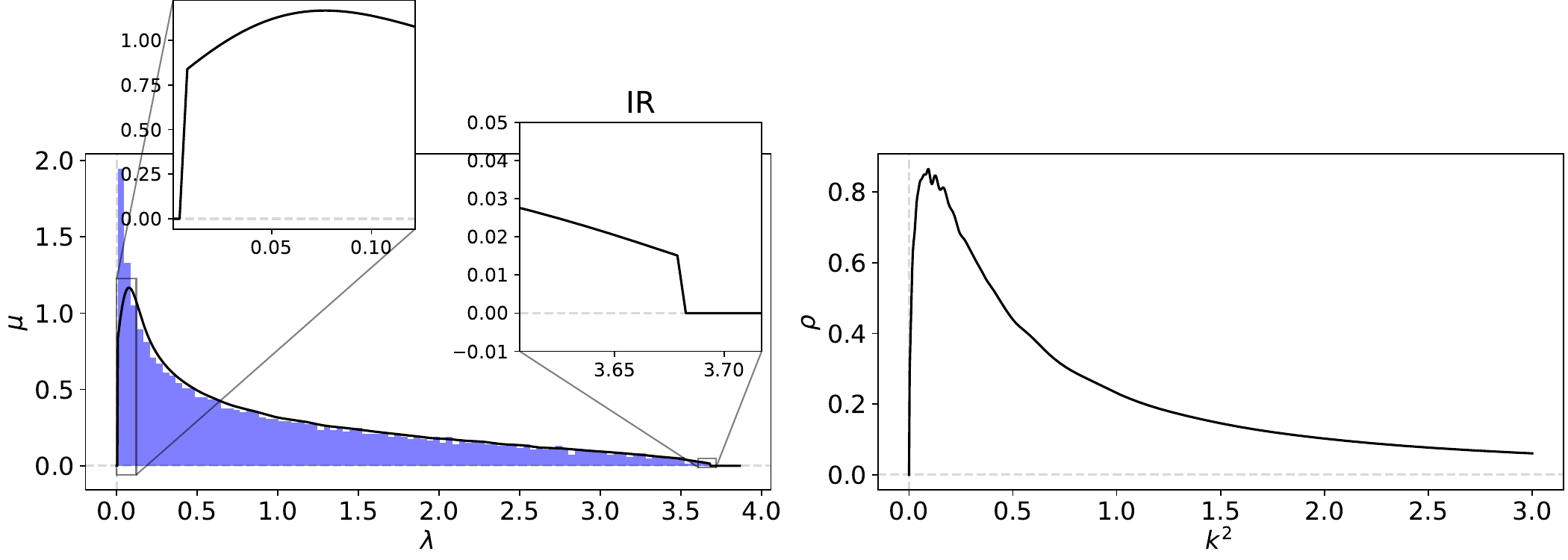}
    \caption{%
        Deep \ir and deep \uv definitions of the eigenvalue distribution (left) and of the momenta distribution (right).
        The analytic \mpdistr distribution is shown on top, some empirical distribution for a modest number of \dof ($N = 2500$, $q = 0.9$) at the bottom.
        The black line is the numerical interpolation used to construct the empirical inverse distribution.
    }\label{fig:figtail}
\end{figure}

\begin{figure}[t]
    \centering
    \includegraphics[width=0.66\textwidth]{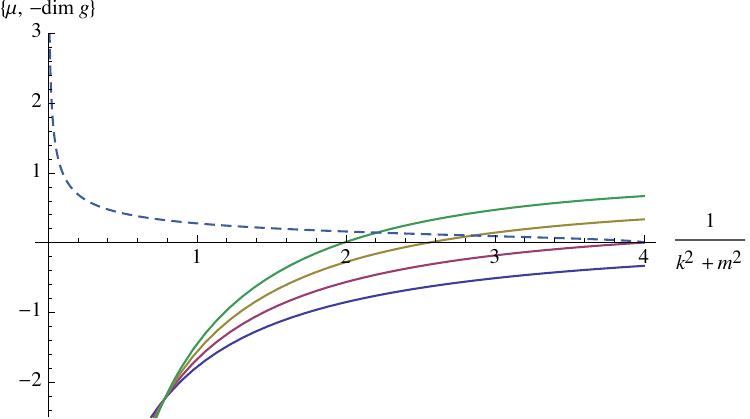}
    \caption{%
        Behaviour of the canonical dimensions for the \mpdistr distribution with $\sigma^2 = q = 1$ (dashed curve).
        We plotted the behaviour of the canonical dimension for $n = 2$ (blue curve), $n = 3$ (purple curve), $n = 4$ (yellow curve) and $n = 5$ (green curve).
    }\label{fig:canonical_dim_mp}
\end{figure}

The \eft looks like an ordinary \emph{equilibrium} Euclidean field theory for some field $\varphi(p)$, which depends on a nearly continuous variable $p \in \mathbb{R}$, with partition function path integral:
\begin{equation}
    Z[j]
    =
    \int \qty[\dd{\varphi}]\,
    \exp \qty(- S\qty[\varphi]+ \sum_p j(-p) \varphi(p) ),
\end{equation}
where the classical action reads:
\begin{equation}
    S\qty[\varphi]
    =
    \frac{1}{2} \sum_p \varphi(p) \qty(p^2 + u_2) \varphi(-p) + U\qty[\varphi],
    \label{eq:classical_action}
\end{equation}
and where $U\qty[\varphi]$ expands in powers of fields and monomials look like ordinary local interaction in momentum space:
\begin{equation}
    U\qty[\varphi]
    =
    \sum_{n=2}^\infty\,
    \frac{u_{2n}}{(2n)! P^{n-1}}\, \sum_{\qty{p_1,\dots,p_n}}
    \delta_{0,\sum_{i=1}^n p_i} \prod_{i=1}^{2n} \varphi(p_i),
    \label{eq:localpotential}
\end{equation}
where $\delta$ is the standard Kronecker delta.
Notice that, with this notation, $u_2 \equiv m^2$ is the bare mass parameter.
In this context, ``nearly continuous'' means that for $N$ and $P$ large enough but finite, only $P$ values are allowed for $p^2$.
They are distributed according to some (\emph{a priori} unknown) distribution $\rho(p^2)$.
Moreover, we assume that $\rho$ converges weakly toward some continuous distribution in the limit $P \to \infty$.
The parameters involved in the definition of the classical action are fixed by the inference condition.
In particular, we impose that the $2$-points correlation function
\begin{equation}
    G(p^2)
    \coloneqq
    \left\langle \varphi(p) \varphi(-p) \right\rangle
    =
    \frac{1}{Z\qty[0]} \int \qty[\dd{\varphi}] e^{-{S}\qty[\varphi]}\, \varphi(p)\varphi(-p),
\end{equation}
matches with the empirical correlation matrix in its eigenbasis.
For the Gaussian model, the correspondence is simply:
\begin{equation}
    \frac{1}{p_{\mu}^2+u_2} = \lambda_\mu,
    \label{eq:def_p2}
\end{equation}
which shows the relation between the generalised momenta $p_{\mu}$ and the empirical eigenvalues $\lambda_\mu$, where $\mu = 0, 1, \dots, P - 1$ and $\lambda_0 \ge \lambda_1 \ge \dots \ge \lambda_{P-1}$.
We define the mass $m^2$ as the inverse of the largest eigenvalue $m^2 = \lambda_0^{-1}$.
The corresponding momenta distribution $\rho_{\text{G}}(p^2)$ can be deduced from the corresponding empirical eigenvalue distribution $\mu_{\text{em}}(\lambda)$.
\begin{remark}
    Note that the definition~\eqref{eq:def_p2} assumes a canonical definition of the concept of ultraviolet (\uv), i.e.\ $p^2 \gg 1$, corresponding to bulk noise, and infrared (\ir), for $p^2 \approx 0$, describing the large eigenvalues.
    The corresponding regions are highlighted on Figure~\ref{fig:figtail}, both for the analytic MP law and for some typical Gaussian realisation.
\end{remark}
A perturbative analysis shows that the Gaussian theory is however unstable.
Figure~\ref{fig:canonical_dim_mp} shows the canonical dimensions (i.e.\ the dominant behaviour of the flow, at linear order, around the Gaussian fixed point) of a \mpdistr distribution of unit variance: asymptotically, in the low energy (large eigenvalues) region, the quartic coupling is relevant.

Beyond the Gaussian theory, the bare propagator receives quantum corrections, which can be absorbed by Dyson's resummation formula:
\begin{equation}
    \begin{split}
        G(p^2)
        & =
        \frac{1}{p^2+u_2}
        +
        \frac{1}{p^2+u_2} \Sigma(p^2) \frac{1}{p^2+u_2}
        +
        \cdots
        \\
        & =
        \frac{1}{p^2+u_2-\Sigma(p^2)},
        \label{eq:quantum_corrections_propa}
    \end{split}
\end{equation}
where $\Sigma(p^2)$ is the \emph{self-energy} function, which encodes all loop corrections to the 2-point function.
In general, the contribution of $\Sigma(p^2)$ is non-trivial and the inference problem $G(p_{\mu}^2) = \lambda_\mu(p_{\mu}^2)$ becomes hard to solve exactly.
Moreover, we expect $\rho(p^2) \neq \rho_{\text{G}}(p^2)$ in general.
Yet, everything simplifies as we focus on the tail of the spectra, in the \ir regime where $p^2 \ll 1$.
In this regime, the standard derivative expansion and local potential approximation (\lpa) works well enough.
We can then use:
\begin{equation}
    G(p^2) \approx \frac{1}{Z\, p^2+m^2_{\text{eff}}},
    \label{eq:field_strength_renorm}
\end{equation}
where $Z \coloneqq 1 - \Sigma^\prime(0)$ is the \emph{field strength renormalization} and $m^2_{\text{eff}} \coloneqq m^2-\Sigma(0)$ the \emph{effective mass}.
In the strict \lpa, the field strength effect can be ignored, as it has been explicitly checked in~\cite{RG2,RG5}.
We then recover exactly the Gaussian correspondence up to a global translation of the mass:
\begin{equation}
    m^2 - \Sigma(0) = \lambda_0^{-1},
\end{equation}
and $\rho_G(p^2) \approx \rho(p^2)$.

\subsection{Functional renormalisation group and local potential approximation}\label{sec:lpa}

Of all the incarnations of Wilson's original idea of \rg, the functional approach is the most practically useful in field theory.
The formalism developed by Wetterich and Morris~\cite{Wett,Morris,Delamotte}, called \emph{effective average action} (\eaa), focuses on effective actions and lends itself better to non-perturbative investigations.
The starting point of the \eaa is to modify the classical action~\eqref{eq:classical_action} adding a scale dependent mass:
\begin{equation}
    \Delta S_k\qty[\varphi]
    =
    \frac{1}{2}\, \sum_p \varphi(p) R_k(p^2) \varphi(-p).
\end{equation}
Let $Z_k[j]$ be the corresponding partition function and $W_k[j] \coloneqq \ln Z_k[j]$ the self energy, the \eaa $\Gamma_k[M]$ is defined as:
\begin{equation}
    \Gamma_k[M] + \Delta S_k[M] = \sum_p j(p) M(-p)-W_k[j],
\end{equation}
where $M(p)$ is the \emph{classical field} in the presence of the source $j(p)$:
\begin{equation}
    M(p) \coloneqq \frac{\partial W_k}{\partial j(-p)}.
\end{equation}
The \emph{regulator} $R_k(p^2)$ is designed such that low momentum modes (with respect to the cut-off $k \in [0,+\infty)$) acquire an effective large mass and decouple from long range physics.
On the contrary, high momentum modes are integrated out.
$\Gamma_k$ looks like the effective action for ``microscopic modes'', with momenta higher than $k$.\footnote{%
    Note that, in contrast with the popular Wilson-Polchinski approaches, the \uv cut-off remains fixed in this approach, and can be surely sent to infinity in general, because flow equations involve only a single loop performed on a restricted window of the momenta.
}
More precisely, $R_k$ is designed such that $\Gamma_k$ interpolates between the classical action and the full effective action $\Gamma$ (the Legendre transform of the free energy $W \coloneqq \ln Z$):
\begin{equation}
    \Gamma_{k=0} = \Gamma,
    \qand
    \Gamma_{k \to \infty} \to S.
\end{equation}
In this article, we use the well-known Litim regulator~\cite{Litim}:
\begin{equation}
    R_k(p^2) \coloneqq \qty(k^2-p^2) \theta\qty(k^2-p^2),
    \label{eq:litim_reg}
\end{equation}
which has the advantage of decoupling the mass from the computation of the canonical dimensions (more details in what follows).
The equation describing how the \eaa changes as $k$ evolves is the \emph{Wetterich equation}:
\begin{equation}
    \dot{\Gamma}_k
    =
    \frac{1}{2}\sum_p\, \dot{R}_k(p^2)\, {\qty(\Gamma^{(2)}_k+R_k)}^{-1}(p,-p),
    \label{eq:wett}
\end{equation}
where $\Gamma^{(2n)}_k$ designates the $2n$-th functional derivative of $\Gamma_k$ with respect to the classical field and the ``dot'' is the derivative with respect to $t \coloneqq \ln k$.
Note that, in this discrete context, $p$ and $k$ are effectively dimensionless.
The Wetterich equation is exact, but cannot be solved exactly in general.
Approximations, usually called ``truncations'', are required.
Notice also that by definition:
\begin{equation}
    \Gamma_{k=0}^{(2)}(p=0)
    \equiv
    m^2_{\text{eff}} = \lambda_0^{-1}.
\end{equation}

Focusing on the \lpa, the truncation we choose for $\Gamma_k$ reads:
\begin{equation}
    \Gamma_k[M]
    \coloneqq
    \frac{1}{2}\sum_p M(p) (p^2+u_2(k)) M(-p) + \mathcal{U}_k[M],
\end{equation}
where boundary conditions are such that $u_2(k \to \infty) = m^2$ and $\mathcal{U}_{k \to \infty}[M] = U[M]$.
Furthermore, the potential $\mathcal{U}_k[M]$ is assumed to be local according to the definition above.
In the strict \lpa, the derivative expansion neglects expansions in power of $p^2$ beyond the leading order in the Gaussian term, and the classical field is assumed to reduce to its $0$-th component $M(p) \approx M \delta_{0,p}$.
This way, the flow equation for $\mathcal{U}_k[M]$ can be deduced from~\eqref{eq:wett}.
For the sake of brevity, we skip the details of the derivation, which can be found in~\cite{RG5}, under the assumptions of strict \lpa and the field strength renormalisation discussed in~\eqref{eq:field_strength_renorm}.
In the continuum limit, we get:
\begin{equation}
    \dot{\mathcal{V}}_k\qty[\chi]
    =
    \frac{1}{2} \int \dd{p^2}\,
    \rho(p^2)\, \dot{R}_k(p^2)
    \frac{k^2}{k^2 + \ipd{\chi} \mathcal{V}_k^\prime\qty(\chi) + 2\chi \mathcal{V}^{\prime\prime}\qty(\chi)}.
\end{equation}
where $N \chi \coloneqq M^2 / 2$ and $\mathcal{V}_k\qty(\chi) \coloneqq \eval{\mathcal{U}_k[M]}_{M^2 = 2 N \chi}$.
Usually, the \rg assumes a global rescaling of the lattice scale before partial integration, and for this reason, it is suitable to work with dimensionless quantities.
In this context, however, there are no dimensions at all.
A notion of dimension can emerge from the behaviour of the flow equation~\cite{RG1,LahocheBeyond}.
Indeed, flow equations for local couplings entering the definition of the effective potential $\mathcal{U}_k[M]$ involve single loops, which, because of the choice of the regulator, requires the integral:
\begin{equation}
    L(k) \coloneqq \int_0^k\, \dd{p}\, p\, \rho(p^2).
\end{equation}

In standard field theory, $\rho(p^2)$ is a power law, and $L(k)$ is essentially a power of $k$.
Hence, the different couplings can be rescaled by a suitable power of $k$ such that flow equations turn out to be an autonomous system.
In this context, $\rho(p^2)$ is not a power law, and the best compromise is to rescale couplings such that the $k$ dependence is relegated to the linear term in the flow equation, starting from ($\bar{X}$ denotes the dimensionless version of the quantity $X$):
\begin{equation}
    u_2 \coloneqq k^2 \bar{u}_2 \Rightarrow \dim_{\tau}\qty(u_2) = 2 \dv{\ln k}{\tau} = 2 t^{\prime},
\end{equation}
where $t^\prime \coloneqq \dv*{t}{\tau}$ and $\tau \coloneqq \ln(L(k))$ (we denote everywhere derivatives with respect to $\tau$ with a ${}^{\prime}$ symbol).
Thus, the linear term defines an intrinsic notion of (scale dependent) dimension.
Using the definitions of the flow equations for the couplings $u_{2n}$ defined from the expansion of the potential, we can deduce the corresponding dimensions.The inrested reader can check~\cite{RG5} for the details of the derivation.
From the point of view of the dimensions, we get:
\begin{equation}
    \dim_{\tau}\qty(u_4)
    =
    -2\qty(\frac{t^{\prime\prime}}{t^\prime}+t^\prime\qty(\frac{1}{2} \dv{\ln \rho}{t} - 1)),
\end{equation}
and, in general,
\begin{equation}
    \dim_{\tau}\qty(u_{2n}) = -\qty(n-2) \dim_{\tau}(u_2) + \qty(n-1) \dim_\tau(u_4).
\end{equation}
Notice that these definitions assume to use $\tau$ rather than $t$ as parameter of the flow and that the scale dependency of canonical dimensions is not a specificity of this approach.
Such an unconventional property has been recovered in a different context recently~\cite{Ben1}.
Figure~\ref{fig:canonical_dim_mp} shows the behaviour of the corresponding canonical dimension for a \mpdistr distribution: following the usual definition, a coupling is said to be \emph{relevant} (i.e.\ increases toward \ir scales) as the dimension is negative, and it is otherwise called \emph{irrelevant}.
Dimensions for $\mathcal{V}_k$ and $\chi$ can be easily deduced:
\begin{equation}
    \dim_{\tau}\qty(\mathcal{V}_k)
    =
    t^\prime \dv{t} \ln \qty( k^2\rho(k^2) {\qty(t^{\prime})}^2 ),
\end{equation}
and
\begin{equation}
    \dim_{\tau}\qty(\chi)
    =
    t^\prime \dv{t} \ln \qty(\rho(k^2) {\qty(t^{\prime})}^2).
\end{equation}

The flow equation for the dimensionless potential $\overline{\mathcal{V}}_k(\overline{\chi})$ (expressed only in terms of dimensionless quantities) is:
\begin{equation}
    {\overline{\mathcal{V}}}_k^\prime\qty[\overline{\chi}]
    =
    -
    \dim_{\tau}\qty(\mathcal{V}_k)\overline{\mathcal{V}}_k\qty[\overline{\chi}]
    +
    \dim_{\tau}\qty(\chi) \overline{\chi} \pdv{\overline{\chi}} \overline{\mathcal{V}}_k\qty[\overline{\chi}]
    +
    \frac{1}{1 + \ipd{\overline{\chi}} \overline{\mathcal{V}}_k \qty[\overline{\chi}] + 2\overline{\chi} \partial^2_{\overline{\chi}} \overline{\mathcal{V}}_k\qty[\bar{\chi}]}.
    \label{eq:potential_flow}
\end{equation}

It is useful to define the notions of asymptotic dimension:

\begin{definition}\label{def:asymptoticdim}
    For a universal analytic distribution $\mu(\lambda)$ (typically \mpdistr) behaving as a power law $\mu(\lambda)\sim{(\lambda_{+}-\lambda)}^{\delta}$ in the vicinity of the larger eigenvalue $\lambda_+$, we call $D_0 \coloneqq 2\delta+2$ the asymptotic dimension of the distribution.
\end{definition}
The typical (empirical) distance $\delta \lambda \coloneqq \vert \lambda_{\text{max}}-\lambda_+\vert$ between the largest eigenvalue and the edge $\lambda_+$ can be estimated from the observation that $\mu(\lambda_{\text{max}})\delta \lambda$ must be of order $\sim 1/P$, the typical separation from which we can distinguish two eigenvalues~\cite{Bouchaud3}.
Hence:
\begin{equation}
    \delta \lambda \sim P^{-\frac{2}{D_0}}.
\end{equation}
For Wigner and \mpdistr, $D_0 = 3$ and the underlying field theory behaves like a three dimensional Euclidean field theory as far as power counting is concerned.
In that limit, the flow becomes autonomous, and can admit almost fixed points that we call \textit{asymptotic fixed points}.
The asymptotic value for the dimensions is easy to compute from the formula.
Assuming a power law behaviour $\rho(k^2) \sim {(k^2)}^\alpha$, we get $\dd{\tau} = (2\alpha+2) \dd{t}$, therefore $t^\prime = {(2\alpha+2)}^{-1}$ and $t^{\prime\prime} = 0$. Then:
\begin{equation}
    \dim_{\tau}\qty(u_4) \to \frac{1-\alpha}{1+\alpha}.
\end{equation}
For $\alpha = 1/2$, we then get for the asymptotic dimension:
\begin{equation}
    \dim_{\tau}\qty(u_4) \to \frac{1}{3} \approx 0.33.
\end{equation}

Let us summarise more precisely some recent conclusions obtained using the equilibrium field theory formalism in the vicinity of the \mpdistr class.
There are two main statements regarding the behaviour of the \rg flow:
\begin{itemize}
    \item purely noisy signals, close enough to the \mpdistr class, are characterised by the relevance of quartic and sextic local couplings in the deep \ir, and the influence of the signal is to make them less relevant~\cite{RG2},
    \item the presence of a signal reduces the size of the symmetric phase, and then induces phase transition with $\mathbb{Z}_2$ symmetry breaking~\cite{RG5}.
\end{itemize}

As discussed in~\cite{RG6}, where out-of-equilibrium stochastic field theory was considered, signal delays the divergence of the potential, thus determining the existence of two distinct regimes:
\begin{itemize}
    \item noisy datasets (low \snr) never reach equilibrium in the \ir in the underlying stochastic field theory, thus breaking ergodicity (restraining to critical coarsening,~\cite{Bray_1994});
    \item the presence of signal (higher \snr) makes it possible to maintain the equilibrium for longer periods of time.
\end{itemize}

\begin{figure}[t]
    \centering
    \includegraphics[width=0.66\textwidth]{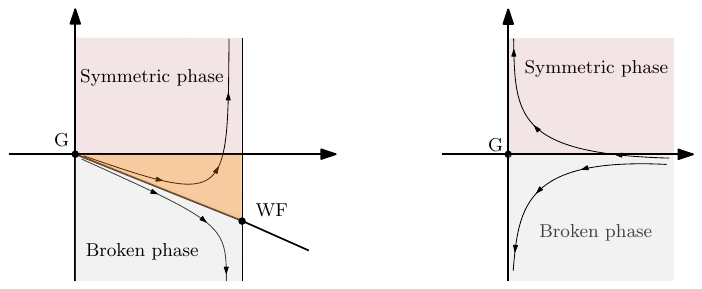}
    \caption{%
    Qualitative behaviour of the \rg flow in the vicinity of the Gaussian fixed point (G) for the $\phi^4_{4-\epsilon}$ field theory for $\epsilon > 0$ (left) and for $\epsilon < 0$ (right).
    As $\epsilon$ decreases, the \wf fixed point reaches the Gaussian one, and the symmetry restoration region (in orange) disappears as $\epsilon$ vanishes.
    }\label{fig:figWF}.
\end{figure}

In what follows, we focus on revisiting the equilibrium statements in the realistic scenario of image analysis.
In particular, we wish to take into account the finite size effects, and quantify the results of the \rg approach defining a consistent \lod.
We shall also present an empirical justification of the phenomenon underlying the detection, which will lead us to propose an interpretation of the signal/noise components seen in the images.

Notice that the signal regime is clearly an expected consequence of the noisy one.
In the standard $\epsilon$-expansion around dimension $4$, for instance, it is well-known that the relevance of the quartic coupling and the existence of an interacting fixed point depend on the sign of $\epsilon$.
As $\epsilon$ is positive (i.e.\ for a dimension smaller than $4$), a fixed point exists: it is the so-called Wilson-Fisher (WF) fixed point, controlling the ferromagnetic second order phase transition.
This fixed point is located in the negative mass region, and collapses toward the Gaussian fixed point as $\epsilon \to 0$, i.e.\ as the dimension reaches the critical dimension.
For $\epsilon \leq 0$, the transition is controlled by the Gaussian fixed point, and the critical line moves along the zero mass axis (see Figure~\ref{fig:figWF}).
The symmetric phase, including symmetry restoration scenario, is then reduced as $\epsilon$ decreases, the symmetry restoration region being cancelled as $\epsilon \to 0$.

Obviously in standard field theory, the collapsing phenomena is a consequence of the still debated dimensional regularisation~\cite{schonfeld2016physical}, especially concerning the physical meaning of non-integer dimensions.
The approaches we consider in this article, in contrast, naturally deal with non-integer dimensions, as already considered in the context of spin glasses~\cite{lahoche2024intriguing,RG5,lahoche2024functional,lahoche2024functional2}.
Increasing the signal strength in the empirical spectrum decreases the effective value for $\epsilon$ at the edge of the spectrum.
Hence, despite the fact that we cannot consider global fixed points in the flow due to the intrinsic dependency of the canonical dimension on the \rg scale, it is possible to consider the asymptotic flow corresponding to the scaling at the edge, and to rely on the influence of the ``informational'' \dof of the flow, due to the change in behaviour.
This phenomenon is based essentially on the conclusions of our previous work, and while they may seem a little obscure to the reader who is not familiar with them, the following will clarify them.

In summary, our approach is entirely based on the Wilsonian renormalisation group philosophy: by systematically integrating out the high-momentum modes corresponding to the bulk noise, we construct an effective field theory that captures the relevant long-range correlations associated with the signal.
This framework relies on the absence of a pre-existing background space.
Instead, the geometry is emergent, with the empirical spectrum itself defining the effective momentum scale and dimensionality.
The detection of signal is thus mapped to the study of relevant perturbations around the noise fixed point in the deep \ir, guaranteed to be robust by the universality of the underlying spectral distribution.
This implies that the characterisation of the signal does not require a detailed knowledge of the underlying microscopic theory, provided the noise belongs to the appropriate universality class.

\section{Data and Empirical Methodology}\label{sec:data_analysis}

For numerical experiments, we rely on the classical Python scientific libraries \texttt{numpy}~\cite{Harris:2020xlr} and \texttt{scipy}~\cite{Virtanen:SciPy10Fundamental:2020}.
Our goal is to numerically solve differential equations of the form:
\begin{equation}
    \dv{f(s, x)}{s} = \mathcal{D}_{\qty{\partial_x, \partial^2_x}}\qty[f](s, x),
\end{equation}
where $\mathcal{D}$ is a differential operator, of second order at most, acting on the function $f$.

The flow equation for the potential~\eqref{eq:potential_flow}, derived in the \lpa approximation, contains the full information of all infinitely many interactions of the theory.
It is, thus, impossible to solve exactly.
Extracting any information requires truncations~\cite{Delamotte,Wett,Morris}, capable of projecting the full flow into a reduced (often finite) dimensional subspace.
In this paper, we essentially focus on the \lpa, which considers only local couplings in the sense of~\eqref{eq:localpotential}, and ignores wave function effects (see~\cite{RG5} for further discussion).
This effect is indeed expected to play a less significant role for dimensions higher than the critical dimension $D_0 = 4$, and we fix the detection threshold at this value (see below) for which the universal behaviour of the flow is expected to be clearly modified (non asymptotic interacting WF like fixed point).
A standard choice is the following truncation (for a uniform field):
\begin{equation}
    \mathcal{V}_k[\chi]
    =
    \frac{u_4(k)}{2!}
    {\qty(\chi - \kappa(k))}^2
    +
    \frac{u_6(k)}{3!}
    {\qty(\chi - \kappa(k))}^3
    +
    \dots,
    \label{eq:truncationpotential}
\end{equation}
Since we focus on spectra in the vicinity of the \mpdistr law, we know (see Figure~\ref{fig:canonical_dim_mp}) that all the local couplings are relevant in the deep \uv, with arbitrary dimension, a fact that seems to invalidate standard truncation schemes of~\eqref{eq:truncationpotential}.
However, we shall fix our \uv scale $\Lambda$ at which we initialise the flow in a mesoscopic regime where only quartic and sextic couplings are relevant\footnote{%
    This argument implicitly assumes that physical trajectories arising from the flow at the true microscopic scale reach the region of phase space that we study.
}.
In the symmetric phase (i.e.\ assuming we expand around $M = 0$), the following truncation is also suitable in the \lpa:
\begin{equation}
    U_k\qty[\varphi]
    =
    \sum_{n=2}^\infty\,
    \frac{u_{2n}}{(2n)! P^{n-1}}\, \sum_{\qty{p_1,\ldots,p_n}}
    \delta_{0,\sum_{i=1}^n p_i} \prod_{i=1}^{2n} M\qty(p_i)
\end{equation}
Assuming again to truncate around sextic interactions, the partial differential equation in~\eqref{eq:potential_flow} can be decoupled in a system of ordinary differential equations (see again~\cite{RG5}):
\begin{equation}
    \begin{cases}
        \bar{u}_2^{\prime} & = - \dim_{\tau}\qty(u_2)\, \bar{u}_2 - 2 \frac{\bar{u}_4}{{\qty(1 + \bar{u}_2)}^2}
        \\
        \bar{u}_4^{\prime} & = - \dim_{\tau}\qty(u_4)\, \bar{u}_4 - 2 \frac{\bar{u}_6}{{\qty(1 + \bar{u}_2)}^2} + 12 \frac{\bar{u}_4^2}{{\qty(1 + \bar{u}_2)}^3}
        \\
        \bar{u}_6^{\prime} & = - \dim_{\tau}\qty(u_6)\, \bar{u}_6 - 60 \frac{\bar{u}_4 \bar{u}_6}{{\qty(1 + \bar{u}_2)}^3} - 108 \frac{\bar{u}_6^3}{{\qty(1 + \bar{u}_2)}^4}
    \end{cases},
    \label{eq:coupl_flow_eqn}
\end{equation}
where the canonical dimensions have been computed explicitly in Section~\ref{sec:lpa}.

Simply replacing the definitions of $t$ and $\tau$ in the previous equations shows that the canonical dimensions do not depend on previous states in the \rg flow.
They only depend on the position in the spectrum $\rho\qty(p^2)$ (or $\rho_G\qty(p^2)$).
Moreover, the flow equations~\eqref{eq:coupl_flow_eqn} can be numerically solved using a finite element approach:
\begin{equation}
    u_{2n}\qty(k^2 - \Delta k^2)
    =
    u_{2n}\qty(k^2) - \Delta k^2\, \mathcal{R}\qty(u_{2n}\qty(k^2), u_{2(n+1)}\qty(k^2)),
\end{equation}
where $\mathcal{R}$ is the right hand side of~\eqref{eq:coupl_flow_eqn}, and $\Delta k^2$ is a finite (small) step on the spectrum.
Notice that the minus sign is due to the direction of the \rg evolution from the ultraviolet (\uv) region towards the deep \ir (i.e.\ from $k^2 > 0$ to $k^2 \to 0$).

\begin{algorithm}[t]
    \caption{Construction of the samples}\label{alg:sample_constr}

    \DontPrintSemicolon
    \SetKwInOut{Input}{input}
    \SetKwInOut{Init}{init}
    \SetKwInOut{Let}{let}
    \SetKwInOut{Output}{output}

    \Input{size of the sample $N > 0$, and ratio $q \in \qty[0, 1]$}
    \Let{$P = \left\lfloor q\, N \right\rfloor$}
    \Input{$\beta > 0$}
    \Input{$Z \in \mathbb{R}^{N \times P}$ where $Z \sim \mathcal{N}\qty(0, \sigma^2)$}
    \Input{an image $S \in {\qty[0, 255]}^{H \times W \times C}$}

    \If{$C > 1$}{$S_{ij} \gets C^{-1}\, \sum\limits_{c = 1}^C S_{ijc}$ \tcp*{convert to B/W}}
    $S \gets \frac{S - \left\langle S \right\rangle}{\sqrt{\mathrm{Var}\qty(S)}}$ \tcp*{standardise the image}
    $S \gets \mathrm{resize}(S) \in \mathbb{R}^{N \times P}$ \tcp*{interpolate to sample size}
    \Let{$X = \beta\, S + Z \in \mathbb{R}^{N \times P}$}
    $\qty(\Sigma, W) = \mathrm{SVD}(X)$ \tcp*{compute singular values and right eigenvectors}
    $\mathrm{E} \gets \mathrm{flatten}\qty(\Sigma^2 / (N - 1))$ \tcp*{convert to covariance eigenvalues}
    $\mathrm{E}^{\prime} = \text{remove\_spikes}\qty(\mathrm{E})$ \tcp*{PCA --- isolated spikes removal}
    $\tilde{\mu}_G \gets \text{histogram}\qty(\mathrm{E}^{\prime})$

    $\mu_G \gets \operatorname{KDE}\qty(\tilde{\mu}_G)$ \tcp*{interpolation by kernel density}

    \Output{$\rho_G \gets \mu_G\qty(\frac{1}{k^2 + m^2} + \lambda_-) / {\qty(k^2 + m^2)}^2$ \tcp*{momenta distribution}}
    \Output{$W^T$ \tcp*{covariance eigenvectors}}
\end{algorithm}

We build the samples for the analysis using a simple additive model for normally distributed noise with a \snr $\beta \ge 0$:
\begin{equation}
    X = \beta\, S + Z,
    \label{eq:additive_model}
\end{equation}
where $Z \sim \mathcal{N}\qty(0, \sigma^2)$ is a $N \times P$ matrix with normally distributed entries, and $S \in \mathbb{R}^{N \times P}$ is the centred signal matrix.
Unless otherwise stated, in our numerical exploration, we use $\sigma^2 = 1$ and
\begin{equation}
    N = \num{2.0e4},
    \quad
    P = \num{1.8e4},
    \quad
    \text{s.t.}
    \quad
    q = \frac{P}{N} = 0.9.
\end{equation}

\begin{figure}[t]
    \centering
    \includegraphics[width=0.15\textwidth]{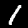}
    \hfill
    \includegraphics[width=0.15\textwidth]{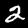}
    \hfill
    \includegraphics[width=0.15\textwidth]{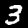}
    \hfill
    \includegraphics[width=0.15\textwidth]{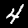}
    \hfill
    \includegraphics[width=0.15\textwidth]{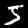}
    \caption{%
        Samples extracted from the MNIST dataset~\cite{LeCun:2010:MNIST} and used for numerical evaluations.
    }\label{fig:exp_images}
\end{figure}

\begin{figure}[t]
    \centering
    \includegraphics[width=0.45\textwidth]{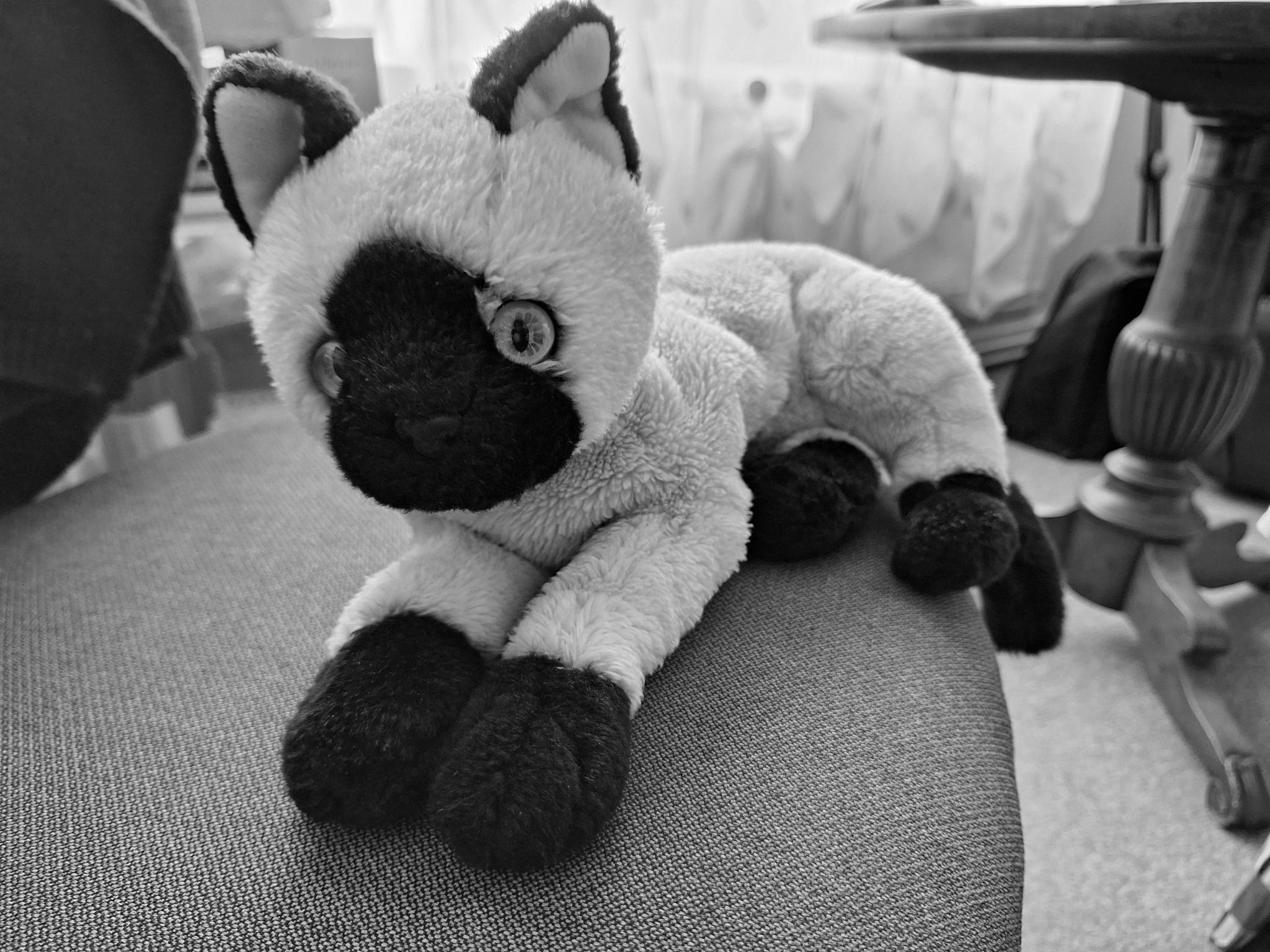}
    \caption{%
        Realistic scenario considered in the analysis: a traditional photo of a (plush) cat with a non trivial background.
        For simplicity, we consider a monochrome version.
    }\label{fig:gianduja}
\end{figure}

\begin{figure}[t]
    \centering
    \includegraphics[width=0.45\textwidth]{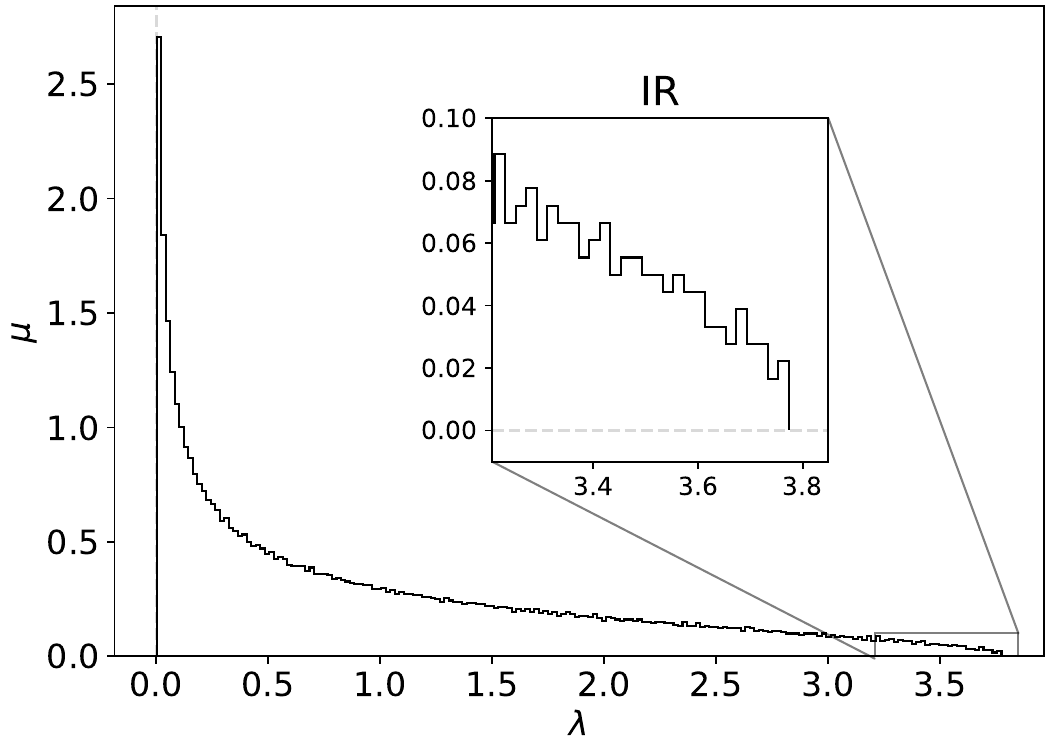}
    \hfill
    \includegraphics[width=0.45\textwidth]{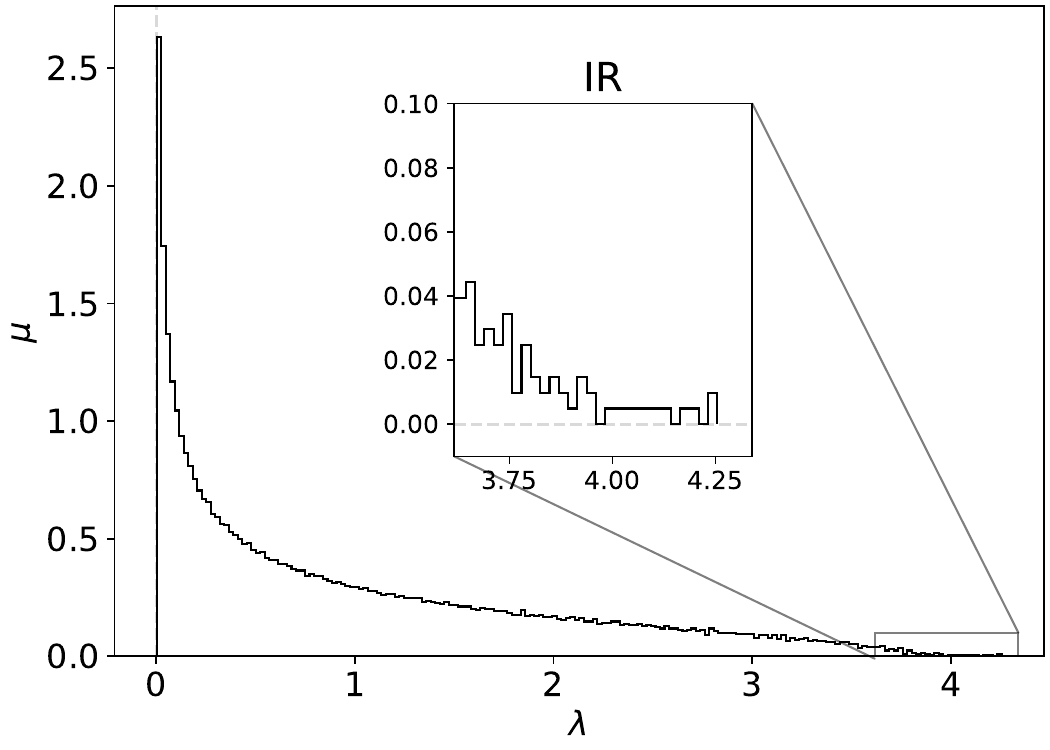}
    \caption{%
        Empirical distribution corresponding to Figure~\ref{fig:gianduja} for $\beta=0$ (no signal, on the left), and for $\beta=0.4$ (on the right).
    }\label{fig:distgianduja}
\end{figure}

The computation of the \rg equations and the canonical dimensions use the outputs of Algorithm~\ref{alg:sample_constr}, which shows the construction of the sample distribution of momenta.
As the analysis deals with finite size objects (images), we need to take into account the natural scale of the empirical distributions.
These objects are defined through a finite and ordered set of eigenvalues whose normalised histograms represent the corresponding distributions $\rho_G(p^2)$.
The use of summary statistics to define the marginal likelihood of the empirical momenta naturally introduces a concept of ``energy step'' in the \rg flow, that is a physical energy difference:
\begin{equation}
    \Delta_{\text{phys}} = P^{-\alpha},
    \label{eq:time_step}
\end{equation}
where $0.5 \le \alpha < 1$ can be fixed by studying the distance between isolated spikes and the bulk distribution of momenta (we fix it to $\alpha = 0.5$ in our numerical experiments).
Under this threshold, eigenvalues start to become isolated, rather than densely populated.
The histogram is thus built using a bin width of $\Delta_{\text{phys}}$, then fitted by a kernel density estimation (KDE) of the distribution to obtain a curve, to be easily manipulated.
The density function of the momenta is then computed by inverting and translating to the origin the probability distribution function of the inverse variable:
\begin{equation}
    \rho_G\qty(k^2)
    =
    \frac{1}{{\qty(k^2 + m_{\text{eff}}^2)}^2}\,
    \mu_G\qty(\frac{1}{k^2 + m_{\text{eff}}^2} + \lambda_-),
\end{equation}
where $m_{\text{eff}}^2$ is the inverse of the largest eigenvalue, as argued in the previous sections.
For our numerical exploration, we use images from the known MNIST dataset~\cite{LeCun:2010:MNIST} for their apparent simplicity and structure, shown in Figure~\ref{fig:exp_images}, and an illustration of a real environment, shown in Figure~\ref{fig:gianduja}.
Figure~\ref{fig:distgianduja} shows the corresponding empirical distribution for different values of the \snr.

For the reasons discussed in the previous paragraph, it also becomes numerically unfeasible to track the evolution of~\eqref{eq:coupl_flow_eqn} or compute the canonical dimensions in the very deep \ir ($k^2 = 0$).
All computations in the following sections stop at an arbitrary energy scale $k^2_{\text{IR}}$, chosen for its closeness to the smallest attainable value $\Delta_{\text{phys}}$ and its distance from possible numerical instabilities.
Numerically, such energy scale has been chosen to be the midpoint
\begin{equation}
    k^2_{\text{IR}}
    =
    \frac{k^2_{*} - k^2_{0.5}}{2},
\end{equation}
where $\rho_G\qty(k^2_{0.5}) = 0.5$ and $k^2_{*} = \underset{k^2}{\text{argmax}}\, \rho_G\qty(k^2)$ have been chosen as sensible definitions of the near \ir zone of the bulk distribution of momenta.
A more formal justification comes from known results on eigenvalue density~\cite{Zinn2}: since all eigenvalues are densely packed inside a compact real-valued interval, variations in the shape of the distribution forcefully propagate across all eigenvalues, from \uv to \ir, and vice versa.
For the purpose of signal detection in nearly continuous spectra, the choice of the energy scale can thus remain arbitrary: effects linked to the presence of signal will be present at any value in the distribution of momenta (or eigenvalues), though measurements are simpler near the \ir region, where most of the changes take place.

Empirically, this implies that an ``absolute'' detection of the signal becomes impossible (not physical): though the values of~\eqref{eq:coupl_flow_eqn} and the canonical dimensions can be deterministically computed for the asymptotic \mpdistr distribution, empirical finite size effects introduce random effects.
However, this makes the mechanism even more interesting, in the light of the discussion on the universality classes discussed in Section~\ref{sec:renorm_group}.
Given an arbitrary energy scale, the values of~\eqref{eq:coupl_flow_eqn} and canonical dimensions can be computed for a \emph{blank} (e.g.\ a signal-less sample in chemometrics) to define a baseline.
Presence of relevant signal, or, in general, any modification in the distribution of eigenvalues, can then be quantified by the \rg equations and the canonical dimensions as a ``distance'' from the background noise.
The functional \rg thus becomes a tool to perform a relative detection of signal, with respect to an arbitrary background distribution.

Since we are interested in the behaviour of nearly continuous spectra, we take into careful consideration the presence of possible isolated spikes in the momenta distribution.
Our framework is explicitly designed to detect extensive-rank signals that remain hidden in the bulk spectrum after standard PCA has removed the isolated outliers (spikes).
We assume the spikes carry the ``easy'' low-rank component of the signal, while our method targets the ``hard'' component that merges with the noise.
As a matter of fact, increasing $\beta$ inevitably weakens localisation of certain eigenvectors containing most of the signal.
The interpretation of this effect will become the object of the study in the final sections of the article.
For the purpose of this study, we truncate the spectrum of momenta to only its continuous distribution by removing the spikes.
This is possible by simply scanning the eigenvalues from \ir to \uv and computing distances of adjacent values.
In turn, this creates an ordered set which can be used to determine the index in the list of eigenvalues corresponding to the beginning of the bulk:\footnote{%
    As already stated in the previous sections, all computations involve essentially dimensionless parameters, such as $\Delta_{\text{phys}}$.
    In this article, we interchangeably use this parameter to compute a distance between eigenvalues and between momenta, since the numerical value defined in~\eqref{eq:time_step} remains unchanged in both cases.
}
\begin{equation}
    \mu_{\text{bulk}}
    =
    \underset{\mu \in \qty[0, P-1]}{\text{argmin}}
    \qty{\max\qty(0, \qty(\lambda_{\mu} - \lambda_{\mu+1}) - \tilde{\Delta}_{\text{phys}})}\,,
\end{equation}
where $\tilde{\Delta}_{\text{phys}}$ might differ from $\Delta_{\text{phys}}$ and depends on the continuum limit we aim to construct.
Motivations coming from random matrix theory invite to consider $\tilde{\Delta}_{\text{phys}} = \mathcal{O}\qty(1/P)$, the typical spacing between two eigenvalues being of order $1/P$ for Wigner matrices for instance.
With this definition, the spectrum surely involves different continuous components, the larger one being called ``the bulk''.
Obviously, this opens the possibility to consider different definitions of the continuum limit, having consequence on the definition of the bulk and the ``spikes''.
Numerical experiments show that a value of $\tilde{\Delta}_{\text{phys}} = P^{-0.8}$ is overall well adapted, and follows the discussion used to define~\eqref{eq:time_step}.
However, we also explored different definitions such as a linear dependence in $\beta$ to artificially overfit the dataset ($\beta$ is usually unknown), with no significant changes to the overall conclusions.

From what was stated previously, we can then reduce the set of eigenvalues to:
\begin{equation}
    \Lambda = \qty{\lambda_{\mu_{\text{bulk}}} \ge \lambda_{\mu_{\text{bulk}}+1} \ge \dots \ge \lambda_P},
\end{equation}
since $\mu_{\text{bulk}} \ge 0$ by construction.
Clearly, the reduced cardinality $\qty|\Lambda| \le P$ might introduce additional finite size effects.
However, for $P \gg 1$, no effects, in addition to those already taken care of with the previous procedure, were detected.
Indeed, for the values of $\beta$ explored here, we experimentally observe only a few tens of spikes, which has a negligible impact on the distribution of $P = \num{1.8e4}$ \dof.
In simpler terms, the procedure amounts to analyse a spectrum of eigenvalues whose spikes have already been considered using the traditional \pca.

Finally, we recall that the universal value of the considered field theory is inherited from the universal character of the random matrix eigenvalue distributions, and we always assume to remain as close as possible.
This proximity can be quantified using standard statistical distance in the literature.
In this paper, we consider the \textit{Kullback-Leibler} (\kl) divergence (which, strictly speaking, is not a metric distance as it does not satisfy the triangle inequality):
\begin{definition}
    For two (probability) distributions $P \colon \mathbb{X} \to [0,1]$ and $Q \colon \mathbb{X} \to [0,1]$ defined on the same probability space $\Omega$, the \kl divergence is defined as:
    \begin{equation}
        \mathrm{D}_{\text{KL}}(P \lVert Q)
        =
        \sum_{x\in \mathbb{X}}\, P(x)\,\log \left( \frac{P(x)}{Q(x)}\right).
        \label{eq:KLdiv}
    \end{equation}
\end{definition}
The divergence between $P$ and $Q$ gives a measure of the number of bits needed to encode some data using distribution $Q$ instead of the $P$.
In this article, we will consider $P$ as the \mpdistr distribution, and $Q$ as the empirical distribution of momenta $\rho_G$ defined previously.
Other definitions of measure could also be considered, such as the Wasserstein distance.
The advantage of the \kl divergence in this context is its intrinsic sensitivity to the \snr.
However, relations exist between the two similarity measures, see for instance~\cite{Belavkin_2018}.
Furthermore, notice that this definition implies that the distributions we compare must necessarily operate in the same set $\mathbb{X}$.
We therefore cannot compare two distributions associated with matrices of different size.
To advance further into the discussion, let us provide the following definition:
\begin{definition}
    If $C$ is defined as in~\eqref{eq:C0def}, is such that:
    \begin{enumerate}
        \item $X$ is a random matrix with unknown distribution, depending on some parameters $(\alpha_1, \alpha_2, \dots,\alpha_k)$;
        \item the empirical spectrum of $C$ converges toward \mpdistr as $P \to \infty$ but with $q$ fixed and finite,
    \end{enumerate}
    then $C$ is in the \mpdistr class.
\end{definition}

In this section, $X$ is a \iid Gaussian matrix with variance $\sigma^2$ and zero mean value.
For $\beta \neq 0$, we need to quantify the ``distance'' with a given sample in the corresponding \mpdistr class (with $\beta = 0$).
We will then have to calculate the \kl divergence each time by choosing the same value of $P$.
It is, however, difficult to quantify the limiting distance at which the universality argument is no longer legitimate.
We would need to be able to quantify at what point the theory is no longer able to reproduce the effective correlations, but since the space of Hamiltonians is, here, an infinite-dimensional functional space, such a task becomes quite difficult.
We plan to address the issue in future work.
We will therefore opt for a pragmatic choice here, fixing implicitly some upper bound for $\beta_M$:
\begin{equation}
    \mathrm{D}_{\text{KL}} < \frac{1}{P} \log \qty(0.1 \sqrt{P}),
    \label{eq:boundentropy}
\end{equation}
which essentially means that the number of bits required to encode data with optimal coding for $Q$ must not exceed $10\%$ of the square root of the total number of degrees of freedom.
Estimating this bound is actually subtle, and we will come back to it in Section~\ref{sec:largebeta}.

\section{Numerical Results and Analytical Validation}\label{sec:numeric}

In this section, we present the numerical validation of the \frg framework, showing the application of the dimensional analysis developed in Section~\ref{sec:lpa}, and its theoretical justification.
Our goal is not merely to report empirical performance, but to provide a consistent physical interpretation of the signal detection process in terms of phase transitions and universality classes.
To this end, our analysis is structured around four key theoretical points, which also represent the main novel contributions of this work:

\begin{enumerate}
    \item \textbf{Detection thresholds:} we establish a hierarchy of thresholds ($\beta_t < \beta_c < \beta_O$) that rigorously define the \lod for extensive-rank signals. Unlike traditional univariate definitions, this field-theoretic criterion is model-agnostic, and relies only on the universal breakdown of the noise sector rather than specific priors on the signal structure. We contrast this regime with the standard \bbp transition~\cite{math3} for isolated spiked models, demonstrating that our field-theoretic approach remains sensitive in the bulk where standard spectral methods fail. In turn, this represents the dual notion of the bimodal connected phase identified by~\cite{Landau2023}.
    \item \textbf{Microscopic mechanism:} we investigate the origin of the transition, linking the macroscopic dimensional crossover to a spontaneous symmetry breaking in the effective potential and a deviation from the universal \emph{Porter-Thomas} statistics of eigenvectors.
    \item \textbf{Stability and variability:} we quantify the intrinsic variability of the method, showing that the signal-induced flow is robust against the aleatoric uncertainty of finite-size random matrix realisations.
    \item \textbf{Formalism and components:} we formalise these observations by defining a ``spectral distance'' based on the canonical dimensions and propose a theoretical interpretation for the cyclic behaviour of the \rg flow as a probe for independent noise components.
\end{enumerate}

The code for reproducibility is available at this \href{https://github.com/thesfinox/frg-signal-detection}{link}.

\subsection{Detection thresholds and dimensional phase transition}\label{sec:mainresults}

\begin{figure}
    \centering
    \begin{minipage}[b]{0.45\textwidth}
        \centering
        {\large Analytic computation (\mpdistr)} \\[0.5em]
        \includegraphics[width=\linewidth]{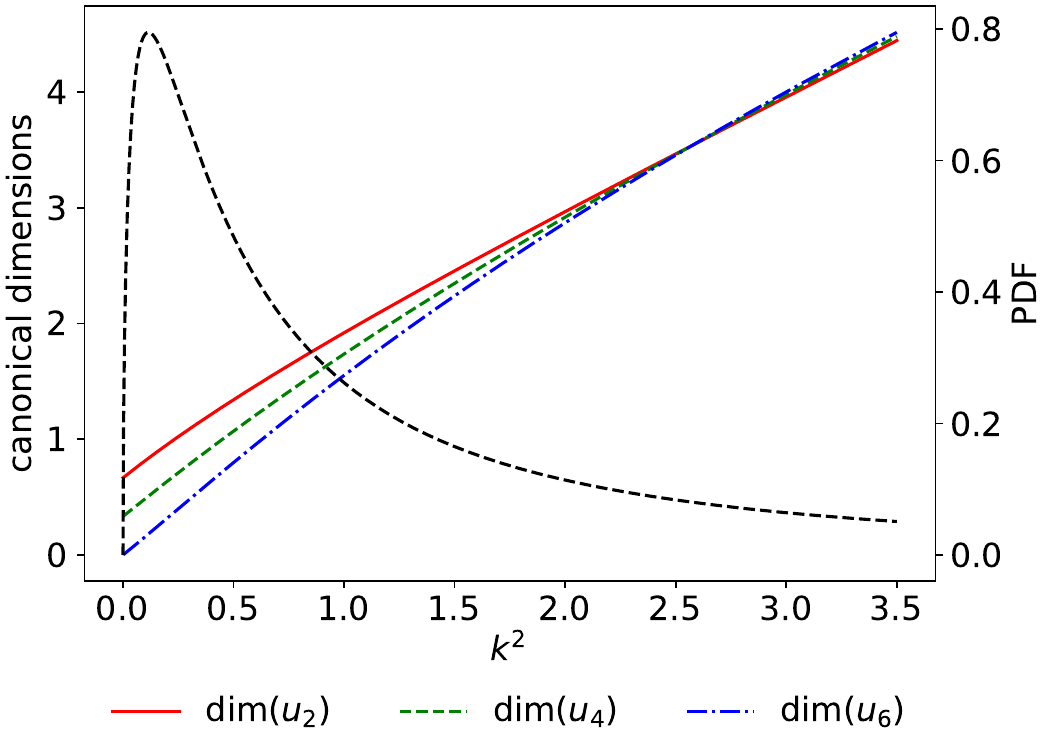}
    \end{minipage}
    \hfill
    \begin{minipage}[b]{0.45\textwidth}
        \centering
        {\large Numerical simulation ($\beta = 0$)} \\[0.5em]
        \includegraphics[width=\linewidth]{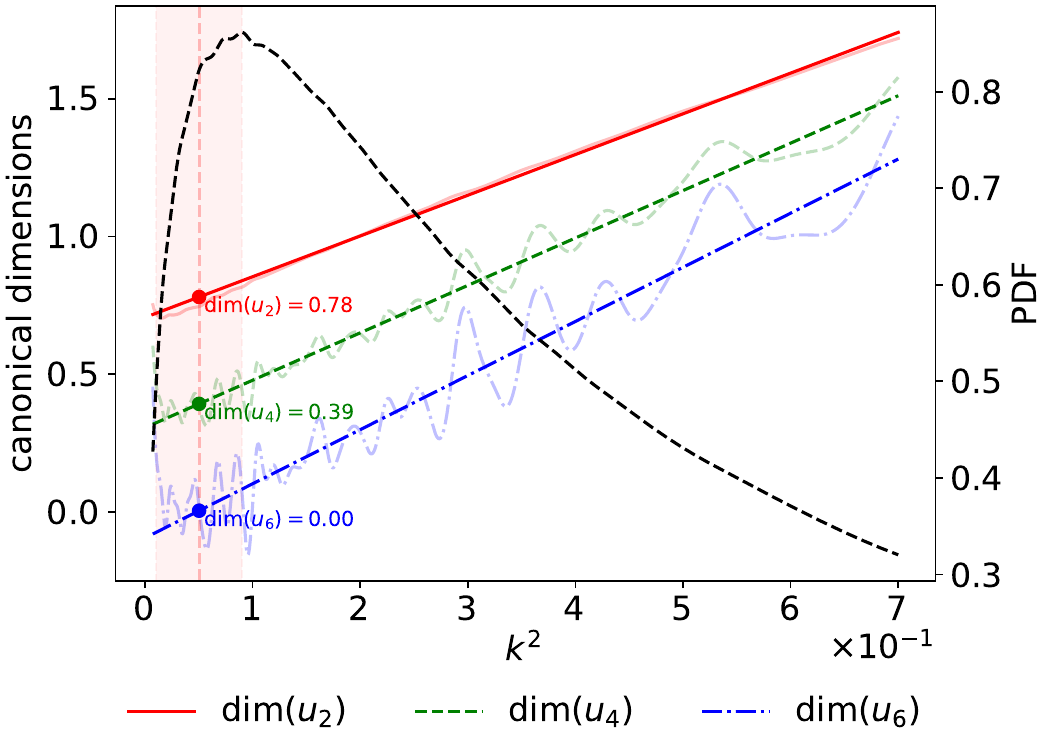}
    \end{minipage}
    \\[1em] 

    \begin{minipage}[b]{0.45\textwidth}
        \centering
        {\large Numerical simulation ($\beta = 0.1$)} \\[0.5em]
        \includegraphics[width=\linewidth]{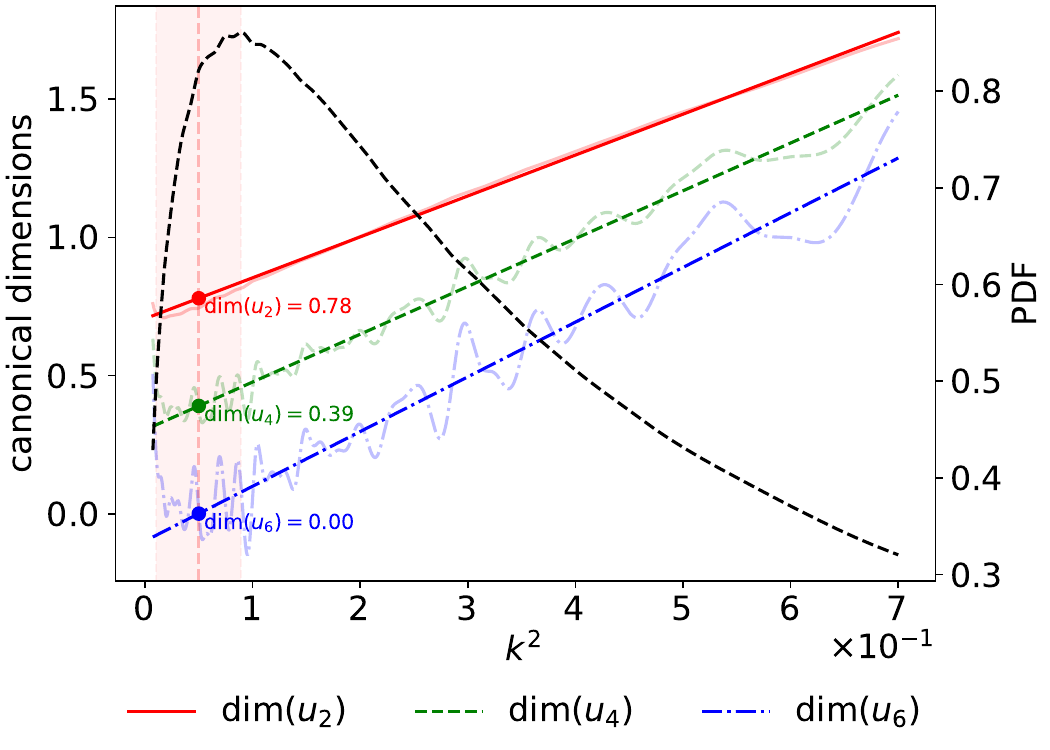}
    \end{minipage}
    \hfill
    \begin{minipage}[b]{0.45\textwidth}
        \centering
        {\large Numerical simulation ($\beta = 0.2$)} \\[0.5em]
        \includegraphics[width=\linewidth]{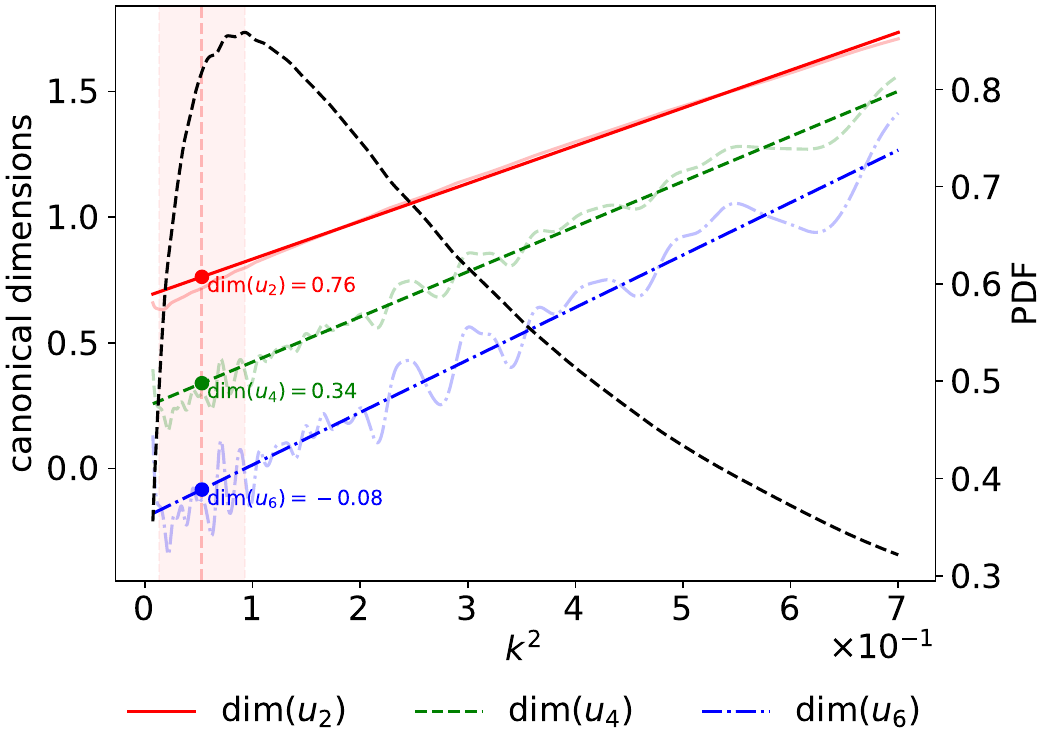}
    \end{minipage}
    \\[1em]

    \begin{minipage}[b]{0.45\textwidth}
        \centering
        {\large Numerical simulation ($\beta = 0.3$)} \\[0.5em]
        \includegraphics[width=\linewidth]{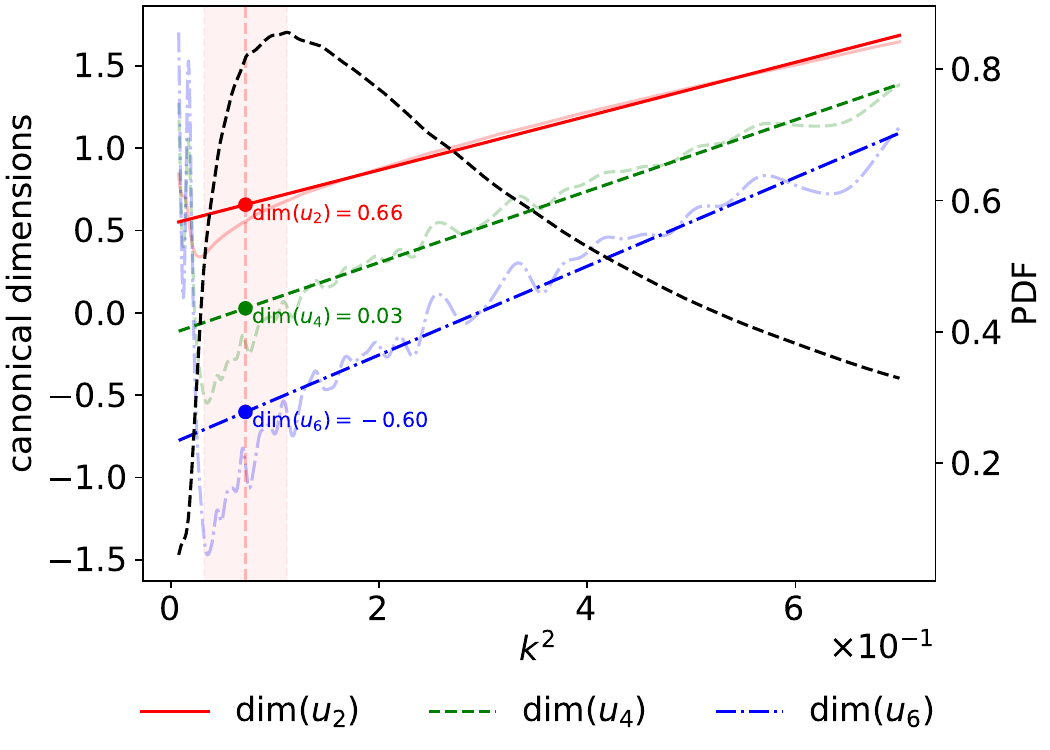}
    \end{minipage}
    \hfill
    \begin{minipage}[b]{0.45\textwidth}
        \centering
        {\large Numerical simulation ($\beta = 0.4$)} \\[0.5em]
        \includegraphics[width=\linewidth]{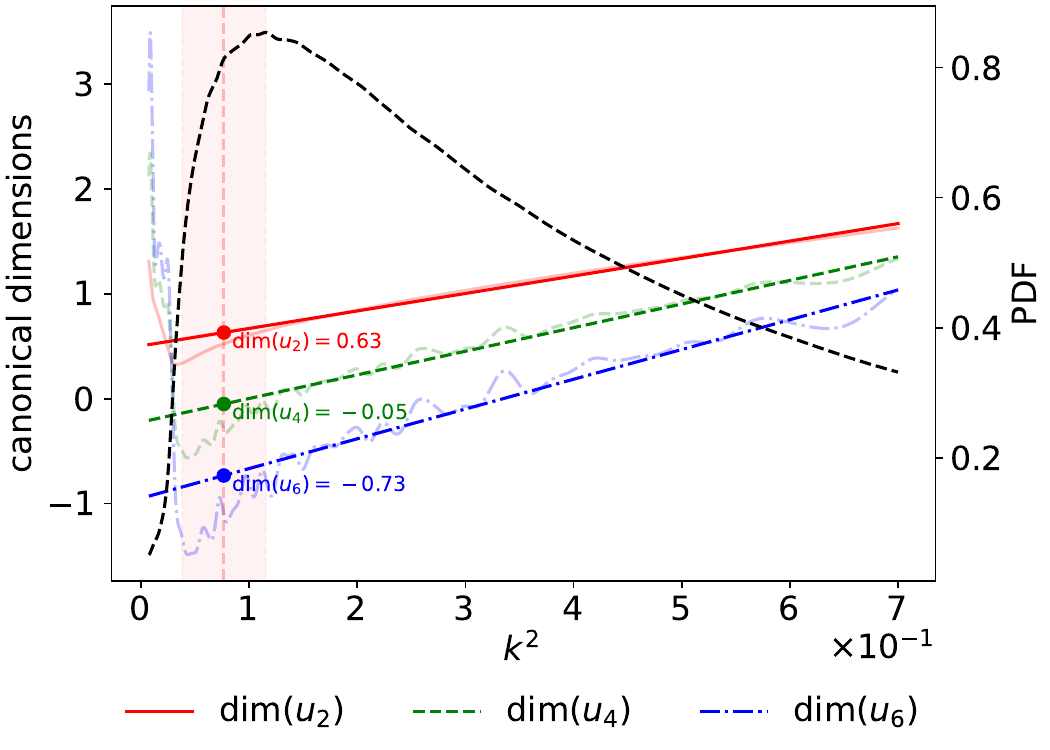}
    \end{minipage}

    \caption{%
        Behaviour of the canonical dimension (specifically $\dim_{k^2}\qty(u_2)$, $\dim_{k^2}\qty(u_4)$, and $\dim_{k^2}\qty(u_6)$) in the $k^2$-space of Figure~\ref{fig:gianduja} for increasing values of \snr $\beta$.
        The first figure in the top left corner provides a comparison with the analytic \mpdistr distribution.
        Values of the canonical dimensions can be read on the left y-axis, while values of the moment distribution are on the right, to be displayed on the same plot.
    }\label{fig:figplotgianduja}
\end{figure}

\begin{figure}
    \centering
    \includegraphics[width=0.66\textwidth]{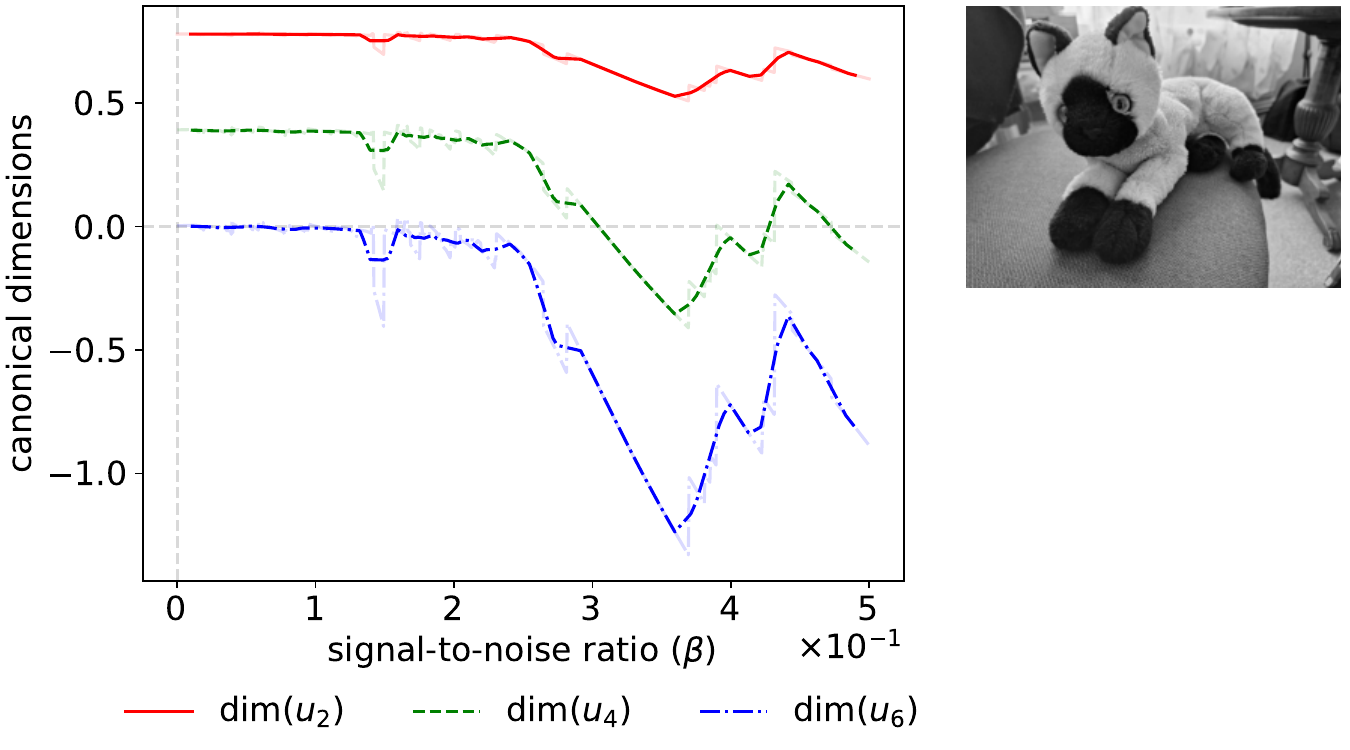}
    \caption{%
    Behaviour of the canonical dimension at the scale $k^2_{\text{IR}}$ with respect to $\beta$.
    The step between consecutive computations along the x-axis is $\Delta \beta = \num{5e-4}$.
    Solid lines show a moving average of width $\Delta \beta_{\text{w}} = \num{2e-2}$, while experimental values are slightly transparent.
    }\label{fig:figplotgianduja2}
\end{figure}

First, we consider the realistic sample corresponding to Figure~\ref{fig:gianduja} as a function of the \snr $\beta$.
The behaviour of the canonical dimension with respect to $k^2$ is shown in Figure~\ref{fig:figplotgianduja} and Figure~\ref{fig:figplotgianduja2}.
Notice that, because of numerical instabilities at the tail of the spectrum, we consider the arbitrary \ir scale $k^2_{\text{IR}}$ defined in the previous section and highlighted on each plot by a vertical dashed red line.

\begin{figure}
    \centering
    \includegraphics[width=0.45\textwidth]{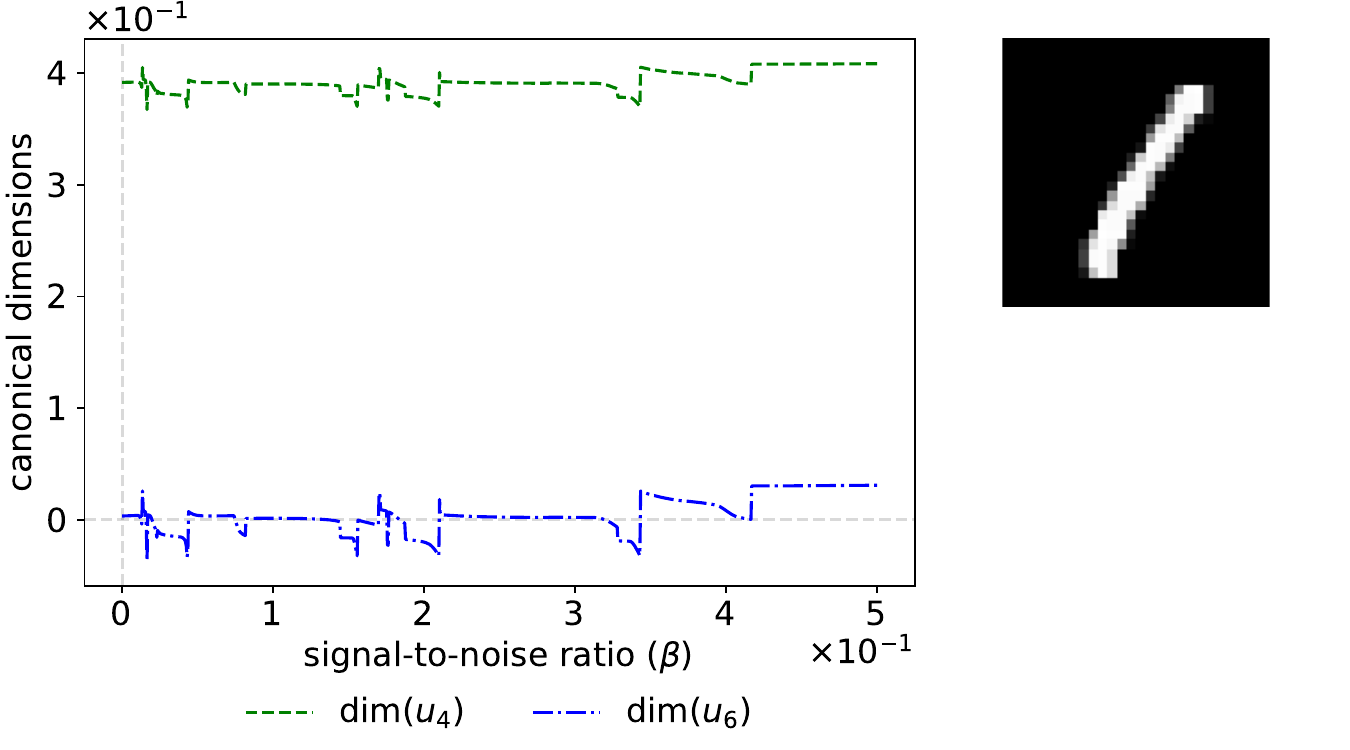}
    \hfill
    \includegraphics[width=0.45\textwidth]{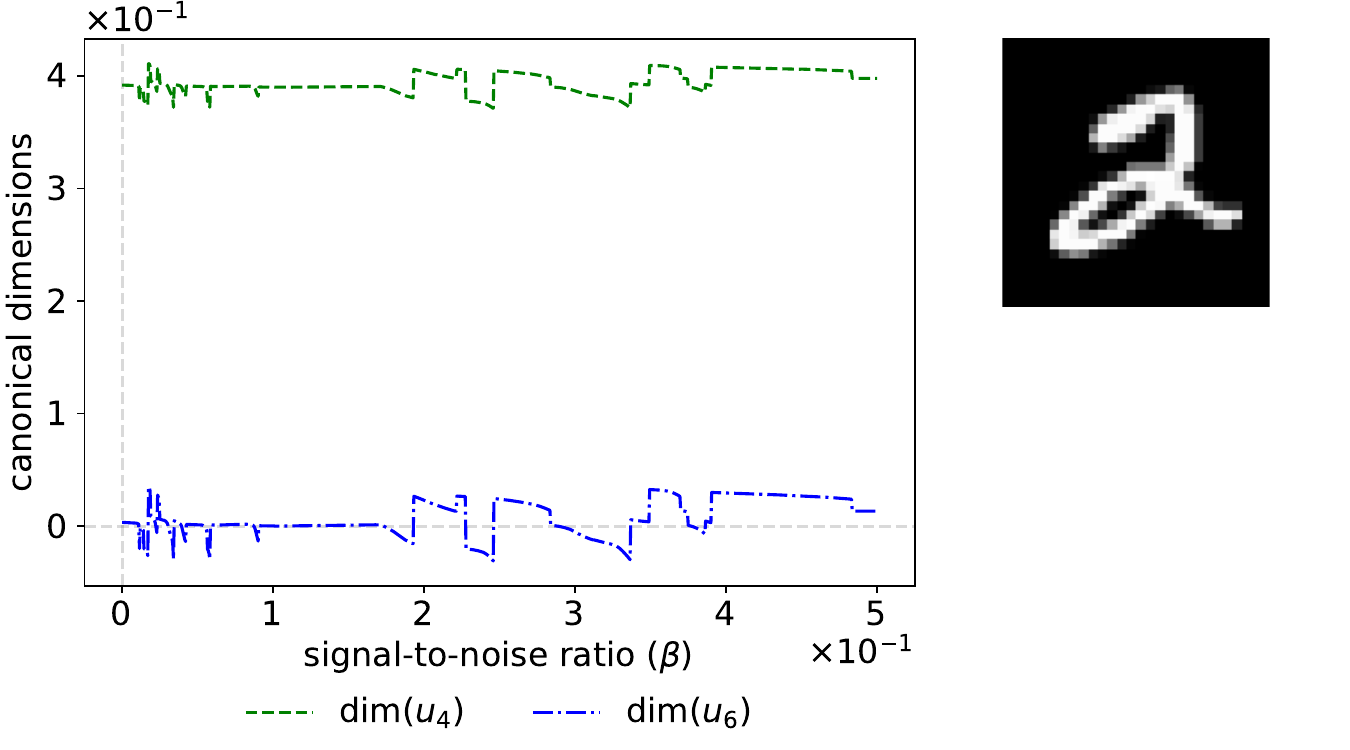}\\[0.5em]
    \includegraphics[width=0.45\textwidth]{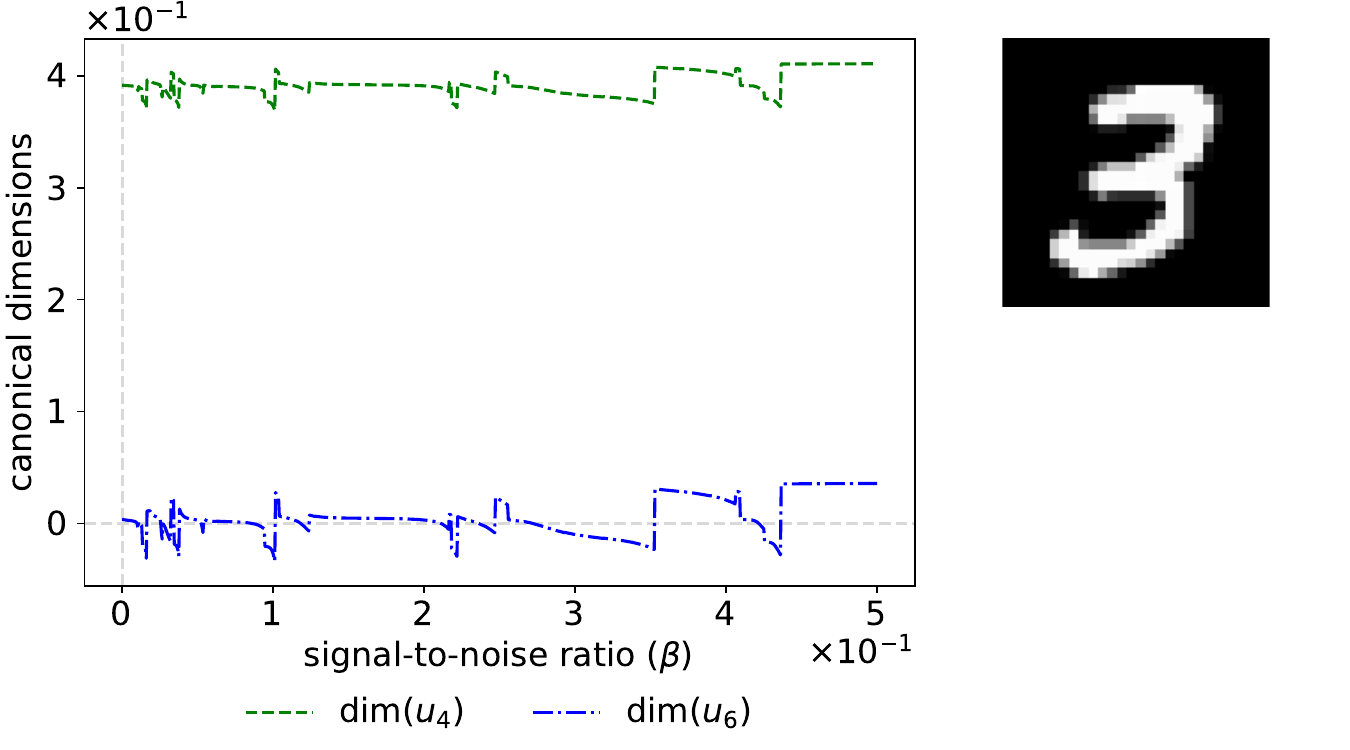}
    \hfill
    \includegraphics[width=0.45\textwidth]{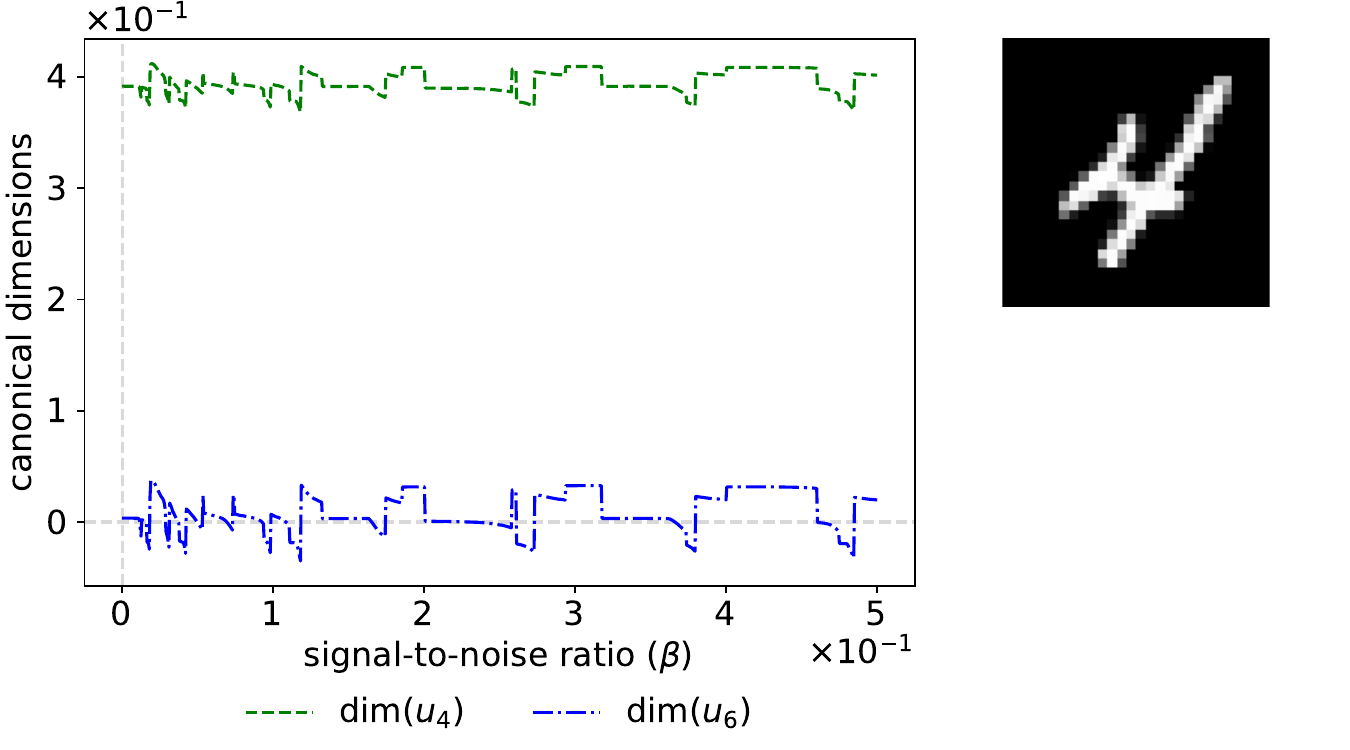}
    \caption{%
        Typical behaviour of the canonical dimension in the MNIST set.
    }\label{fig:mnistplot}
\end{figure}

In Figure~\ref{fig:figplotgianduja} we show the global behaviour of the canonical dimensions by increasing progressively the \snr magnitude $\beta$.
In all cases, the dashed black curve represents the empirical inverse distribution $\rho(k^2)$, except for the first figure of the left corner which is for the analytic \mpdistr distribution.
We can observe some important features:

\begin{enumerate}
    \item the values of the canonical dimensions fluctuate, and these fluctuations increase with the rank $n$ of the interactions (essentially by a factor $(n-1)$);
    \item the best linear interpolations (ignoring the end points of the dataset, corrupted by some numerical instabilities due to some residual spike) agree with the analytic predictions for $\beta = 0$;
    \item the plots illustrate the \emph{rigidity property} of the distribution: the canonical dimension is essentially unaffected as $\beta$ remains small enough, but changes significantly at a certain value that we identify as the \lod $\beta_t$ (for this example, $\beta_t \approx 0.15$).
\end{enumerate}

\begin{table}[t]
    \caption{Summary of the detection thresholds identified by the \frg flow.}\label{tab:thresholds}
    \centering
    \begin{tabular}{@{}cp{0.6\textwidth}@{}}
        \toprule
        \textbf{Symbol} & \textbf{Physical Interpretation}                                                                                                                    \\
        \midrule
        $\beta_t$       & \textbf{Limit of detection (\lod).} The ``rigidity'' threshold where canonical dimensions first deviate from the noise baseline.                    \\
        $\beta_c$       & \textbf{Critical threshold.} The value where the canonical dimension of $u_4$ vanishes ($\dim\qty(u_4) = 0$), marking the effective critical point. \\
        $\beta_O$       & \textbf{Optimal threshold.} The first minimum of $\dim\qty(u_4)$ corresponding to the maximum signal contrast before cyclic effects appear.         \\
        \bottomrule
    \end{tabular}
\end{table}

It is instructive to compare these detection thresholds with the standard \bbp phase transition, which governs the detection of isolated spikes (low-rank signals) in random matrices~\cite{math3}.
In the standard spiked covariance model, the \bbp transition occurs when the signal strength is sufficient to push the largest eigenvalue out of the \mpdistr bulk
For the additive noise model considered here ($X = \beta S + Z$) with $q = P / N$, the critical \bbp threshold is given by $\beta_{\text{BBP}} = q^{1/4}$~\cite{Bouchaud3}.
Using our simulation parameters ($N=\num{2.0e4}$, $q=0.9$), we find $\beta_{\text{BBP}} \approx 0.97$.
Our detection threshold $\beta_t \approx 0.15$ is significantly lower ($\beta_t \ll \beta_{\text{BBP}}$).
This confirms that the \frg framework detects the signal while it is still deeply buried within the bulk of the spectrum, corresponding to the ``bimodal connected phase'' identified recently in the extensive spike model as the emergence of a bimodal distribution of the spectrum~\cite{Landau2023}.

Two other detection thresholds can be defined and motivated from physics.
The first one, the \emph{critical detection threshold} $\beta_c$, is the value at which the asymptotic canonical dimension of $u_4$ at the scale $k^2_{\text{IR}}$ vanishes (i.e.\ the local critical dimension is exactly $4$).
In the example, $\beta_c \approx 0.32$.
The third and last threshold we define is the \emph{optimal threshold} $\beta_O$, defined as the first minimum for $\dim_{\tau}\qty(u_4)$ \emph{below} $\beta_c$.
In the example, $\beta_O \approx 0.37$.
In general $\beta_t \le \beta_c < \beta_O$.
See Table~\ref{tab:thresholds} for a summary of the detection thresholds.

Figure~\ref{fig:figplotgianduja2} illustrates specifically the behaviour of the canonical dimension at the scale $k^2_{\text{IR}}$, and enables us to visualise pragmatically the transition between two different regimes.
In the \emph{rigid regime} $\beta \in \qty[0,\beta_t]$, the \ir canonical dimensions remain essentially constant, up to fluctuations due to the intrinsic variability of the noise around the analytic asymptotic spectra (see Section~\ref{sec:sectionVS}).
Then, variations become larger.
The physical interpretation in terms of \rg is immediate: a strong enough signal makes the flow Gaussian, entirely driven by the flow of the mass which remains the only relevant parameter.
Let us recall that the physical mass is the inverse of the largest eigenvalue $\lambda_0$. Thus, in the immediate vicinity of the Gaussian point, the presence of a signal makes the Gaussian theory, of variance $\lambda_0^{-1}$ a good approximation of the effective behaviour of the microscopic \dof in the spectrum of the correlation matrix.
Our analysis focuses specifically on the couplings $u_2$, $u_4$, and $u_6$, as they represent the only relevant or marginal interactions for the \mpdistr universality class in the deep \ir~\cite{RG5}.
Within the \rg framework, all higher-order interactions ($u_{2n}$ for $n > 3$) are strictly irrelevant: they are rapidly suppressed as the flow evolves toward the \ir and thus carry no significant information regarding signal-induced deformations.

\begin{figure}[t]
    \centering
    \includegraphics[width=0.66\textwidth]{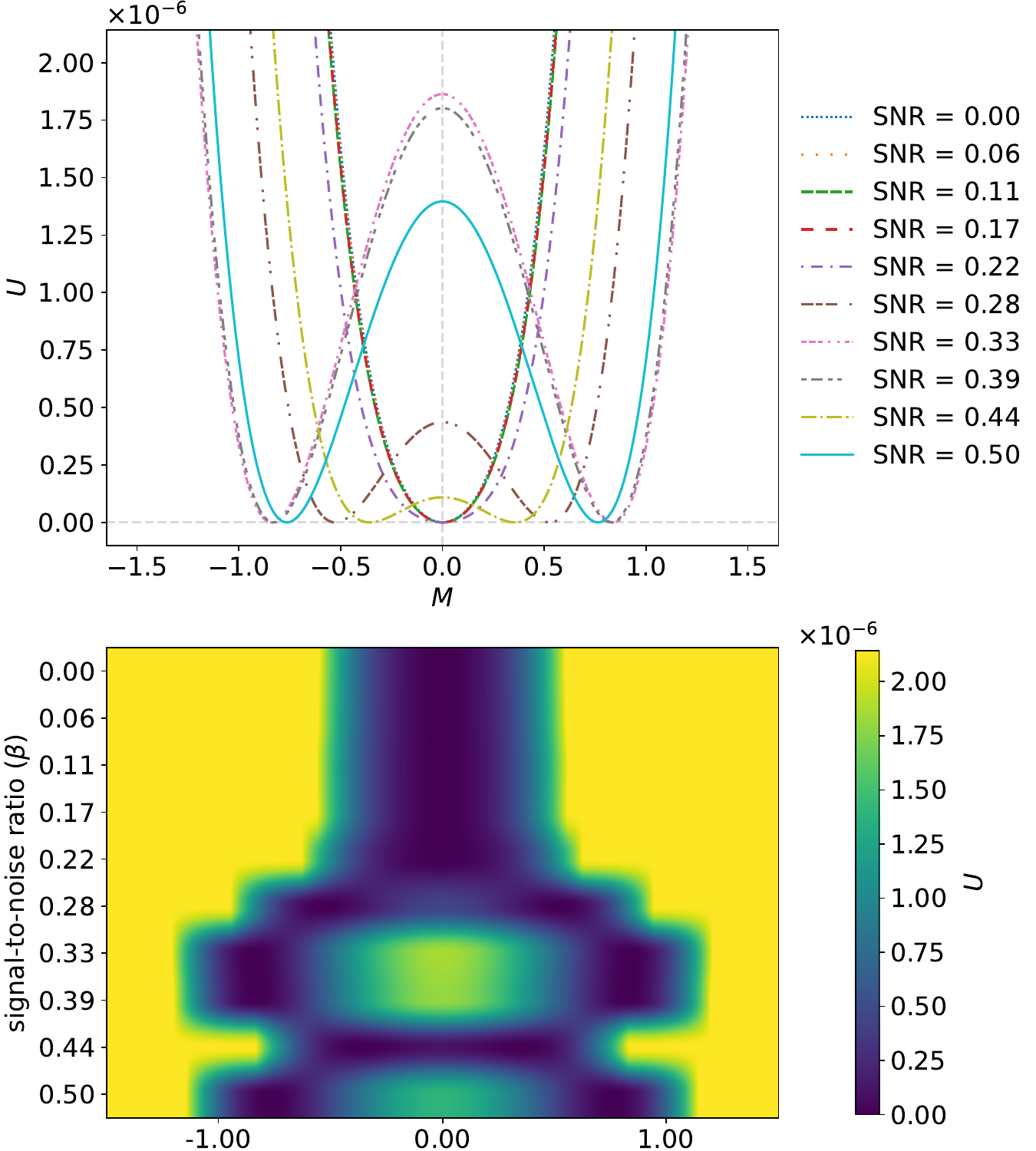}
    \caption{%
        Illustration of the symmetry breaking scenario for larger \snr ($\beta$) using the dataset in Figure~\ref{fig:gianduja}.
        The figure shows the behaviour of the effective potential in the \ir, for the initial conditions at the mesoscopic scale $\Lambda$: $\bar{u}_2(\Lambda) = \num{-8.24e-6}$, $\bar{u}_4(\Lambda) = \num{2.70e-6}$ and $\bar{u}_6 = \num{1.73e-6}$.
    }\label{fig:plotpot1}
\end{figure}

Note that for $\beta > \beta_O$, the canonical dimensions increase again, before decreasing further.
The quartic coupling may even become relevant again, before its canonical dimension becomes negative again.
This cyclic behaviour and its interpretation will be discussed in Section~\ref{sec:largebeta}.
For now, we shall focus on the neighbourhood of $\beta_O$, where our approximations concerning the flow, and in particular the \lpa, seem physically justified.

\begin{figure}
    \centering
    \includegraphics[width=0.66\textwidth]{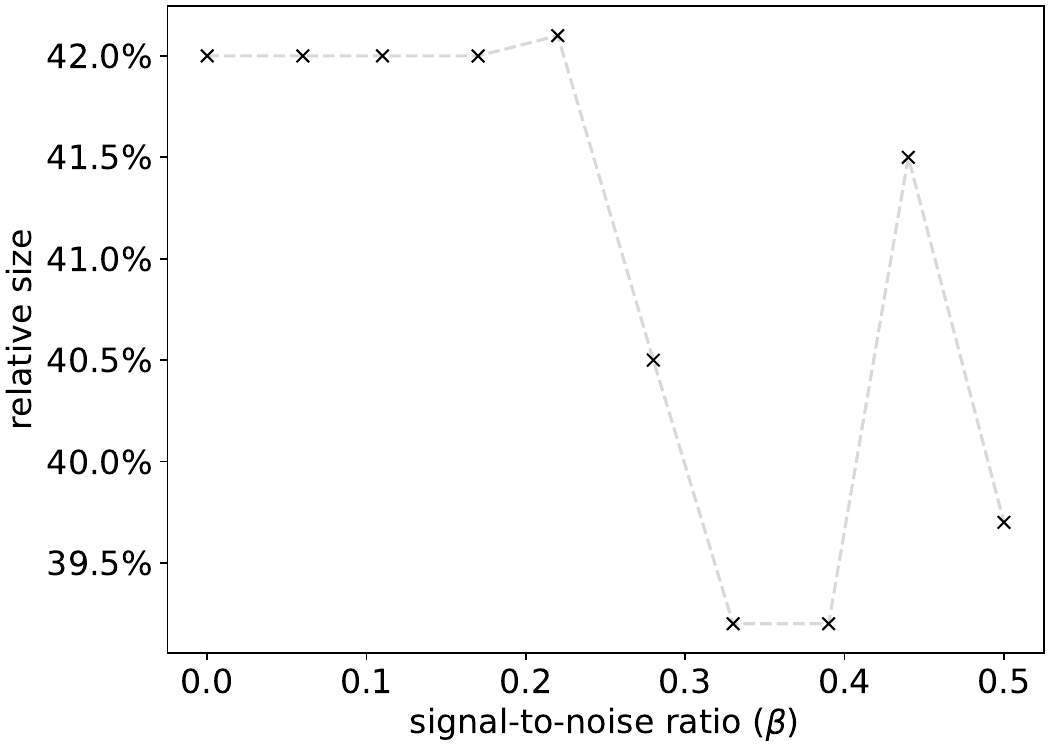}
    \caption{%
        Behaviour of the size (relative to the number of sampled initial conditions) of the symmetric phase with respect to $\beta$.
    }\label{fig:plotpot2}
\end{figure}

The behaviour of the canonical dimension in Figure~\ref{fig:figplotgianduja2} can be compared with the canonical dimension for the handwritten digits in Figure~\ref{fig:mnistplot}.
As well-known in literature, these samples are usually simple enough in nature for their features to be captured by algorithms such as \pca.
As stated, no clear signal is thus expected in the bulk associated to these samples, since the isolated spike capture almost all the relevant information.
However, we might expect to detect some remnants of the nearly continuous deformation, due to the presence of a non trivial signal, as shown by the variation in behaviour of the sextic coupling, though the quartic coupling remains relevant.

Up to this point, we only considered properties around the Gaussian fixed point, which is also the only global fixed point because of the scale dependency of the canonical dimensions.
Now, let us focus on trajectories that are initially quite close to the Gaussian point, but are still far away for non-Gaussian effects to appear, without compromising the reliability of our approximations.\footnote{%
    Numerically, we sample \num{2.5e3} points using a \emph{Latin Hypercube Sampling} scheme in the box $\qty{\qty(\bar{u}_2, \bar{u}_4, \bar{u}_6) ~\mid~ \bar{u}_n \in \qty[-\num{1e-5}, \num{1e-5}] ~ \forall n = 2, 4, 6}$ at a mesoscopic scale $\Lambda$.
}
In particular, we deal with the overall shape of the potential, which is a feature that is more robust to approximations than the values of the couplings themselves.
The results are shown in Figure~\ref{fig:plotpot1} and Figure~\ref{fig:plotpot2}.
Figure~\ref{fig:plotpot1} shows the evolution of the potential at the scale $k^2_{\text{IR}}$ as a function of the \snr, with initial conditions in the symmetry restoration region (orange region in Figure~\ref{fig:figWF}).
In complement, Figure~\ref{fig:plotpot2} shows the evolution of the size of the region where symmetry is restored as a function of $\beta$, and we see that the largest variations follow exactly those of the canonical dimension.
The size of the symmetric phase is measured numerically by sampling initial conditions for the potential in the local approximation and counting the fraction of trajectories that flow towards a symmetric minimum ($\mathbb{Z}_2$ invariant) versus a broken symmetry minimum.
The transition shown in Figure~\ref{fig:plotpot1}, which associates the presence of a signal with a breaking of the $\mathbb{Z}_2$ symmetry, is the consequence of the modification of the shape of the empirical distribution of eigenvalues in the \ir, and therefore has a dimensional origin.
Unlike the standard spontaneous symmetry breaking in $\varphi^4$ models, which is driven by a thermal mass parameter (temperature), this transition is driven by a change in the effective dimensionality of the system itself.
As the signal deforms the spectrum, the scaling of the couplings changes, effectively shifting the system from a symmetric phase (noise-dominated, $D \approx 3$) to a broken phase (signal-dominated, $D > 4$).
We will thus define it as a \emph{dimensional symmetry breaking}, and to our knowledge this is the only case recorded in the literature.

\begin{figure}
    \centering
    \includegraphics[width=0.66\textwidth]{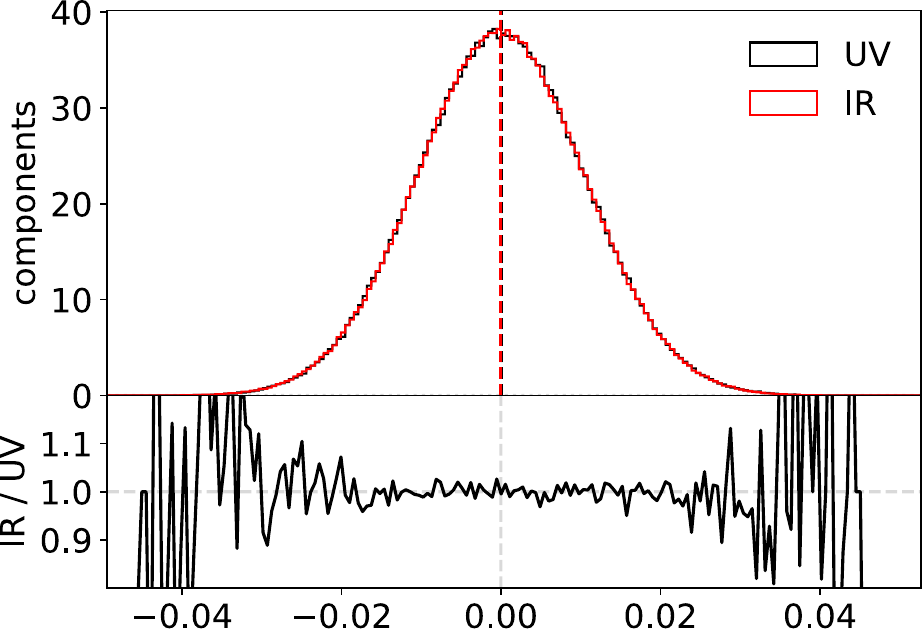}
    \caption{%
        Distribution of the eigenvector components in the \uv (small eigenvalues, 100 eigenvectors), and in the \ir (large eigenvalues, 100 eigenvectors) for $\beta = 0$ (pure noise).
    }\label{fig:distbeta0}
\end{figure}

To conclude this presentation of the results near the detection threshold, let us consider the statistical properties of eigenvectors.
First, consider eigenvectors in the \mpdistr class.
For purely noisy data, eigenvectors
\begin{equation}
    u_\lambda \coloneqq (u_\lambda^1,u_\lambda^2,\dots, u_\lambda^P),
\end{equation}
are \emph{delocalised} with entries not greater than $\simeq 1$~\cite{bouferroum:hal-00835504,bogomolny2017modification}.
Moreover, the corresponding rotation eigenmatrix is asymptotically fully Haar distributed on the group $\mathrm{O}(N)$, for large $N$.
Without additional information, the distribution of the components, $s = u_{(\mu)}^{i}$, as $i$ varies, can be well estimated by the maximum entropy distribution satisfying the constraint $\sum_i {(u_\lambda^i)}^2 = N$.
The corresponding maximum entropy distribution is the \emph{Porter-Thomas} distribution~\cite{Porter1956}:
\begin{equation}
    p(s) = \frac{1}{\sqrt{2\pi}}\exp(-\frac{s^2}{2}).
\end{equation}
This behaviour is confirmed empirically for $\beta = 0$, as Figure~\ref{fig:distbeta0} shows (we consider the eigenvectors corresponding to the 100 smaller eigenvalues for \uv, and those at the \mpdistr mass scale $k^2 = \qty{(\lambda_+ - \lambda_-)}^{-1}$ for the \ir, in order to compare different values of $\beta$ consistently).
The distribution agrees with the Porter-Thomas maximal entropy estimator, seemingly confirming, as it is well-known, that the true distribution is no more structured than the Porter-Thomas distribution.

\begin{figure}
    \centering
    \begin{minipage}{0.4\textwidth}
        \centering
        {\large $\beta = 0.2$} \\[0.3em]
        \includegraphics[width=\linewidth]{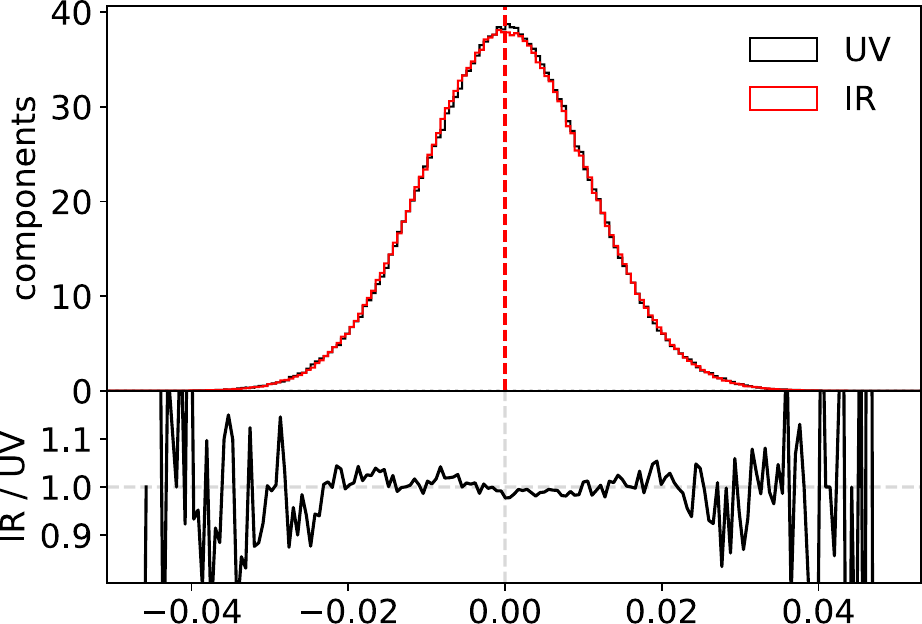}
    \end{minipage}
    \hfill
    \begin{minipage}{0.4\textwidth}
        \centering
        {\large $\beta = 0.3$} \\[0.3em]
        \includegraphics[width=\linewidth]{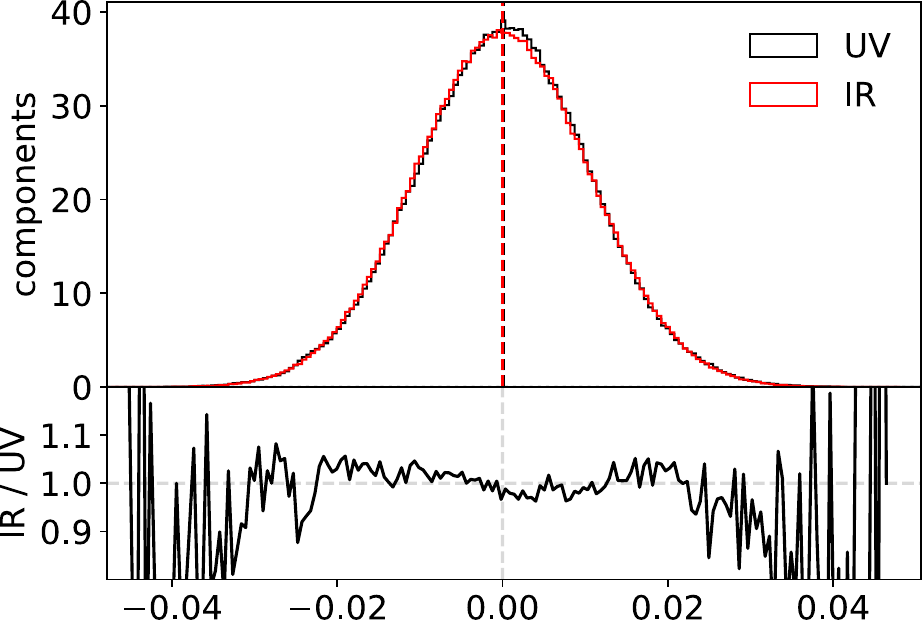}
    \end{minipage}
    \\[1em]

    \begin{minipage}{0.4\textwidth}
        \centering
        {\large $\beta = 0.4$} \\[0.3em]
        \includegraphics[width=\linewidth]{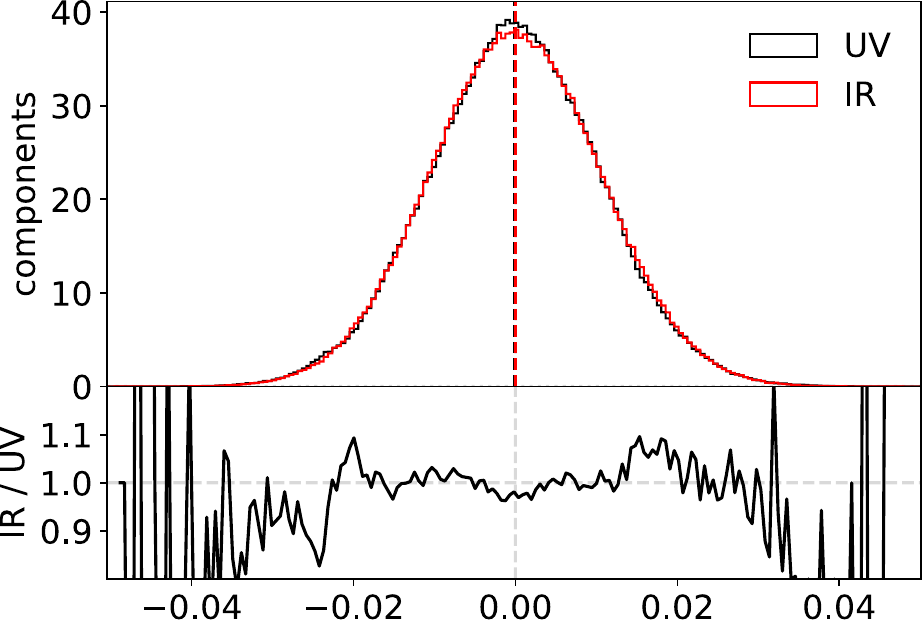}
    \end{minipage}
    \hfill
    \begin{minipage}{0.4\textwidth}
        \centering
        {\large $\beta = 0.5$} \\[0.3em]
        \includegraphics[width=\linewidth]{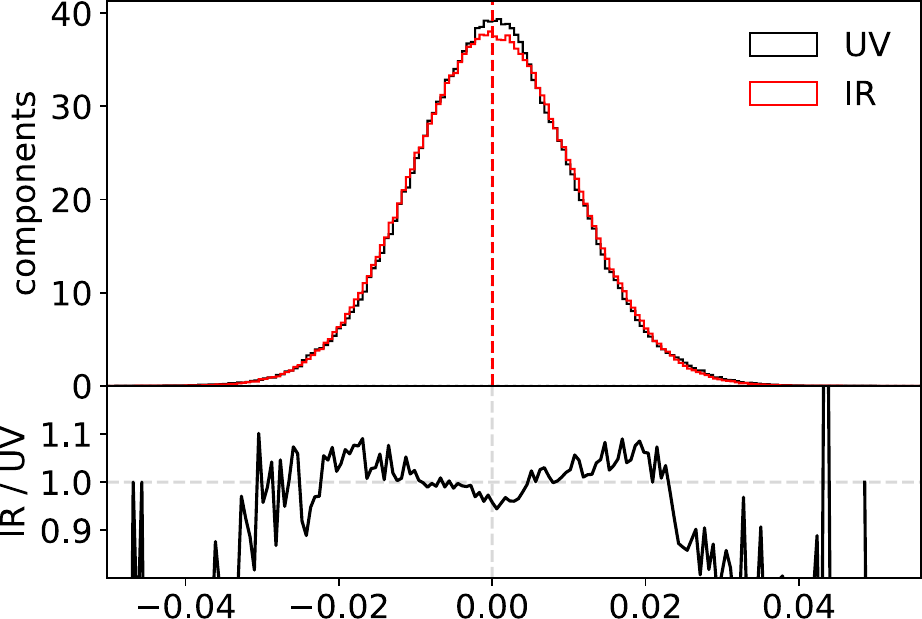}
    \end{minipage}

    \caption{%
        \uv and \ir distributions of eigenvector components for different values of $\beta$.
    }\label{fig:distbetabeta}
\end{figure}

\begin{figure}
    \centering
    \includegraphics[width=\textwidth]{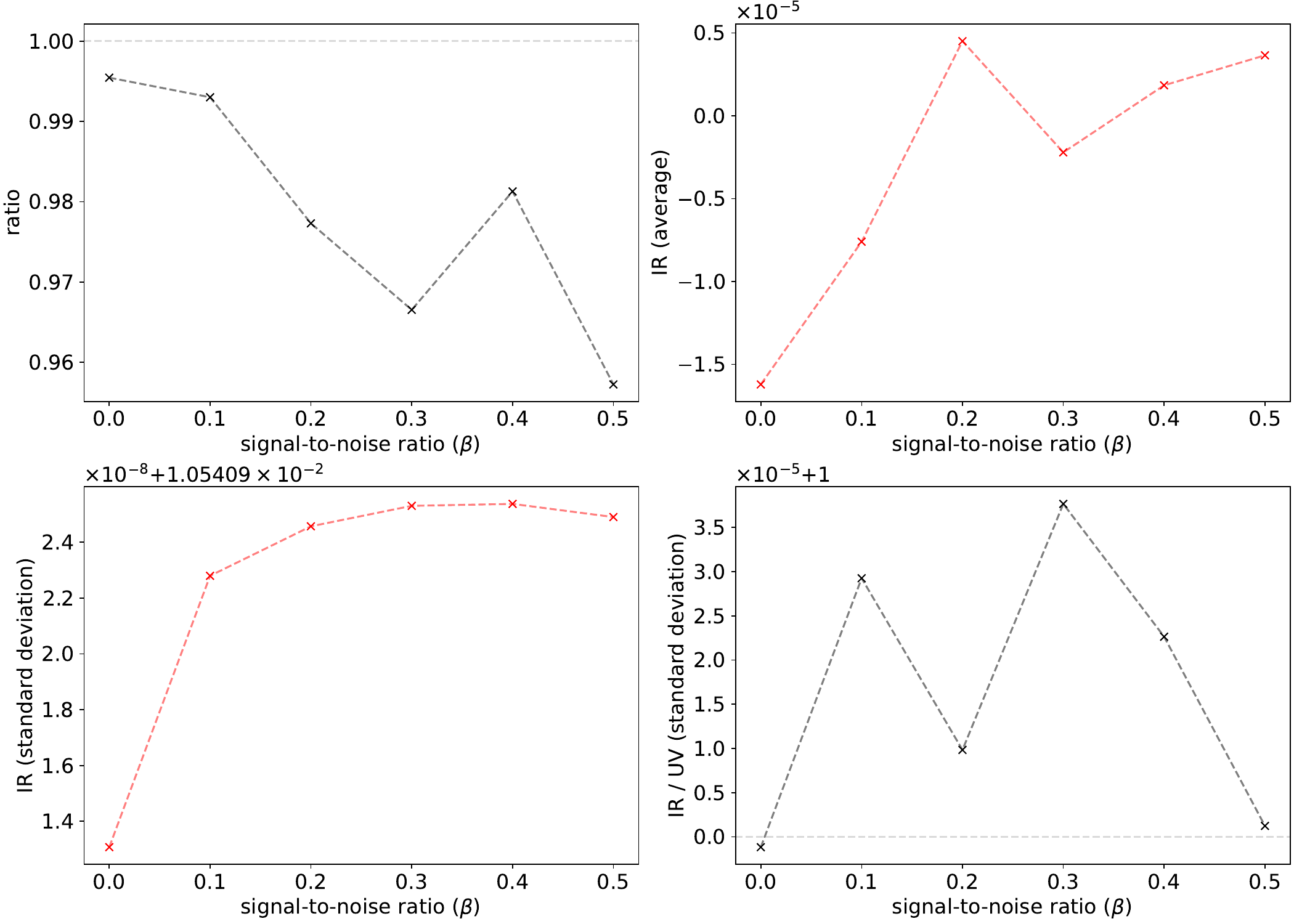}
    \caption{%
        Summary of relevant statistical features of eigenvectors distributions (the value of the ratio plot at the origin, the mean value and standard deviation of the \ir components, the ratio of the standard deviation of \ir and \uv components).
    }\label{fig:distbetabeta2}
\end{figure}

Figure~\ref{fig:distbetabeta} illustrates the statistic of eigenvectors for increasing values of the \snr, the relevant properties being summarised in Figure~\ref{fig:distbetabeta2}.
These results clearly show that a change in the statistics occurs for values of $\beta$ associated with significant changes in the canonical dimensions and justify the definitions of $\beta_t$, $\beta_c$ and $\beta_O$ as strong indicators of the presence of a signal.
The major change concerns the standard deviation of the distribution, which increases with the signal intensity, and the ratio of the IR and UV standard deviations marks a peak at each minimum of the canonical dimensions, and in particular near the value of $\beta_O$.
Furthermore, a shift in the mean is also associated with the presence of a signal.
This numerical result agrees with the theoretical result~\cite{bogomolny2017modification}, which however concerns a single spike.

\subsection{Intrinsic variability}\label{sec:sectionVS}

\begin{figure}
    \centering
    \includegraphics[width=0.4\textwidth]{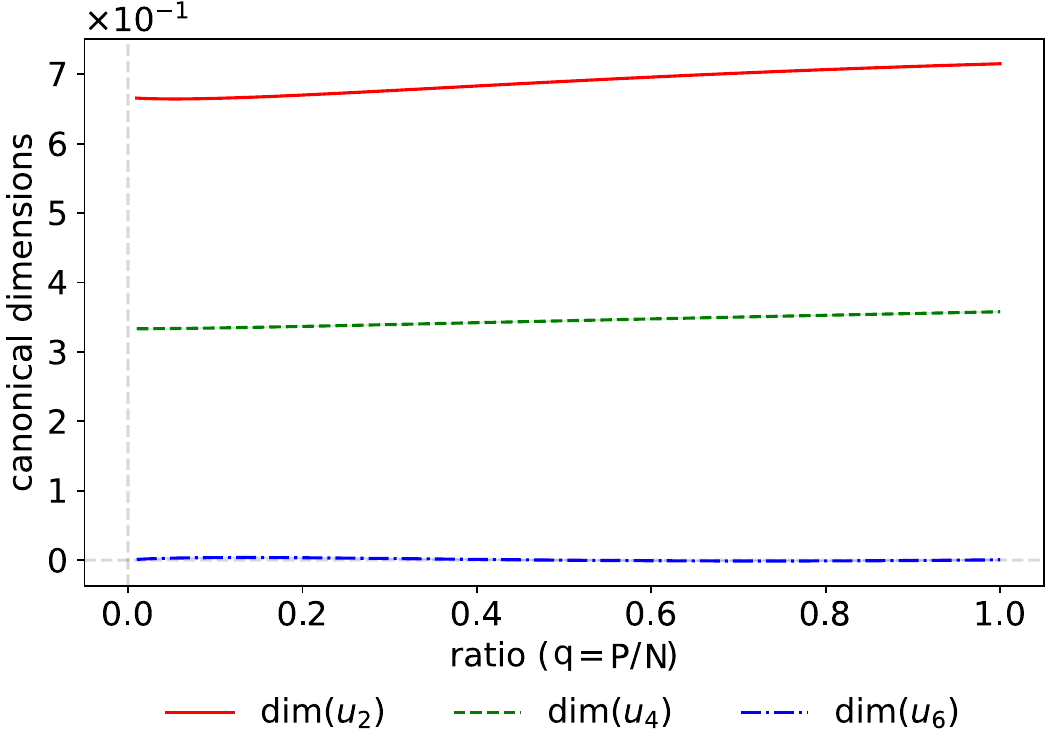}
    \hfill
    \includegraphics[width=0.4\textwidth]{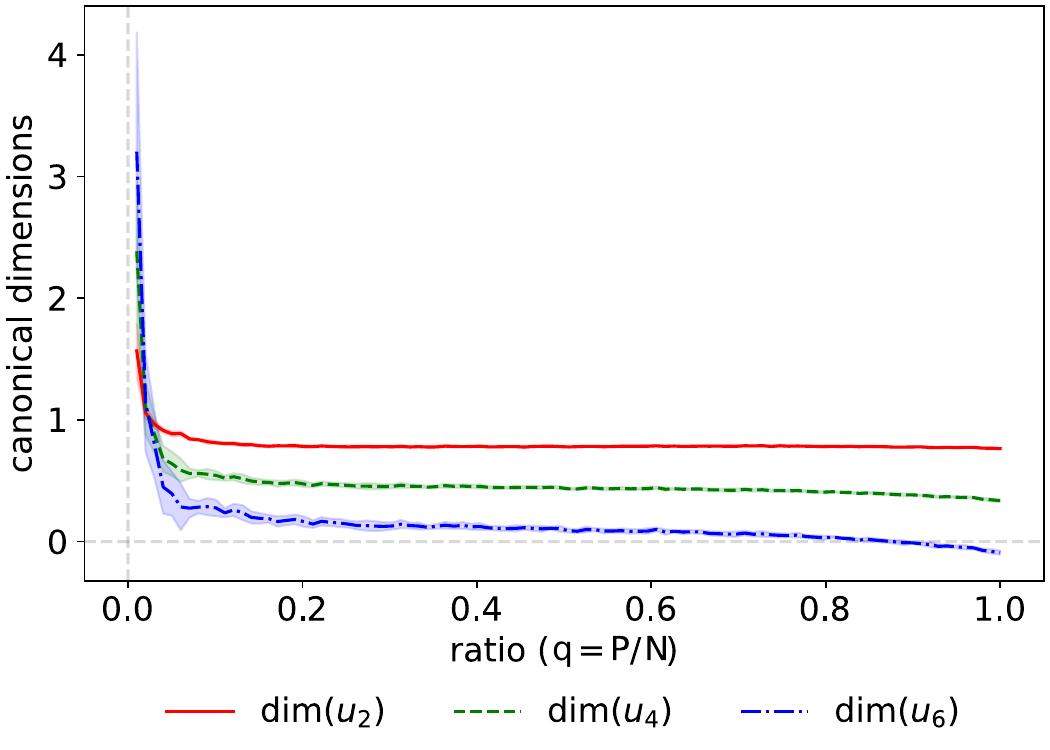}
    \caption{%
        (right) Behaviour of the empirical canonical dimension in the \ir with respect to $q$ ($N$ fixed).
        (left) Same behaviour using the analytic \mpdistr law.
    }\label{fig:figPvar}
\end{figure}

\begin{figure}
    \centering
    \includegraphics[width=0.4\textwidth]{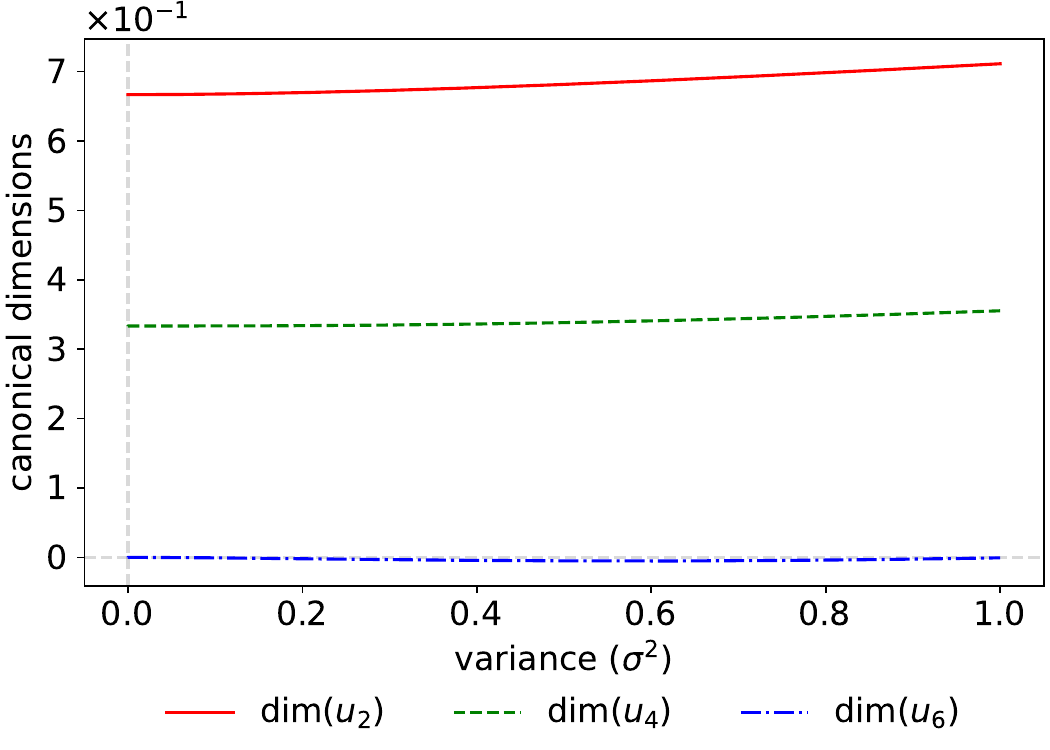}
    \hfill
    \includegraphics[width=0.4\textwidth]{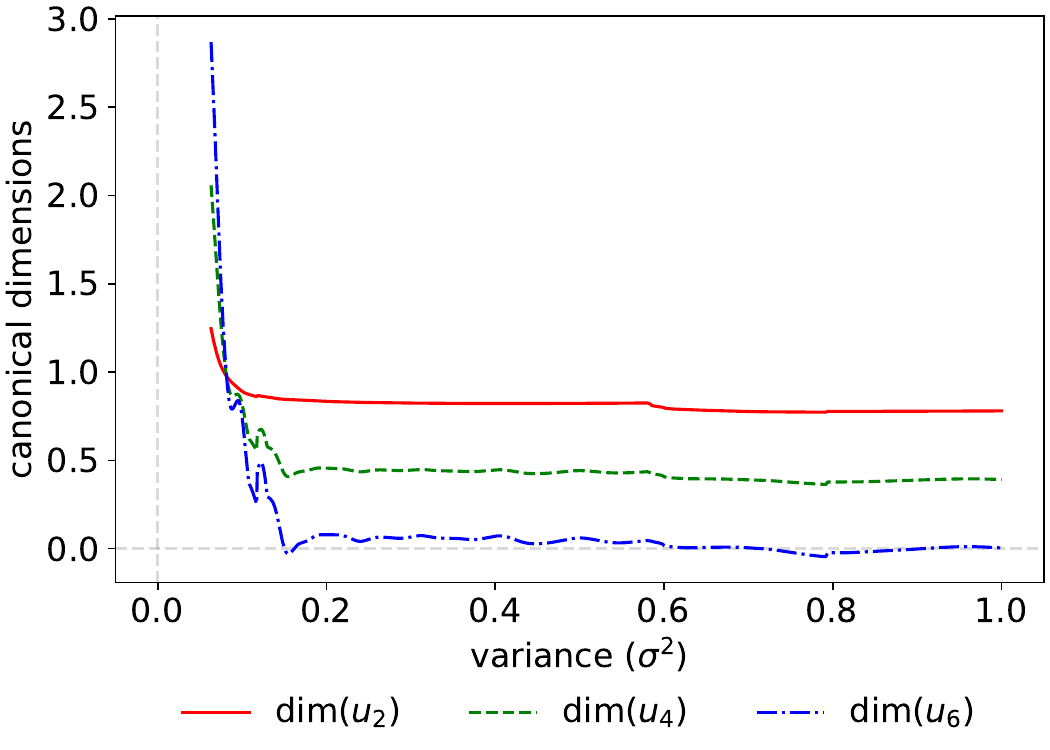}
    \caption{%
        (right) Behaviour of the empirical canonical dimension in the \ir with respect to the variance $\sigma^2$.
        (left) Comparison with the analytic predictions derived from the \mpdistr law with varying variance.
    }\label{fig:deviation}
\end{figure}

As noted in Section~\ref{sec:mainresults}, the influence on the values of the canonical dimensions of intrinsic fluctuations in the data is several orders of magnitude smaller than those induced by the signal.
More precisely, this actually fixes a single particular detection threshold that we could call \emph{variability threshold}, which we will discuss in more detail in Section~\ref{sec:secdistance}, where we will propose a specific criterion to evaluate it.
Here, we simply propose a comparison of typical variability-related effects with those induced by the signal, in the neighbourhood of $\beta_O$.


In this context, ``variability'' refers to the aleatoric uncertainty arising from finite-size effects in the random matrix realisations. Even for a fixed set of macroscopic parameters ($N$, $P$, $q$, $\sigma^2$), different microscopic realisations of the noise matrix $Z$ induce fluctuations in the empirical eigenvalue spectrum. Quantifying this baseline variance is essential to distinguish genuine signal-induced symmetry breaking from statistical noise.

\begin{figure}
    \centering
    \includegraphics[width=0.66\textwidth]{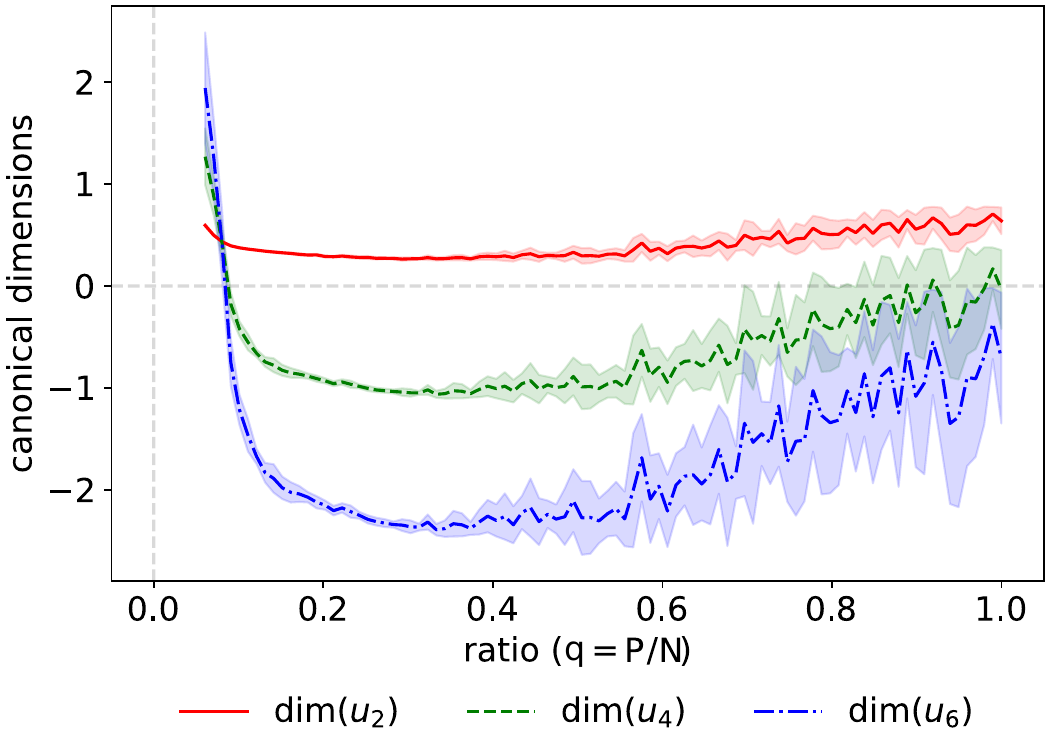}
    \caption{%
        Behaviour of the empirical canonical dimension in the \ir with non-zero \snr ($\beta=1.25$) with respect to $q$ (keeping $N = \num{2e4}$ fixed) using the dataset in Figure~\ref{fig:gianduja}.
    }\label{fig:Pvarbetaneq0}
\end{figure}

The results concerning our data are summarised in Figure~\ref{fig:figPvar}, Figure~\ref{fig:deviation} and Figure~\ref{fig:Pvarbetaneq0}.
In particular, Figure~\ref{fig:figPvar} and Figure~\ref{fig:deviation} illustrate and quantify the variability of the canonical dimensions with respect to the realisation in the \ir for different choices of the distribution parameters, such as the value of $q$ (averaged over 100 random realisations each time, keeping $N$ fixed) and the variance of the distribution.
Figure~\ref{fig:Pvarbetaneq0} shows the behaviour of the canonical dimension as a function of $q$ (averaged over 100 realisations) for a different value of $\beta > 0$ (keeping $N$ fixed and large).
This result illustrates unambiguously that the magnitude of the effects (especially the sign of the dimension of the quartic coupling) due to the presence of the signal is far larger than that coming from intrinsic fluctuations of the data.

Random matrix theory allows us to define quantitatively the typical size of fluctuations induced by finite sample effects, particularly on the tail of the spectrum.
Indeed, it is well-known that the scale of fluctuation for the top eigenvalue is given by the Tracy-Widom distribution, and is typically $\sim P^{-2/3}$~\cite{Bouchaud3}.
Hence, signal-induced effects on the power counting become comparable in magnitude as fluctuation for $\beta \simeq \beta_0 \coloneqq P^{-2/3}$.
Once again, it is important to understand that this is an intrinsic limitation of our approach.
For the values of $P$ considered in this work, we find $\beta_0 \approx \num{2.6e-3}$, far enough from the typical detection scale $(\beta \sim 0.3)$.
However, this observation has no absolute value, the detection scale being fixed by the dataset itself.
This separation of scales ($\beta_t \gg \beta_0$) confirms that the dimensional phase transition observed at $\beta_t \approx 0.15$ is a robust macroscopic effect, distinct from the finite-size fluctuations of the spectral edge.

\subsection{An attempt at formalisation}\label{sec:secdistance}

We can further formalise these definitions by motivating a new notion of distance between distributions, inspired by our numerical experiments.
In this section, we formalise the comparison between the empirical signal and the noise background by defining a set of spectral distances.
These quantities act as order parameters, measuring the deviation of the empirical renormalization group flow from the universal \mpdistr baseline.

We first define the notions of \mpdistr \emph{distribution proxy}.
Let $\mu(\lambda)$ be some empirical spectrum and let $\mu_{\Delta_{\text{phys}}}(\lambda)$ be the bulk distribution (whose definition depends on $\Delta_{\text{phys}}$), and let $\lambda_{\pm}$ be the edge values of the spectrum. Note that, in what follows, a pessimistic estimation for $\lambda_+$ is given as the first global minimum (in the IR region, disregarding local fluctuations) of $\dim u_4$ in the bulk.
\begin{definition}
    Let $\mathcal{D}$ be the adherent set of \mpdistr distributions $\nu_{\sigma_*^2,q}\equiv \nu(q)$ (see~\eqref{MPdistribution}) with ratio $q$ and variance $\sigma_*^2=\frac{\lambda_+-\lambda_-}{4 \sqrt{q}}$.
\end{definition}
The \mpdistr distributions in the \emph{adherent set} have edge bounds $\lambda_\pm$, and we define the \emph{distance} $G_\mathcal{D}(\mu,\nu)$ as:
\begin{definition}
    Let $\mu$ be the empirical spectral density and $\nu \in \mathcal{D}$ be a reference \mpdistr distribution.
    The direct Gaussian\footnote{%
        It is ``Gaussian'' since it uses the Gaussian power counting.
    } distance $G_\mathcal{D}(\mu,\nu)$ between $\mu$ and $\nu$ is defined as:
    \begin{equation}
        G_\mathcal{D}\qty(\mu,\nu(q)) \coloneqq \max_{(\lambda_-,\lambda_+)} \, \Big\vert  \dim_{\tau}\qty(u_4)\big\vert_{\mu} - \dim_{\tau}\qty(u_4)\big\vert_{\nu} \, \Big\vert.
    \end{equation}
\end{definition}
\begin{definition}
    The \mpdistr distribution proxy $\nu_{*}(\lambda)\in \mathcal{D}$ is the analytic \mpdistr distribution such that:
    \begin{equation}
        G(\mu,\nu_{*})=\min_{q} \max_{(\lambda_-,\lambda_+)} \, \Big\vert  \dim_{\tau}\qty(u_4)\big\vert_{\mu} - \dim_{\tau}\qty(u_4)\big\vert_{\nu} \, \Big\vert.
    \end{equation}
\end{definition}
This definition in particular implies that:
\begin{equation}
    \frac{\mathrm{d}}{\mathrm{d} q}\, G_{\mathcal{D}}(\mu,\nu(q)) \Big\vert_{\nu=\nu_{*}}=0.
\end{equation}
Moreover, notice that $G(\mu,\nu_{*})$ is also a good definition of the distance between $\mu$ and the set $\mathcal{D}$, and we define the \emph{direct concordance index} $\eta(\mu)$ of the distribution $\mu$ as:
\begin{equation}
    \eta(\mu) \coloneqq  G(\mu,\nu_{*}).
\end{equation}

In other words, the adherent set $\mathcal{D}$ represents the topological closure of the family of \mpdistr distributions consistent with the observed spectral edges.
We further introduce the concordance index $\eta(\mu)$ as a global measure of the agreement between the empirical and reference \rg flows.

We moreover define the \emph{direct absolute global adherence} $\zeta(\mu)$ of the distribution $\mu$ as:
\begin{equation}
    \zeta(\mu) \coloneqq \min_\lambda \, \Big\vert  \dim_{\tau}\qty(u_4)\big\vert_{\mu} - \dim_{\tau}\qty(u_4)\big\vert_{\nu_*} \, \Big\vert.
\end{equation}
Both $\eta(\mu)$ and $\zeta(\mu)$ quantify the proximity and the fluctuations around the \mpdistr distribution proxy.
These global quantities are not necessarily the most relevant for the signal detection problem, however.
Indeed, we have seen in the previous sections that in use cases, fluctuations are more significant in the \uv, while signal effects and large deviations mainly affect the \ir properties.
Furthermore, we will see in Section~\ref{sec:largebeta} that the sign of the difference has an informational meaning.
For these reasons, we propose the following definitions:
\begin{definition}
    We define the local direct concordance index at scale $\lambda$, $\eta_{<\lambda}(\mu)$ and the direct relative adherence $\zeta_{<\lambda}(\mu)$ of the distribution $\mu$ as:
    \begin{equation}
        \eta_{<\lambda}(\mu) \coloneqq
        \max_{(\lambda,\lambda_+)} \, \Big\vert  \dim_{\tau}\qty(u_4)\big\vert_{\mu} - \dim_{\tau}\qty(u_4)\big\vert_{\nu_*} \, \Big\vert,
    \end{equation}
    \begin{equation}
        \zeta_{<\lambda}(\mu)\coloneqq \min_{(\lambda,\lambda_+)} \, \Big( \dim_{\tau}\qty(u_4)\big\vert_{\nu_*} - \dim_{\tau}\qty(u_4)\big\vert_{\mu} \, \Big).
    \end{equation}
\end{definition}
The second quantity in particular is sensitive to the \emph{relative sign}.
From the previous analysis, we know that the sign must be positive if a signal is present in the spectrum.

\begin{figure}[t]
    \centering
    \includegraphics[width=\textwidth]{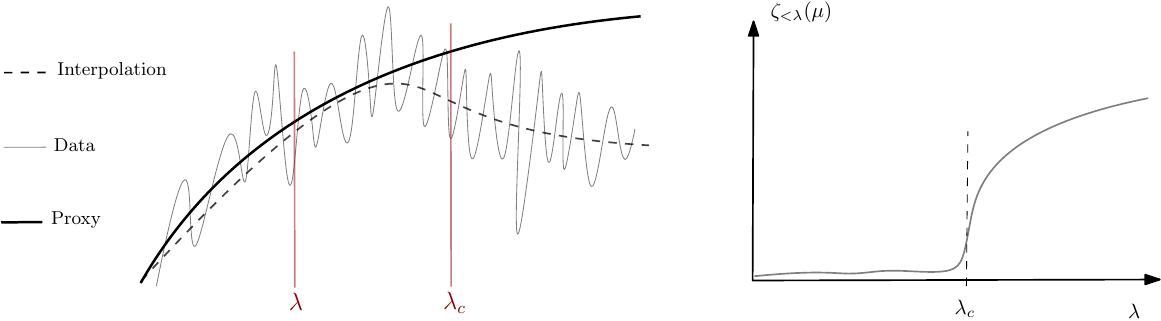}
    \caption{%
        Qualitative illustration of the way the definition $\zeta_{<\lambda}(\mu)$ works.
    }\label{fig:illustration}
\end{figure}

\begin{figure}[t]
    \centering
    \begin{minipage}[b]{0.45\textwidth}
        \centering
        {\large $\beta = 0.21$} \\[0.5em]
        \includegraphics[height=0.19\textheight]{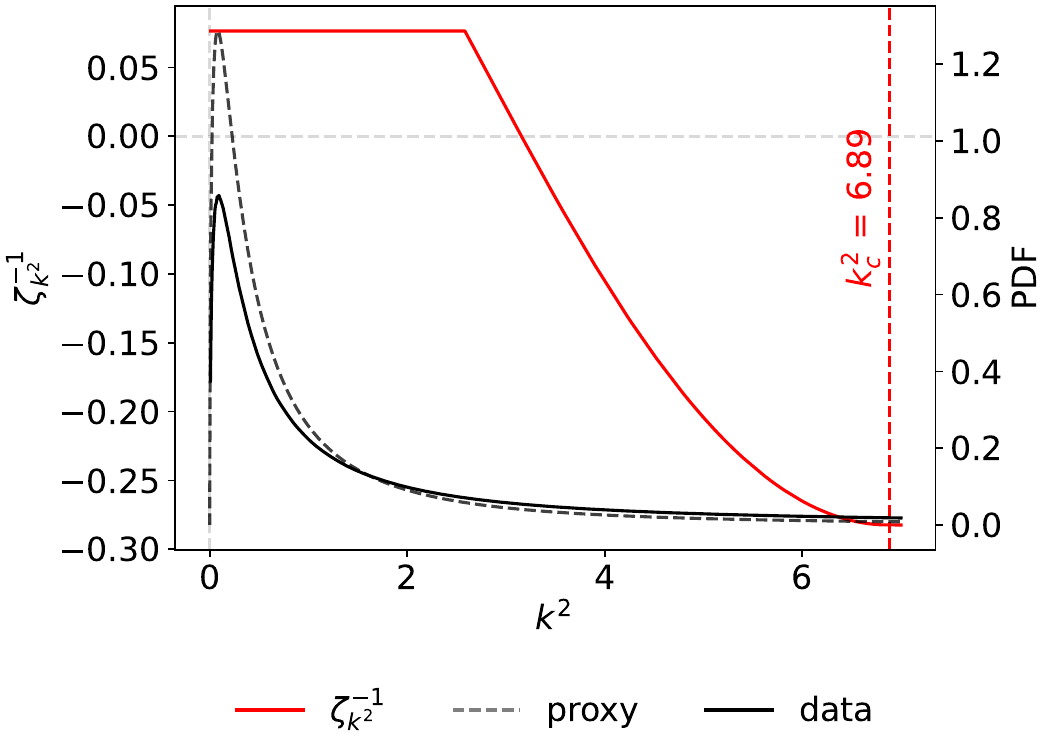}
    \end{minipage}
    \hfill
    \begin{minipage}[b]{0.45\textwidth}
        \centering
        {\large $\beta = 0.37$} \\[0.5em]
        \includegraphics[height=0.19\textheight]{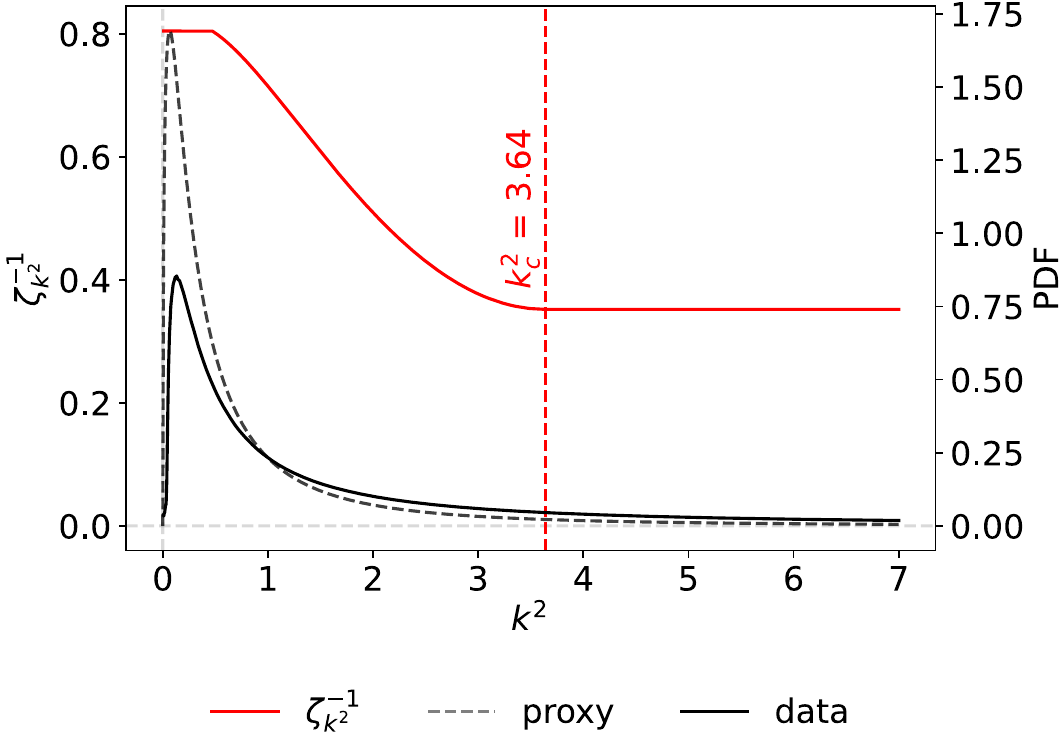}
    \end{minipage}
    \\[1em] 

    \begin{minipage}[b]{0.45\textwidth}
        \centering
        {\large $\beta = 0.50$} \\[0.5em]
        \includegraphics[height=0.19\textheight]{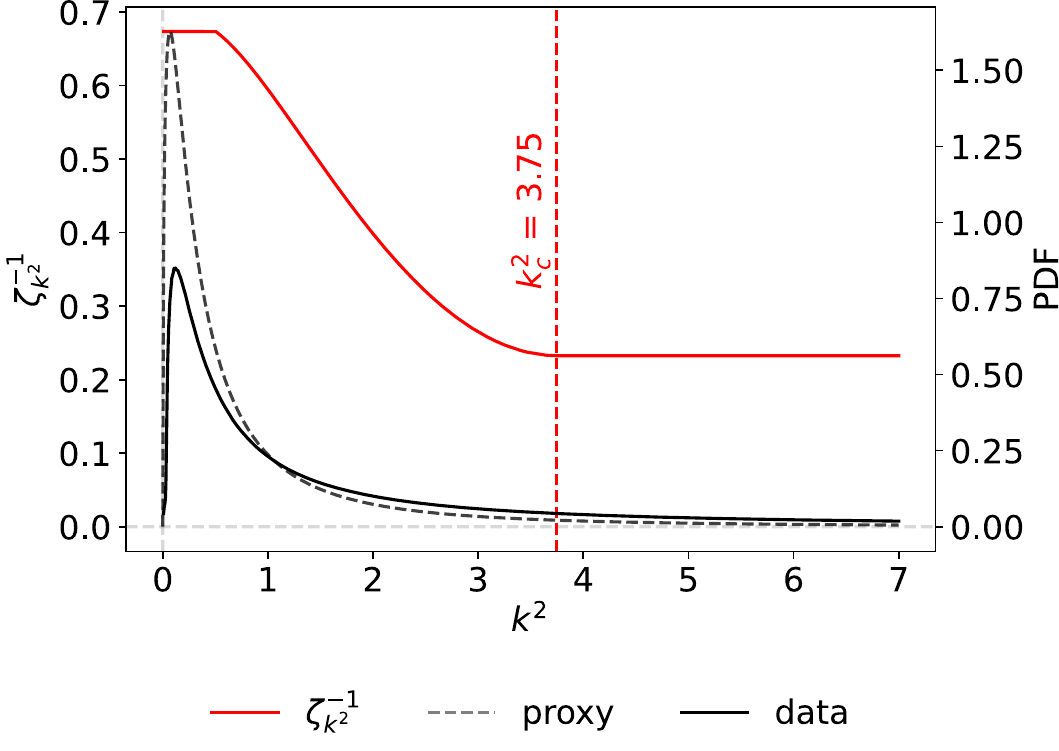}
    \end{minipage}

    \caption{%
        Values of the local inverse adherence (axis on the left, red curve) and spectra (axis on the
        right) for different values of the \snr $\beta$.
    }\label{fig:local_adherence}
\end{figure}

This definition fits the idea we have of statistical ensemble fluctuations.
Indeed, we expect (and the empirical results of the previous sections explicitly show this) that the fluctuations oscillate around the proxy, which we observe at sufficiently high energy \uv scales (for fairly small $\lambda$).
So if we choose $\lambda$ in this region, $\zeta_{<\lambda}(\mu)$ will be essentially zero.
Conversely, when the global trend drags the empirical spectrum far enough from the proxy, so that the probability that the fluctuations cross the curve vanish at a certain value $\lambda_c$, $\zeta_{<\lambda}(\mu) \neq 0$ as long as the signal scale is well above the typical fluctuation scale (see the discussion in Section~\ref{sec:sectionVS}).
Figure~\ref{fig:illustration} illustrates this idea qualitatively, with the function $\zeta_{<\lambda}(\mu)$ becoming significantly different from zero from $\lambda > \lambda_c$.
This $\lambda_c$ is thus a way to approximate the decoupling cut $\Lambda$ where the empirical distribution starts to strongly differ from the proxy.
Another more pragmatic way to construct this cut, discussed in~\cite{RG3,RG4}, is to define a $\lambda_c^\prime$ at the point where $\dim_{\tau}\qty(u_4) = 0$, and these two values are generally different, although they represent a significant deviation from the class of \mpdistr distributions.
We will therefore set:
\begin{equation}
    \Lambda= \min (\lambda_c,\lambda_c^\prime).
\end{equation}

Finally, let us also mention the possibility of defining a notion of absolute Gaussian distance, which we will call \emph{inverse Gaussian distance}.
This idea exploits the fact that by construction, the passage from the empirical distribution $\mu$ to $\rho$ essentially masks the limits of the distribution, since in the continuous limit, the interval $(\lambda_-,\lambda_+)$ is mapped on $(0,+\infty)$, opening the possibility of comparing distributions associated with different supports:
\begin{definition}
    Let $\rho_1(p^2)$ and $\rho_2(p^2)$ be two inverse distributions.
    We define the local inverse Gaussian distance at scale $k^2$, $g_{k^2}(\rho_1,\rho_2)$, and the local inverse adherence $\zeta_{k^2}^{-1}(\rho_1)$ as:
    \begin{equation}
        g_{k^2}(\rho_1,\rho_2) \coloneqq \max_{(0,k^2)} \, \Big\vert  \dim_{\tau}\qty(u_4)\big\vert_{\rho_1} - \dim_{\tau}\qty(u_4)\big\vert_{\rho_2} \, \Big\vert,
    \end{equation}
    \begin{equation}
        \zeta_{k^2}^{-1}(\rho_1)\coloneqq  \min_{(0,k^2)} \, \Big( \dim_{\tau}\qty(u_4)\big\vert_{\rho_*} - \dim_{\tau}\qty(u_4)\big\vert_{\rho_1} \, \Big).
    \end{equation}
    where $\rho_*$ is the inverse of the proxy for $\rho_1$.
\end{definition}
Figure~\ref{fig:local_adherence} shows the behaviour of $k^2_c$ (i.e.\ the quantity corresponding to $\lambda_c$ for the momenta distribution $\rho$) for different values of $\beta$ (we used the realistic image in Figure~\ref{fig:gianduja}).
The \mpdistr distribution proxy $\rho_*$ has been computed using the inverse distribution $\rho$, in order to best match the empirical momenta distribution in the explored window $k^2 \in [0, 7]$, explored numerically.
As visible in the plots, for values of $\beta$ corresponding to the presence of the most intense signal (see Figure~\ref{fig:figplotgianduja}), the empirical distribution decouples from the \mpdistr distribution proxy already at \uv scales with an intense $\zeta^{-1}_{k^2}$.
The distribution for values of \snr corresponding to weaker signals seemingly decouple farther in the \uv, though the intensity of the local inverse adherence remains smaller and possibly close to simple statistical fluctuations.
Finally, the asymptotic values of $\zeta_{k^2}^{-1}$  in the \uv do not vanish as a consequence of the support of the eigenvalue distribution being mapped from $\qty[\lambda_-, \lambda_+]$ to $[0, +\infty)$ when considering momenta: the distance between the distribution and its proxy are spanned over an infinitely long interval.


Finally, we note that while the dimension of the quartic coupling $u_4$ provides a robust order parameter in the \ir, higher-order couplings (like $u_6$) are more sensitive to local fluctuations.
The canonical dimensions depend on the logarithmic derivative of the spectral density ($\dv{\ln \rho}{t}$).
Consequently, these dimensions can become ill-defined or singular in regions where the spectral density vanishes (spectral gaps) or changes abruptly.
This makes high-order couplings more sensitive probes of the \uv microstructures.

\subsection{Estimating the independent components for data noises}\label{sec:largebeta}

\begin{figure}[t]
    \centering
    \includegraphics[width=0.45\textwidth]{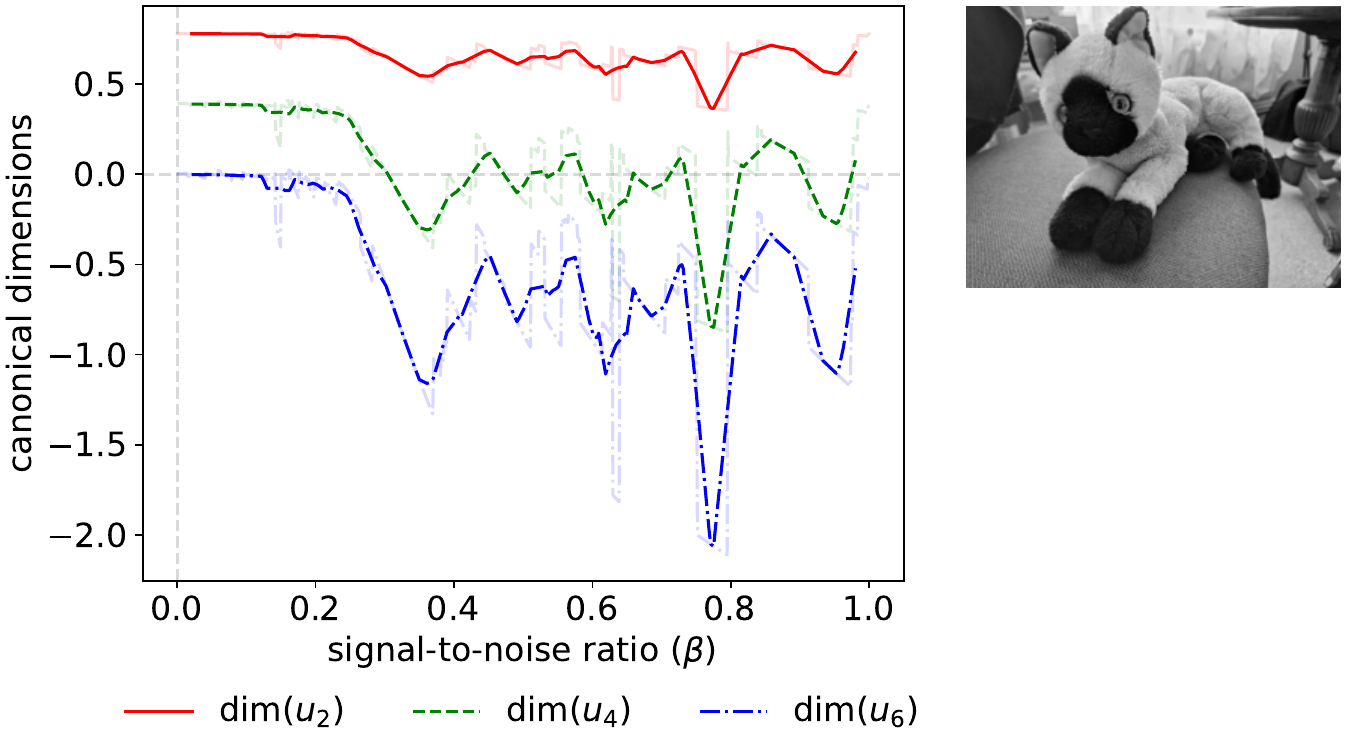}
    \hfill
    \includegraphics[width=0.45\textwidth]{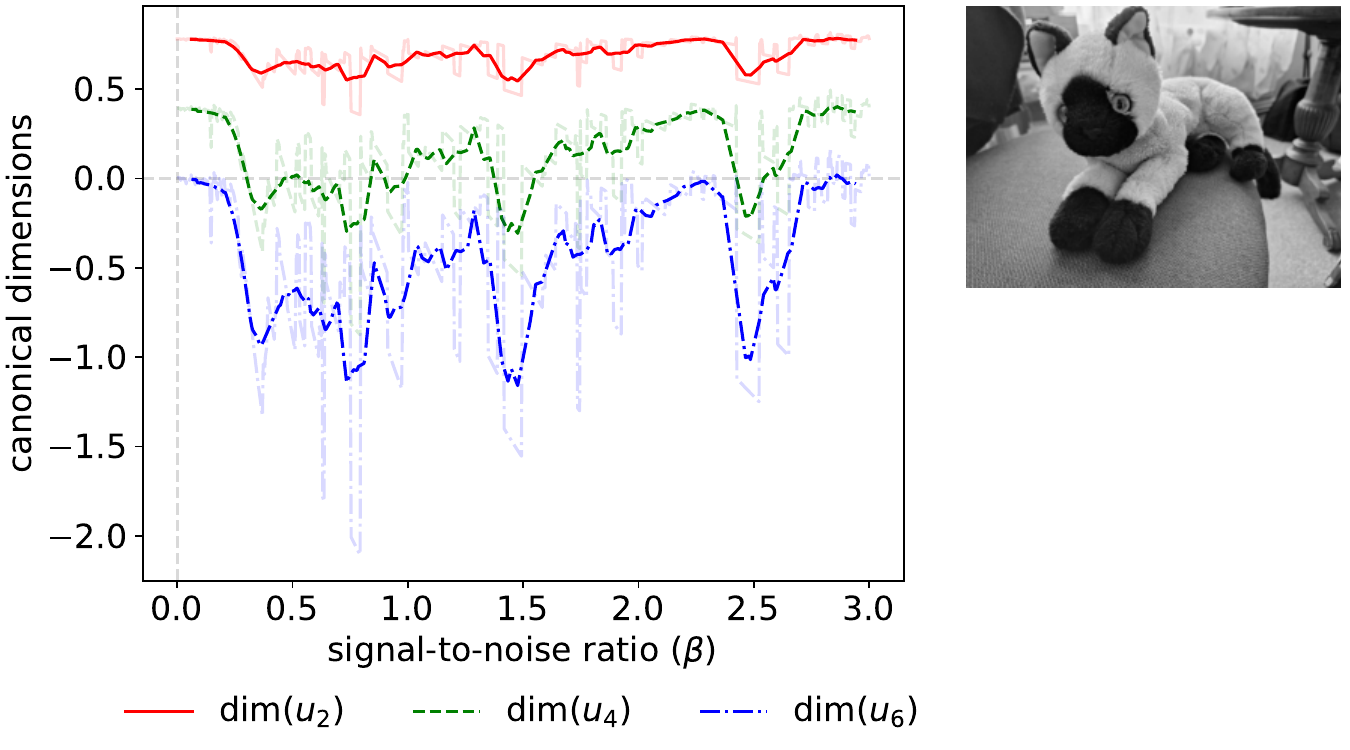}\\
    \includegraphics[width=0.45\textwidth]{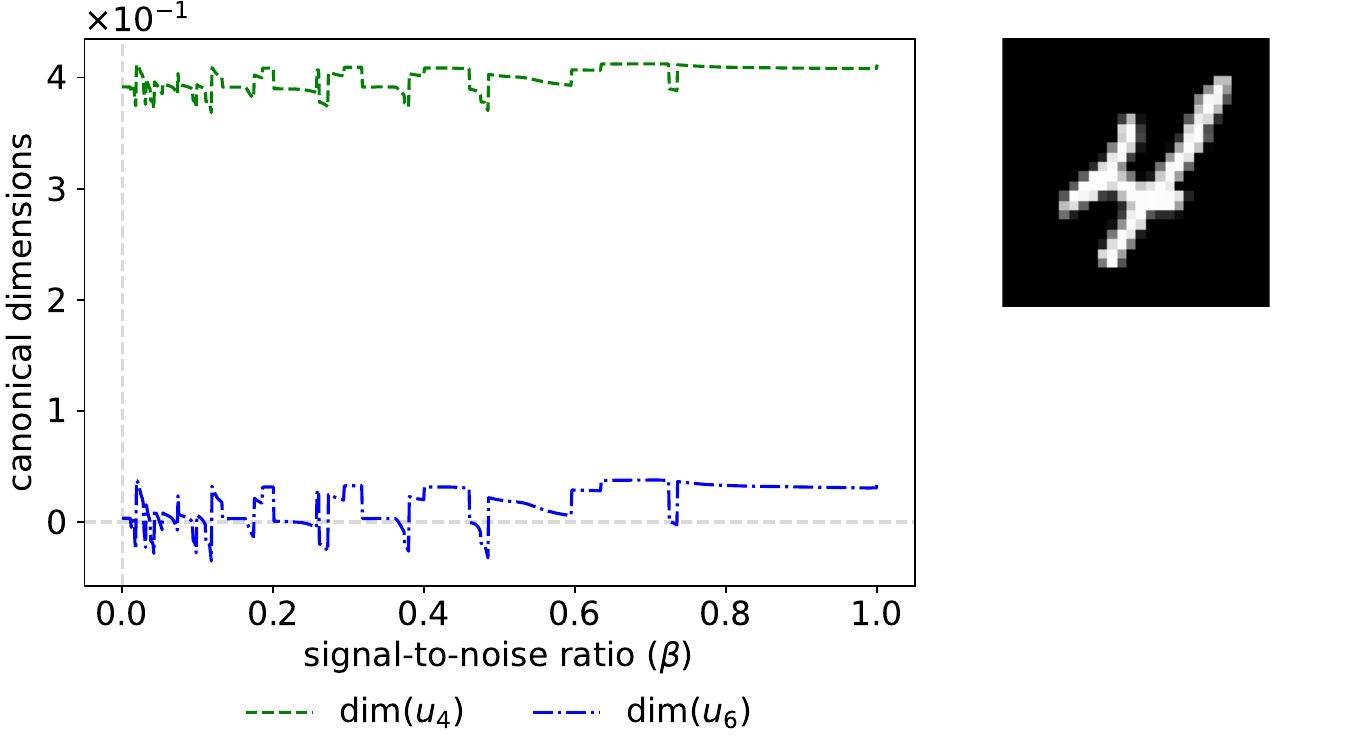}
    \hfill
    \includegraphics[width=0.45\textwidth]{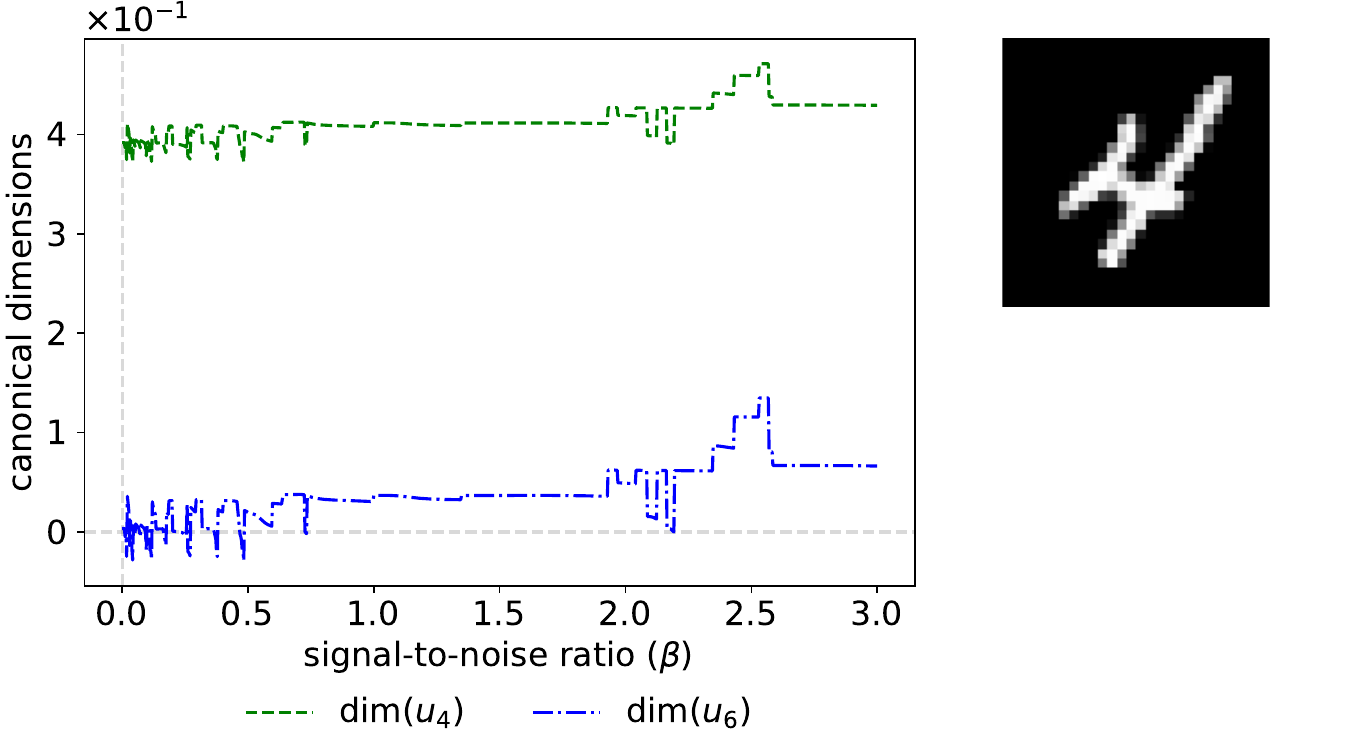}
    \caption{%
        Values of the canonical dimensions for a realistic image and a handwritten digit.
    }\label{fig:can_dim_mnist}
\end{figure}

The behaviour previously observed numerically for the symmetric phase region can be quantified by looking at other markers of the presence of signal.
In Section~\ref{sec:mainresults}, we discussed the existence of a cyclic phenomenon, and we return to that in this section.
Figure~\ref{fig:can_dim_mnist} shows the canonical dimensions at scale $k^2_{\text{IR}}$ for a realistic image and for one of the handwritten digits:
\begin{enumerate}
    \item for the latter, the conclusions follow those we gave in Section~\ref{sec:mainresults}: the dimension of $u_4$ never vanishes, and oscillates around the analytic value given by the proxy. According to our criteria, no signal is therefore quantifiable in the spectrum, which seems confirmed by the statistical properties of the eigenvectors. The isolated spikes capture most of the information, though the remnants are nonetheless detectable in the remaining bulk;
    \item for the realistic picture in Figure~\ref{fig:can_dim_mnist}, on the other hand, the behaviour is more interesting: after passing the first threshold $\beta_O$, the canonical dimensions increase again, up to a new maximum, then decrease again, and this phenomenon occurs following irregular cycles, producing a series $\qty{\beta_O^{(1)},\beta_O^{(2)},\dots, \beta_O^{(M_0)}}$ for some $M_0$. This phenomenon continues up to a certain \snr limit $\beta_L$, from which the canonical dimensions resume their oscillations around the analytic values given by the \mpdistr distribution proxy. At this scale, the Gaussian matrix $Z$ decouples from the signal.
\end{enumerate}

One possible interpretation of these two distinct regimes might be related to the definition of the noise (background) distribution already contained in the image.
As we define our additive model~\eqref{eq:additive_model}, the image used as $S$ can be further decomposed as:
\begin{equation}
    S = S_0 + \sum\limits_{i = 1}^M \tilde{S}_i\qty(\omega_i).
    \label{eq:signal_decomp}
\end{equation}
The decomposition models a hierarchy of contributions to the overall system: $S_0$ can be considered as the primary signal (possibly composed by multiple spikes and nearly continuous spectra of eigenvalues), while $\qty{\tilde{S}_i}_{i = 1, 2, \dots, M}$ account for systematic effects (e.g.\ sensor response, systematic uncertainties, background gradients, etc.).
Parameters $\omega_i$ can be seen as \emph{confounders}, connected to the presence of these spurious sources.
We thus propose to consider~\eqref{eq:signal_decomp} in~\eqref{eq:additive_model} and define a new refined extensive-rank model:
\begin{equation}
    Y
    =
    \beta S_0 + \qty(Z + \beta \sum\limits_{i = 1}^M \tilde{S}\qty(\omega_i))
    =
    \beta S_0 + \tilde{Z}_M\qty(\beta),
\end{equation}
where the ``new'' background distribution depends crucially on $\beta$ and the number of \emph{noise components}.
Consequently, we propose to estimate the number of independent noise components $M$ by counting the number of distinct cycles $M_0$ observed in the canonical dimension flow as a function of $\beta$.
Each cycle corresponds to a subset of eigenvalues exiting the bulk, marking the decoupling of a specific ``confounder'' layer from the noise background.

\begin{figure}[t]
    \centering
    \includegraphics[width=0.45\textwidth]{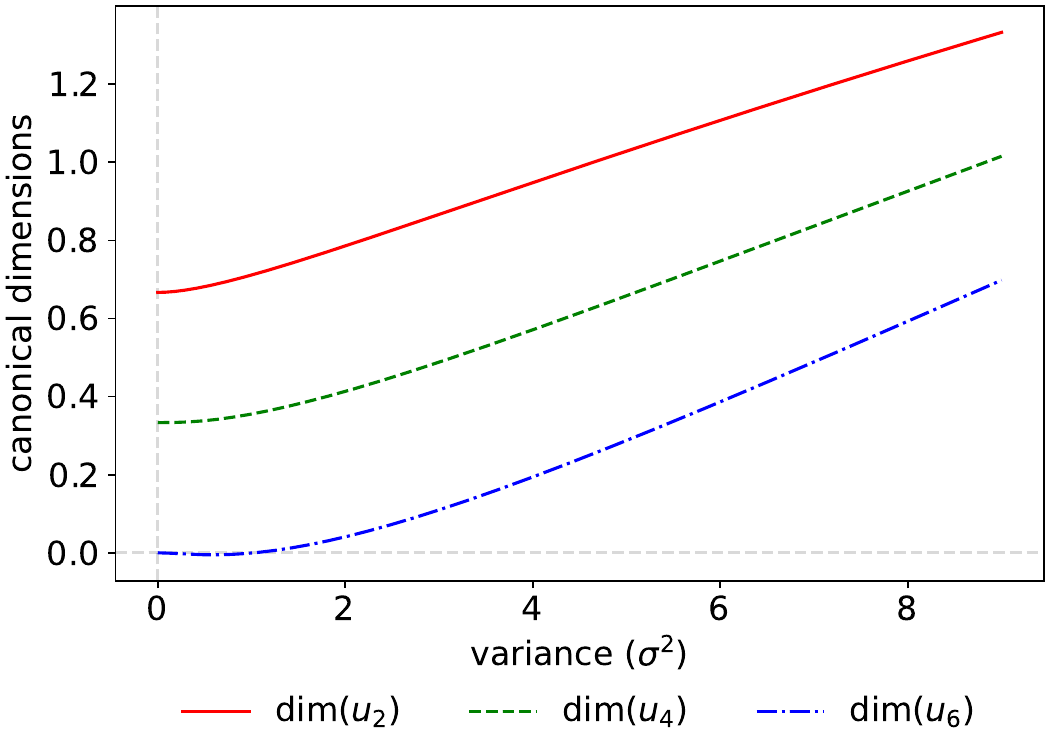}
    \quad
    \includegraphics[width=0.45\textwidth]{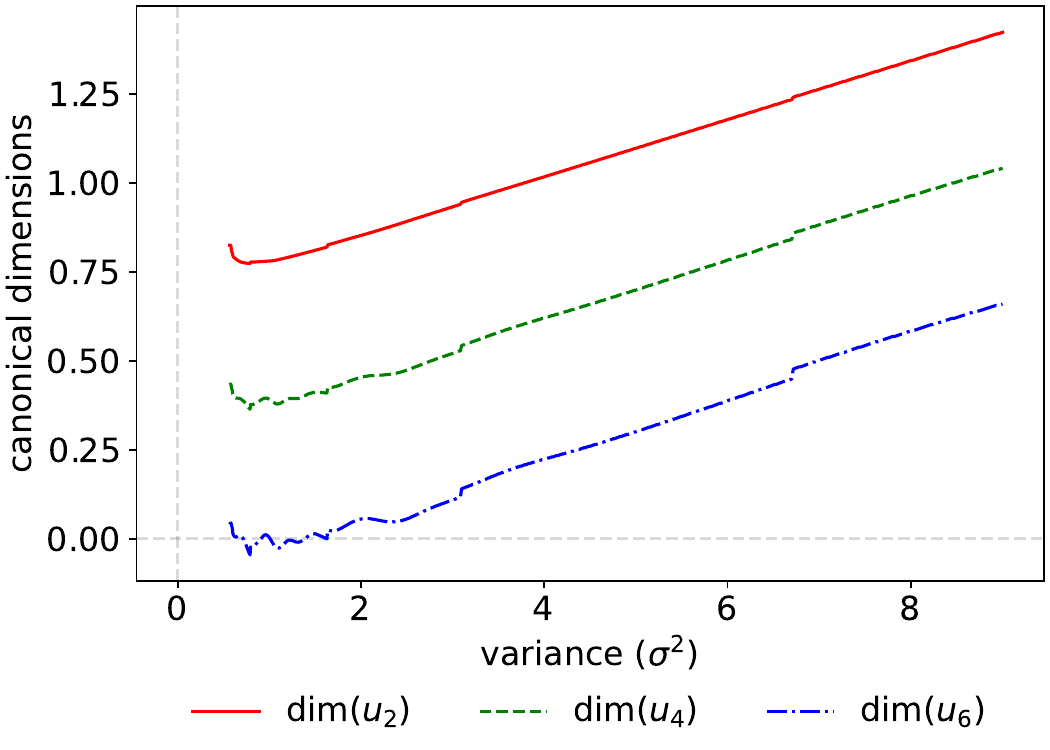}
    \caption{%
    Canonical dimensions at the scale $k^2_{\text{IR}}$ as a function of the variance parameter, in the \mpdistr distribution (left) and for an empirical sample of the \mpdistr distribution (right).
    }\label{fig:can_dim_var}
\end{figure}

Figure~\ref{fig:can_dim_mnist} shows an estimation of the presence of the different $\tilde{S}_i < S_0$.
Since we consider only the bulk distribution of eigenvalues, and not the spikes, starting from a sensible value of $\beta$, the presence of constant values represents the presence of a different source of confounding variables.
This observation is enforced by the fact that the canonical dimensions might become more irrelevant when the signal source is actually normally distributed, that is $\beta > 0$ is large enough that all spikes are no longer in the bulk distribution of eigenvalues: this boils down to an additive model~\eqref{eq:additive_model}, where the eigenvalue distribution of $S$ follows an empirical \mpdistr distribution with variance $\sigma^2$.
This implies $\operatorname{Var}\qty(X) = 1 + \beta^2 \sigma^2 > 1$.
The net effect on the canonical dimensions is a delay in the descent of the value, as shown in Figure~\ref{fig:can_dim_var}, explaining why we observe that canonical dimensions increase as spikes exit the bulk.
This is particularly visible in the case of the handwritten digit, which only contains very weak remnants of the signal.
The realistic image seems, more naturally, to rejoin the values of a usual \mpdistr distribution.

The mechanism can be summarised as follows, as a function of the \snr $\beta$:
\begin{enumerate}
    \item for low values of $\beta$, only the largest spikes of $S_0$ exit the empirical bulk distribution (close enough to the \mpdistr distribution), and we can detect the presence of signal only if the bulk distribution is affected;
    \item at a given value of $\beta$, all spikes of $S_0$ exit the bulk distribution, leaving only the distribution of the eigenvalues related to $\tilde{S}_{i \ge 1}$;
    \item still increasing $\beta$, we encounter values for which the largest spikes of $\tilde{S}_i$ (usually, low rank, hence only few spikes) become detectable by \pca, leaving only a weak intensity distribution inside that of the original $Z$;
    \item for large ranges of values of $\beta>\beta_L$, these low intensity distributions decouple from the bulk, and remain undetectable, thus only affecting the global behaviour of the bulk distribution of $Z$ (see the following for additional details);
    \item sudden changes in the behaviour of the canonical dimensions are related to groups of spikes exiting the bulk distribution according to 3.\ and 4.\ in this list.
\end{enumerate}
The mechanism is therefore related to the presence of multiple independent sources of noise in the data, each characterised by a different exit threshold from the bulk distribution.
In turn, this generalises the notion of ``bimodal connected phase'' in~\cite{Landau2023} to a multimodal phase, where each mode corresponds to a different independent noise component.
In this scenario, the estimator $M_0$ provides a (probably pessimistic) estimate for the integer $M$ quantifying the number of intrinsic sources of noise in the data.
A determination of $M$ remains difficult, as it would theoretically imply to scan for all possible values of the \snr, which, however, is not bounded by an upper value at this stage:

\begin{equation}
    M_0\leq M.
\end{equation}

However, let us return to the discussion about the relative entropy threshold in~\eqref{eq:boundentropy}, as this is not an increasing function of $\beta$, but rather an oscillating function:
\begin{remark}
    The validity of this multi-component detection relies on a specific stability criterion.
    For each identified component $i$, the detection is considered valid only if the relative entropy (i.e.\ the \kl divergence) remains bounded by~\eqref{eq:boundentropy} throughout the entire dimensional crossover interval $[\beta_t^{(i)}, \beta_O^{(i)}]$.
    This ensures that the system remains within the basin of attraction of the reference universality class long enough for the dimensional transition to be physically meaningful.
\end{remark}

\section{Conclusion and Open Issues}\label{sec:conclusions}

This work establishes the \frg as a robust framework for analysing the spectral geometry of extensive-rank signals, a regime where standard outlier detection fails.
The \frg proves to be a well-established and powerful method for probing the structure of data in realistic scenarios, where signals are not isolated spikes but are mixed with the noise bulk.
Our approach leverages rigorous theoretical arguments supported by extensive numerical evidence to characterise the detection mechanism in this nearly continuous regime.
Specifically, we have defined a set of canonical dimensions whose stability serves as a sensitive order parameter for the underlying spectral deformation.
We showed that these dimensions undergo a sharp crossover at a critical signal-to-noise ratio $\beta_t$, identifying the onset of the ``bimodal connected phase'' predicted by random matrix theory~\cite{Landau2023}.
This detection occurs well below the standard \bbp threshold, confirming that the \frg is capable of recovering signal information ``in the bulk'' before any spectral gap opens.
In this sense, our work provides a field-theoretic dual description of the topological phase transition associated with extensive-rank signals.

Our results are not merely numerical artefacts but are grounded in a consistent physical picture.
We first established the theoretical framework, identifying the relevant operators for the universality class of the noise (Marchenko-Pastur), and then provided numerical support for these claims.
Crucially, our findings do not depend on arbitrary truncation choices.
As proved in previous works~\cite{RG5,RG6}, the universality class of the noise dictates that only the quartic ($u_4$) and sextic ($u_6$) couplings are relevant or marginal in the \ir.
Including higher-order interactions (e.g., $u_8$ or $u_{10}$) would be redundant, as these operators are irrelevant in the \rg sense and thus cannot serve as effective indicators of the transition.
This ensures that the dimensional phase transition we observe is a robust feature of the effective field theory, not a byproduct of the approximation scheme.

Furthermore, we provided a comprehensive set of theoretical arguments to verify that the \frg indeed detects the presence of signal as a deformation in the infrared: the stability of the Wilson-Fisher fixed point, the contraction of the symmetric phase volume, and the deviation of eigenvector statistics from the universal Porter-Thomas distribution.
We supported these theoretical claims with multiple independent numerical probes.
The consistent cyclic behaviour of the flow in the presence of complex noise also allowed us to propose a heuristic criterion for estimating the number of independent noise components, offering a novel tool for interpreting different sources in complex datasets.

In the future, efforts should be invested in extending this formalism to other universality classes.
Moreover, the field theory we are considering has a rare feature: the 2-point function is known exactly from the data, which suggests considering an inverse flow formalism to reconstruct the bare action~\cite{Berman_2024,NNQFT1}.
This becomes particularly interesting for possible applications in \ai, as recent work~\cite{masuki2025generativediffusionmodelinverse} on generative diffusion models hints at deep connections between inverse \rg flows and data generation.
Theoretical efforts should also focus on capturing the global momentum dependence of vertex functions, which could refine the detection thresholds further, as power counting arguments seem to suggest.
Finally, recent work~\cite{lahoche2024functional,lahoche2024functional2} suggests that a matrix field theory incorporating non-locality could emulate the matrix vacuum itself, potentially removing the need to postulate a reference distribution and paving the way for a fully self-contained field theory of data.

\section*{Acknowledgements}
The authors acknowledge support from the COMETA COST Action \href{https://www.cost.eu/actions/CA22130/}{CA22130}.

\section*{CRediT Author Statement}
\textbf{Riccardo Finotello} Conceptualization, Data curation, Formal analysis, Investigation, Methodology, Software, Validation, Visualization, Writing --- original draft, Writing --- review \& editing; %
\textbf{Vincent Lahoche} Conceptualization, Formal analysis, Investigation, Methodology, Validation, Writing --- original draft, Writing--- review \& editing; %
\textbf{Dine Ousmane Samary} Conceptualization, Formal analysis, Investigation, Methodology, Validation, Writing --- original draft, Writing --- review \& editing.

\section*{References}
\printbibliography[heading=none]

\end{document}